\documentclass[aps,pra,groupedaddress,nofootinbib,longbibliography]{revtex4-1}
\pdfoutput=1 

\usepackage[utf8]{inputenc}
\usepackage{amsmath}
\usepackage{amsfonts}
\usepackage{amssymb}
\usepackage{color}
\usepackage{graphicx}
\usepackage{siunitx}
\usepackage[capitalize]{cleveref} 
\usepackage[version=4]{mhchem}
\usepackage{float}
\usepackage{subfig}
\usepackage{mathrsfs}
\usepackage{comment}
\usepackage{tikz}
\usetikzlibrary{positioning}

\graphicspath{{./figures/}} 

\begin{document}

\title{A data-based, reduced-order, dynamic estimator for reconstruction of non-linear flows exhibiting limit-cycle oscillations}
\author{Juan Guzm\'an-I\~nigo} 
\email[]{j.guzman-inigo@imperial.ac.uk}
\affiliation{Dept. of Mechanical Engineering, Imperial College, London SW7 2AZ, UK}
\author{Markus Sodar}
\affiliation{Dept. of Aeronautics, Imperial College, London SW7 2AZ, UK}
\author{George Papadakis}
\email[corresponding author, ]{g.papadakis@imperial.ac.uk}
\affiliation{Dept. of Aeronautics, Imperial College, London SW7 2AZ, UK}
\begin{abstract}
We apply a data-based, linear dynamic estimator to reconstruct the velocity field from measurements at a single sensor point in the wake of an aerofoil. In particular, we consider a NACA0012 airfoil at $Re=600$ and $\ang{16}$ angle of attack. Under these conditions, the flow exhibits a vortex shedding limit cycle. A reduced order model (ROM) of the flow field is extracted using proper orthogonal decomposition (POD). Subsequently, a subspace system identification algorithm (\texttt{N4SID}) is applied to extract directly the estimator matrices from the reduced output of the system (the POD coefficients). We explore systematically the effect of the number of states of the estimator, the sensor location, the type of sensor measurements (one or both velocity components), and the number of POD modes to be recovered. When the signal of a single velocity component (in the stream wise or cross stream directions) is measured, {the reconstruction of the first two dominant POD modes} strongly depends on the sensor location. We explore this behaviour and provide a physical explanation based on the non-linear mode interaction and the {spatial distribution of the modes}. When however, both components are measured, the performance is very robust, and is almost independent of the sensor location when the optimal number of estimator states is used. {Reconstruction of the less energetic modes is more difficult, but still possible.}
\end{abstract}

\maketitle

\section{\label{sec:Introduction}Introduction}


Flow reconstruction (or estimation) from limited measurements has a very wide range of applications, for example in active flow control (drag reduction or enhanced mixing and reaction), cardiovascular flows (extraction of shear stress patterns that are known to affect the behaviour of endothelial cells and the inception of atherosclerosis), optimisation of energy extraction etc. The area is vast and below only some characteristic approaches/directions are briefly sketched, in order to place the present contribution in context. 

Due to the large spatial dimensionality of 2D and 3D flows, reduced order models that extract the dominant structures have been naturally employed for flow reconstruction. For example, models based on Proper Orthogonal Decomposition (or POD) \cite{holmes_lumley_berkooz_rowley_2012} have been widely applied. Using a set of sensors embedded in the flow, the objective is to compute the coefficients of POD modes so as to minimise the error (usually defined as the $\mathcal{L}_2$-norm) between the true and estimated flow field. Willcox \cite{Willcox_2006} applied an extension of POD, called Gappy POD \cite{Everson_Sirovich_1995} to find the optimal placement of the sensor points. Yildrim et al (2009) \cite{YILDIRIM_el_al_2009} compared Gappy POD as well as a more intuitive approach (placement of sensors at the peaks of modes) and found that the latter works very well. \citet{SEMAAN_2017} computed optimal sensor placements using machine learning. Other reduced order modelling strategies, such as the resolvent analysis \cite{mckeon2010critical} have been used for estimation of laminar (but also turbulent) flows using limited measurements \cite{beneddine2017unsteady, gomez2016estimation, Thomareis_Papadakis_2018, illingworth_monty_marusic_2018}. More recent trends use sparse representation and compressive sensing \cite{Brunton_et_al_2014, Manohar_et_al_2018}. The linear stochastic estimation (LSE) method of Adrian can also estimate the velocity field at one point given observations (such as velocity \cite{Adrian_1979} or velocity and deformation tensor \cite{adrian_moin_1988}) at nearby points. The connection between LSE and POD is detailed in \cite{holmes_lumley_berkooz_rowley_2012}. 

All the previous approaches are considered as static estimators. The coefficients of the reduced order models (or libraries in the case of sparse representation) are obtained by solving a minimisation problem (usually based on least squares, but other norms have also been employed), however the history (dynamic) effect of the coefficients is not considered. For dynamic estimation, the Kalman filter is the standard approach for linear problems (refer to the book of Kailath et al \cite{Kailath_Hassibi_Sayed_2000} for a very lucid and detailed exposition). The filter assumes that the linear system matrices are known, and formulates an optimisation problem that minimises $\mathcal{L}_2$-norm between the true and the estimated output. The formulation results into  an algebraic Riccati equation that provides the gain matrix of the estimator. This matrix together with the available measurements are used to force the linear system and extract the estimated states that best reconstruct the output. {This approach was used for instance by Gong et al.~\cite{gong2019model} to successfully estimate the unsteady flow past a cylinder}. The extended Kalman filter can be used for estimation of non-linear problems \cite{Stengel_1994}.

In many practical problems however, the system matrices are unknown, and only input-output data are available. In such cases, a different approach based on system identification algorithms \cite{ljung1999system} can be applied. These algorithms extract the unknown coefficients 
of underlying mathematical models using only the available input-output data. {System identification has been successfully used to the describe the dynamics between one sensor (upstream measurement) to another sensor (downstream measurement) in noise amplifier flows and for linear dynamics. Those models were used to design compensators to effectively estimate and 
reduce the perturbation field in such flows~\cite{herve2012physics,juillet2013control,juillet2014experimental,gautier2014feed}.  
\citet{guzman2014dynamic}\cite{guzman2016recovery} extended the previous approaches to capture the dynamics between upstream measurements and the {\it{entire}} perturbation field in a laminar 
boundary layer. To do this, the authors applied the \texttt{N4SID} algorithm \citep{van94}, 
an algorithm from the subspace identification family~\cite{qin06}. This group of algorithms extract 
the matrices of a (linear) state-space estimator model, using only the available input-output data.} Most importantly, it can also return the optimal number of states of the estimator. The algorithm has been applied successfully to many industrial problems (see examples in the book \cite{Overschee_deMoor_1996}) but its performance on non-linear fluid mechanics problems has yet to be explored (at least to the best of the authors' knowledge). The deep connection between  \texttt{N4SID} and the Kalman filter is also explored in \cite{Overschee_deMoor_1996}.  

Dynamic estimation that retains the non-linearity of the underlying process is much more difficult, but progress has been made recently. Loiseau et al \cite{loiseau2018sparse} use sparse identification to extract an observer that captures the non-linear dynamics of growth and saturation of instabilities in the wake of a circular cylinder. The estimator uses measurements of the lift coefficient and a 2 equation, non-linear, dynamical system is derived. Another approach is based on lifting the original non-linear dynamical system in a higher dimensional space where the evolution is approximately linear (in an uncontrolled system this procedure amounts to numerical approximation of the Koopman operator associated to the non-linear dynamics, see \cite{korda2018linear}). The discrete empirical interpolation method has also been used to nonlinear model order reduction \cite{chaturantabut2010nonlinear, fosas2016nonlinear}. 

While the above non-linear identification approaches are very promising, they are not as well developed as linear ones. The central objective of the present paper is to assess the performance of the linear dynamical identification algorithm \texttt{N4SID} in the flow around a NACA 0012 airfoil. We consider the estimation of the fully developed vortex shedding state, i.e.\ the limit-cycle, using a single sensor  placed at different locations in the wake. 

The paper is organised as follows. In section \ref{sec:Theoretical framework} we provide the theoretical framework of the identification, section \ref{sec:flow configuration} describes the flow configuration, while results are presented 
in sections \ref{sec:results} and \ref{sec:parameters}. We conclude in section \ref{sec:conclusions}.






\section{\label{sec:Theoretical framework}Theoretical framework}
In this section we present a linear model to estimate the dynamics of a fully  non-linear flow. The equations governing a reduced-order model based on the perturbation dynamics around the time-average (mean) flow are introduced and subsequently used to justify the structure of the data-based, linear dynamic estimator. Finally, a linear system identification algorithm is introduced that can determine efficiently the matrices of the estimator directly from the observed input-output data via a statistical learning process.

\subsection{Reduced Order Model based on the perturbation dynamics about a mean flow}\label{sec:ROM}
We consider the incompressible Navier-Stokes equations
\begin{equation}
\frac{\partial \mathbf{u}}{\partial t} + \mathbf{u}\cdot \nabla \mathbf{u} + \nabla p - \frac{1}{Re} \nabla^2 \mathbf{u}=0, \qquad
\nabla \cdot \mathbf{u}=0, 
\label{eq:Navier-Stokes}
\end{equation}
\noindent where $\mathbf{u}$ is the velocity vector, $p$ the pressure and $Re$ the Reynolds number. The flow variables are decomposed into a time-average (denoted by overbar) and a fluctuating part (denoted by prime); e.g. $\mathbf{u}= \overline{\mathbf{u}} + \mathbf{u}'$. Applying the time-averaging operation to Eq.~\eqref{eq:Navier-Stokes} results in
\begin{equation}
\overline{\mathbf{u}}\cdot \nabla \overline{\mathbf{u}} + \nabla \overline{p} - \frac{1}{Re} \nabla^2 \overline{\mathbf{u}}= - \overline{\mathbf{u}'\cdot \nabla \mathbf{u}'}, \qquad
\nabla \cdot \overline{\mathbf{u}}=0,
\label{eq:Mean Navier-Stokes}
\end{equation}
which governs the steady, mean flow. The equations governing the fluctuation part are obtained by subtracting Eq.~\eqref{eq:Mean Navier-Stokes} from Eq.~\eqref{eq:Navier-Stokes} to give
\begin{subequations}
\begin{align}
\frac{\partial \mathbf{u}'}{\partial t}  + \overline{\mathbf{u}}\cdot \nabla \mathbf{u}' + 
\mathbf{u}'\cdot \nabla \overline{\mathbf{u}} + \nabla p' - \frac{1}{Re} \nabla^2 \mathbf{u}'&=\overline{\mathbf{u}'\cdot \nabla \mathbf{u}'} - \mathbf{u}'\cdot \nabla \mathbf{u}', \\
\nabla \cdot \mathbf{u}'&=0, \label{eq:Fluctuation divergence}
\end{align}
\label{eq:Fluctuation Navier-Stokes}
\end{subequations}
\noindent which has been arranged so that the left-hand side is linear in the fluctuation variables. Treating the nonlinearities as a forcing term, Eq.~\eqref{eq:Fluctuation Navier-Stokes} can be rewritten in the classical state-space form
\begin{equation}\label{eq:State Space}
\mathscr{Q} \frac{\partial \mathscr{X}}{\partial t} + \mathscr{A} \mathscr{X} = \mathscr{F},
\end{equation}
where
\begin{subequations}
\begin{gather}
\mathscr{X} = \begin{pmatrix} \mathbf{u}' \\ p' \end{pmatrix},             \\
\mathscr{Q} = \begin{pmatrix} \mathscr{I} & 0 \\ 0 & 0 \end{pmatrix},      \\
\mathscr{A} = \begin{pmatrix} \overline{\mathbf{u}}\cdot \nabla () + 
() \cdot \nabla \overline{\mathbf{u}} - Re^{-1} \nabla^2 & \nabla ()        \\
\nabla \cdot () & 0 \end{pmatrix},                                          \\
\end{gather}
\label{eq:State Space components}
\end{subequations}
and
\begin{equation}
\mathscr{F} = \begin{pmatrix}
  \overline{\mathbf{u}'\cdot \nabla \mathbf{u}'} - \mathbf{u}'\cdot \nabla \mathbf{u}'
  \\ 0
\end{pmatrix}.
\end{equation}

We now introduce a divergence-free orthonormal basis $\left\lbrace\mathbf{\Phi}_i\right\rbrace_{1...m}$, satisfying 
\begin{equation}
\nabla \cdot \mathbf{\Phi}_i = 0 \qquad \text{and} \qquad \left\langle  \mathbf{\Phi}_i(\boldsymbol{x}), \mathbf{\Phi}_j(\boldsymbol{x})\right\rangle=\delta_{ij}, \qquad i = 1,2,...,m
\label{eq:orthonormal}
\end{equation}
where $\left\langle \cdot, \cdot \right\rangle$ is an appropriate scalar product (to be defined later) and $\delta_{ij}$ is the Kronecker delta. The velocity field can then be projected onto this basis according to
\begin{subequations}
\begin{align}
\mathbf{u}'(\boldsymbol{x}, t) &= \sum_{i=1}^{m} y_i(t) \mathbf{\Phi}_i(\boldsymbol{x}) \qquad\quad i = 1,2,...,m \\
y_i(t) &= \left\langle \mathbf{\Phi}_i(\boldsymbol{x}), \mathbf{u}'(\boldsymbol{x}, t)\right\rangle.
\end{align}
\label{eq:projection}
\end{subequations}
We define a state vector given by the first $k$ coefficients of the previous expansion \ $\mathbf{Y} = [y_1 , y_2 , ..., y_k ]^{\top}$ (where $^{\top}$ denotes the transpose) with a corresponding 
reduced basis $\mathsf{U} = [\mathbf{\Phi}_1 , \mathbf{\Phi}_2, ..., \mathbf{\Phi}_k ]$. Finally, we perform a Galerkin projection of Eq.~\eqref{eq:State Space} onto the subspace spanned by $\mathsf{U}$ to obtain
\begin{subequations}
\begin{gather}
\frac{\rm{d} \mathbf{Y}}{\rm{d} t} + \mathsf{A}' \mathbf{Y}(t) = \mathbf{F}'(t) + \boldsymbol{\epsilon}(t),
\end{gather}
where
\begin{gather}
\mathsf{A}'_{ij}=\left\langle \mathbf{\Phi}_i, \mathscr{A} \mathbf{\Phi}_j\right\rangle, \quad
\mathbf{F'}_i = \left\langle \mathbf{\Phi}_i, \mathscr{F}\right\rangle \quad \text{and} \quad
\epsilon_i=\sum_{j=k+1}^{m}y_j \left\langle \mathbf{\Phi}_i, \mathscr{A} \mathbf{\Phi}_j\right\rangle.
\end{gather}
\label{eq:tc state space}
\end{subequations}
\noindent Note that $\boldsymbol{\epsilon}$ represents the error due to the truncation of the expansion ($k<m)$.  

\subsection{Equations of the linear dynamic estimator}\label{sec:estimator}
We now consider a flow estimator of the form
\begin{subequations}
\begin{align}
\frac{\rm{d} \mathbf{X}_e}{\rm{d} t} &+ \mathsf{A}_s' \mathbf{X}_e(t) =  \mathsf{L}'\mathbf{s}(t), \\
\mathbf{Y}_e(t) &= \mathsf{C}' \mathbf{X}_e(t),\\
\mathbf{s}(t) &= \mathsf{C}_s' \mathbf{Y}(t) + \mathbf{g}(t),
\end{align}
\label{eq:tc estimator}
\end{subequations}
where $\mathbf{X}_e \in \mathbb{R}^{N_x}$ is the estimator state vector, $\mathbf{Y}_e \in \mathbb{R}^{k}$ is the estimator output and $\mathbf{s} \in \mathbb{R}^{p}$ contains measurements from $p$ sensor points placed in the flow, $\mathbf{s}=\left[ s_1, s_2, ..., s_p\right]^\top$ . In the present study, we employ sensors that are linear with respect to the velocity field, i.e.\ $s_j(t) = \mathscr{C}_j \mathbf{u}(t)+ g_j(t)$, where $g_j(t)$ is the  noise that corrupts the measurements. The linear operator $\mathscr{C}_j$ is projected onto the reduced-order basis as 
$\mathbf{C}_{s,ji} = \left\langle \mathbf{\Phi}_i, \mathscr{C}_j \right\rangle$. 

We seek to construct the estimator so that the approximate output
$\mathbf{Y}_e(t)$ is as close as possible to the true output $\mathbf{Y}(t)$. Intuitively, this will be achieved if the measurements from the sensor $\mathsf{L}'\mathbf{s}(t)$ correctly represent the non-linear term $\mathbf{F}'(t).$ Note that the sensor may also account for the truncation error
$\boldsymbol{\epsilon},$ if this is non-negligible. 

The equation governing the estimation error, defined as $\mathbf{Z}= \mathbf{Y} - \mathbf{Y}_e$, is obtained by subtracting Eq.~\eqref{eq:tc estimator} from Eq.~\eqref{eq:tc state space} to yield 
\begin{subequations}
\begin{gather}
\frac{\rm{d} \mathbf{Z}}{\rm{d} t} + \mathsf{C}'\mathsf{A}_s'\mathsf{C}'^{\dagger} \mathbf{Z}(t)
= \mathbf{F}^{\text{err}}(t),
\end{gather}
with
\begin{gather}
\mathbf{F}^{\text{err}}(t)
= \mathsf{A}^{\text{err}} \mathbf{Y}(t) + \mathbf{F}'(t) - \mathsf{C}'\mathsf{L}'\mathbf{g}(t) + \boldsymbol{\epsilon}(t), \qquad \text{and} \qquad \mathsf{A}^{\text{err}} = \left( \mathsf{C}'\mathsf{A}_s'\mathsf{C}'^{\dagger}
-\mathsf{A}'-\mathsf{C}'\mathsf{L}'\mathsf{C}_s'\right).
\end{gather}
\label{eq:error}
\end{subequations} 
The relation $\mathbf{X}_e=\mathsf{C}'^{\dagger}\mathbf{Y}_e$ was used to obtain Eq.~\eqref{eq:error} where $\mathsf{C}'^{\dagger}$ denotes the Moore-Penrose generalised matrix inverse~\cite{penrose1955generalized}. Eq.~\eqref{eq:error} shows that the term $\mathbf{F}^{\text{err}}$ must be small for the error to be small and that the eigenvalues of $\mathsf{A}_s'$ and $\mathsf{A}^{\text{err}}$ must the stable.

With the remaining article pertaining to system identification methods, it
is more convenient to express the estimator in a discrete-time framework. In
the discrete-time domain, the mapping of the state-vector $\mathbf{Y}_e$ 
from time $t$ (index $n$) to $t + \Delta t$ (index $n + 1$) reads
\begin{subequations}
\begin{align}
\mathbf{X}_e(n+1) &= \mathsf{A}_s \mathbf{X}_e(n) + \mathsf{L}\mathbf{s}(n), 
\\
\mathbf{Y}_e(n) &= \mathsf{C} \mathbf{X}_e(n), 
\\
\mathbf{s}(n) &= \mathsf{C}_s \mathbf{Y}(n) + \mathbf{g}(n),
\end{align}
\label{eq:td estimator}
\end{subequations}
\noindent where matrix ${\mathsf{L}} = \int_0^{\Delta t} \exp[-{\mathsf{A}}'_s(\Delta t-\tau)] {\mathsf{L}}' \; d\tau$ is associated with the discrete driving term, ${\mathsf{A}}_s = \exp(-{\mathsf{A}}'_s\Delta t)$
denotes the evolution matrix over a time interval $\Delta t,$
${\mathsf{C}}={\mathsf{C}}'$ and $\mathsf{C}_s=\mathsf{C}_s'$ (see also ~\cite{antoulas05}).

\subsection{System identification based on subspace techniques}\label{sec:SysIdent}
\begin{figure}
  \centering
  \begin{tikzpicture}[scale=0.8]  

\node at (-7,8.5) {Step 1:};
\node at (-7,6.4) {Step 2:};
\node at (-7,3.9) {Step 3:};
\draw[gray] (0,5.2)-- (0,1);


\node at (-3,0.5) {Learning dataset};     
\node at (2.5,0.5) {Validation dataset};   

\draw [red] (3,6) -- (4,6);
\draw [dashed,blue] (3,5.5) -- (4,5.5);
\node at (2,6) {\scriptsize Experiment};
\node at (1.6,5.5)   {\scriptsize Model};

\draw [rounded corners, thick,fill=green!20](-2,7) rectangle (0,8); 
\node at (-1,7.5) {System};
\draw [-latex](-3,7.5) -- (-2,7.5);
\node at (-2.5,7.8) {$u$};
\draw [-latex](0,7.5) -- (1,7.5);
\node at (0.5,7.8) {$y$};

        
    \draw[->] (-6,4.5) -- (4,4.5) node[right] {$t$}; 
    \draw[->] (-6,4.5) -- (-6,5.5) node[left] {$y(t)$};
    
        \draw[color=red,samples = 100, domain=0:10, variable=\x]
     plot (\x-6,{(0.5*sin((\x*2*pi) r)
                 +0.7*sin((\x*2*pi/2) r)
                 +0.4*sin((\x*2*pi/3) r)
                 +    sin(((\x)*2*pi/4) r)
                 +0.5*sin((\x*2*pi/10) r))/2 + 4.5})  node[right,above]{$y(t)$}; 
                 
                  \draw[color=blue,dashed,samples = 100, domain=0:6, variable=\x]
     plot (\x-6,{(0.5*sin((\x*2*pi) r)
                 +0.7*sin((\x*2*pi/2) r)
                 +0.4*sin((\x*2*pi/3) r)
                 +    sin(((\x)*2*pi/4) r)
                 +0.5*sin((\x*2*pi/10) r) 
                 +0.1*sin((\x*2*pi/20) r)
                 +0.05*sin((\x*2*pi*1.5) r))/2 + 4.5}); 
             
    \draw[->] (-6,2) -- (4,2) node[right] {$t$}; 
    \draw[->] (-6,2) -- (-6,3) node[left] {$y(t)$};
    
        \draw[color=red,samples = 100, domain=0:10, variable=\x]
     plot (\x-6,{(0.5*sin((\x*2*pi) r)
                 +0.7*sin((\x*2*pi/2) r)
                 +0.4*sin((\x*2*pi/3) r)
                 +    sin(((\x)*2*pi/4) r)
                 +0.5*sin((\x*2*pi/10) r))/2 + 2}); 
      

                  \draw[color=blue,dashed,samples = 100, domain=6:10, variable=\x]
     plot (\x-6,{(0.5*sin((\x*2*pi) r)
                 +0.7*sin((\x*2*pi/2) r)
                 +0.4*sin((\x*2*pi/3) r)
                 +    sin(((\x)*2*pi/4) r)
                 +0.5*sin((\x*2*pi/10) r) 
                 +0.3*sin((\x*2*pi/20) r)
                 +0.08*sin((\x*2*pi*1.5) r))/2 + 2}) node[right,above]{${y}_e(t)$};

\end{tikzpicture}
  \caption{Procedural steps of system identification techniques. Step 1: Numerical 
  simulations/experiments are run and input $(u)$ and output data $(y)$ are acquired. Step 2: The model coefficients are computed by maximising the fit between the output of the system and the prediction of the model for part of the available data (training or learning data set). Step 3: The performance of the model is assessed on a different dataset (testing or validation dataset).}
  \label{fig:sysident}
\end{figure}
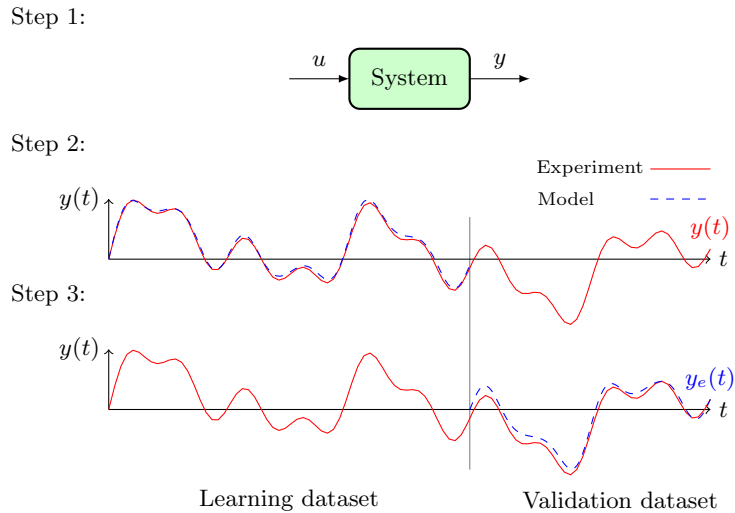


We seek to obtain a linear, time-invariant (LTI) multiple-input-multiple-output (MIMO) system, such as the one given in Eq.~\eqref{eq:td estimator}, from a sequence of observed input-output data. Subspace identification algorithms can accomplish this task. More specifically, such algorithms consider the state-space representation of a stochastic linear system written in the following {\itshape{process form}}
\begin{subequations}
  \begin{align}
    x(n+1) &= {\mathcal{A}} x(n) + {\mathcal{B}} u(n) + w(n)\\
    y(n)   &= {\mathcal{C}} x(n) + {\mathcal{D}} u(n) + v(n)
  \end{align}
  \label{eq:ProcessForm}
\end{subequations}
where $y(n) \in \mathbb{R}^{N_y}$, $x(n) \in \mathbb{R}^{N_x} $, $u(n)
\in \mathbb{R}^{N_u} $, $w(n) \in \mathbb{R}^{N_x} $, $v(n) \in
\mathbb{R}^{N_y} $ are the system output, state, input, state noise,
and output measurement noise, respectively. The matrices ${\mathcal{A}}$, ${\mathcal{B}}$, ${\mathcal{C}}$ and ${\mathcal{D}}$ have appropriate dimensions. The noise covariances are defined as 
\begin{equation}
  {\cal{E}} \left\{ \begin{pmatrix} w(j) \\ v(j) \end{pmatrix}
    \begin{pmatrix} w(i) \\ v(i) \end{pmatrix}^\top \right\} = \begin{pmatrix}
    {\mathcal{Q}} & {\mathcal{S}} \\
    {\mathcal{S}}^\top & {\mathcal{R}}
  \end{pmatrix} \delta_{ij}
\end{equation}
where ${\cal{E}}\{x\}$ stands for the expectation operator.  

Subspace identification methods extract the system matrices ${\mathcal{A}}$, ${\mathcal{B}}$, ${\mathcal{C}}$ and ${\mathcal{D}}$, as well as the covariance matrices ${\mathcal{Q}}$, ${\mathcal{S}}$, and ${\mathcal{R}}$ from a set of input-output measurements. A comprehensive description of these methods is given in the review paper of~\citet{qin06} and the book of~\citet{Overschee_deMoor_1996}. In this study, the \texttt{N4SID}
algorithm~\citep{van94, van95} implemented in \texttt{MATLAB} has been used 
to obtain all the matrices. Eq.~\eqref{eq:ProcessForm} intends to 
mimic Eq.~\eqref{eq:td estimator} and, thus, the matrix $\mathcal{D}$ and 
the noise components $w(n)$ and $v(n)$ are set to zero. The input $u$
corresponds to sensor measurements $\mathbf{s}$ and the output $y$ to the coefficients of the selected modes $\mathbf{\Phi}_i$, which are taken to be the POD modes. 

The \texttt{N4SID} algorithm works in two steps. In the first step, the state sequence $x(n)$ as well as the order of the system $N_x$, are extracted from the input-output data. During this step the (extended) observability matrix, with rank equal to $N_x$, is also obtained. The crucial component in this step is the singular value decomposition of a weighted matrix. In the second step, the matrices ${\mathcal{A}}$, ${\mathcal{B}}$ and ${\mathcal{C}}$, are calculated (to within a similarity transformation) from the system states $x(n)$ by least squares. We note that the order $N_x$ can be specified by the user, or extracted directly from the data (we explore both options later in Sec.~\ref{sec:parameters}). There are other subspace identification methods, which are related with \texttt{N4SID}. This relation is established by the unifying identification theorem \cite{Overschee_deMoor_1996, van95}. 

The application of the aforementioned system identification technique to our problem  requires three procedural steps which are depicted in  Fig.~\ref{fig:sysident}. In the first step, numerical simulations of the full problem are carried out and the inputs (velocity measurements at selected points in the wake) and outputs (coefficients of POD modes) of the estimator are extracted. In the second step, a sub-sample of the extracted data set, referred to as the learning dataset, is processed to determine the system matrices of the identified model. In a third step, a different part of the data set, known as the testing dataset, is used to drive the identified system, and the output $y_e(t)$ produced by the estimator is compared to the measured true output $y(t)$. 

\section{Flow configuration and numerical approach}\label{sec:flow configuration}

The general formalism outlined in the previous section is applied to a flow configuration that is now detailed. We consider the two-dimensional flow around a NACA0012 aerofoil with chord $c$, that forms an angle of attack $\alpha$ with the uniform approaching velocity $U_{\infty}$. We choose $c$ and $U_{\infty}$ as the reference scales for length and velocity respectively. The Reynolds number, defined as $Re=U_{\infty}c/\nu$, is set to $Re=600$ and $\alpha=\ang{16}$. These conditions correspond to an unsteady laminar flow that reaches a limit cycle (the critical Reynolds number at this $\alpha$ is $Re_c \approx 400$, refer to \cite{zhang2016biglobal}).

\begin{figure}
\centering
\subfloat{\includegraphics[width=0.49\textwidth]{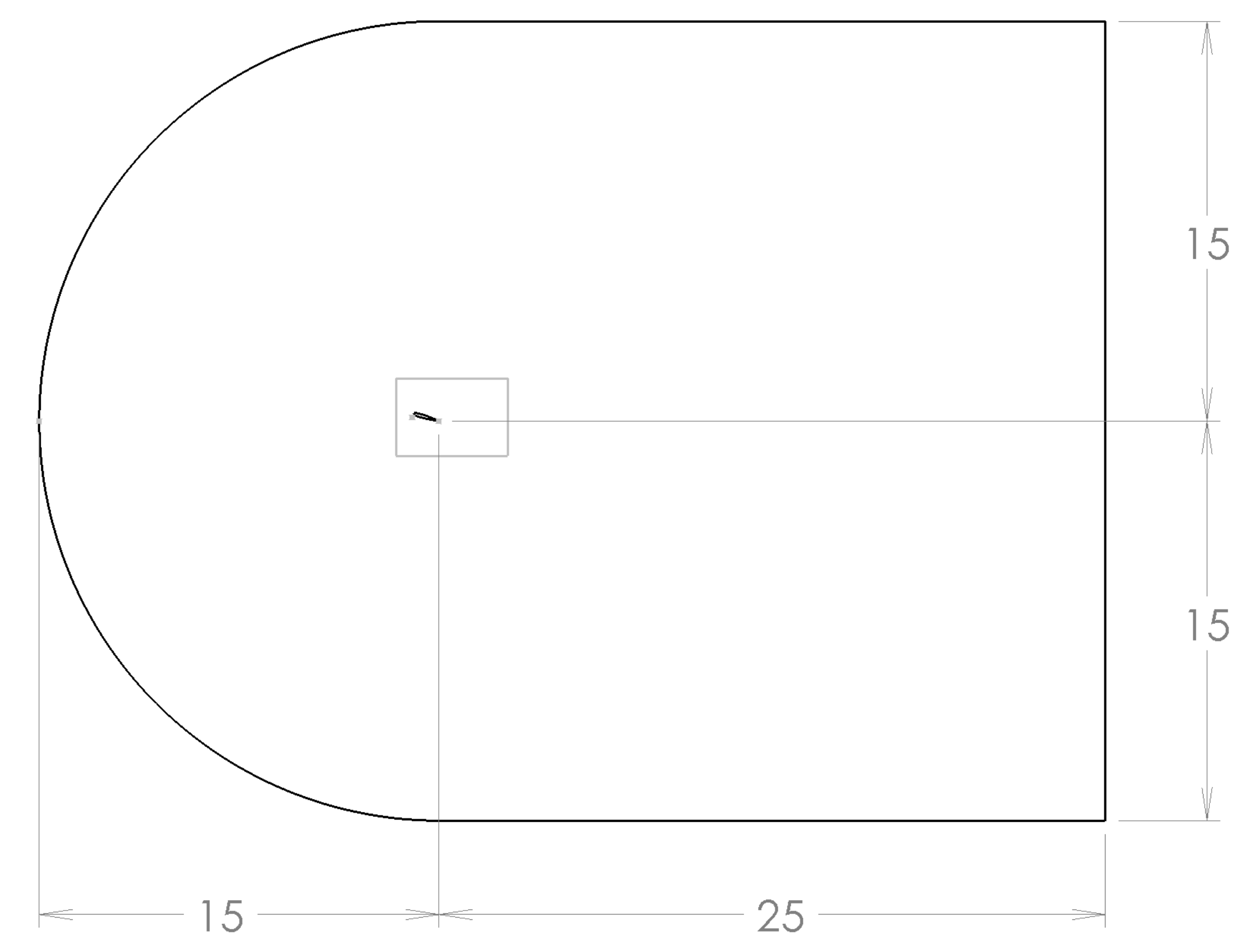}}
\subfloat{\includegraphics[width=0.49\textwidth]{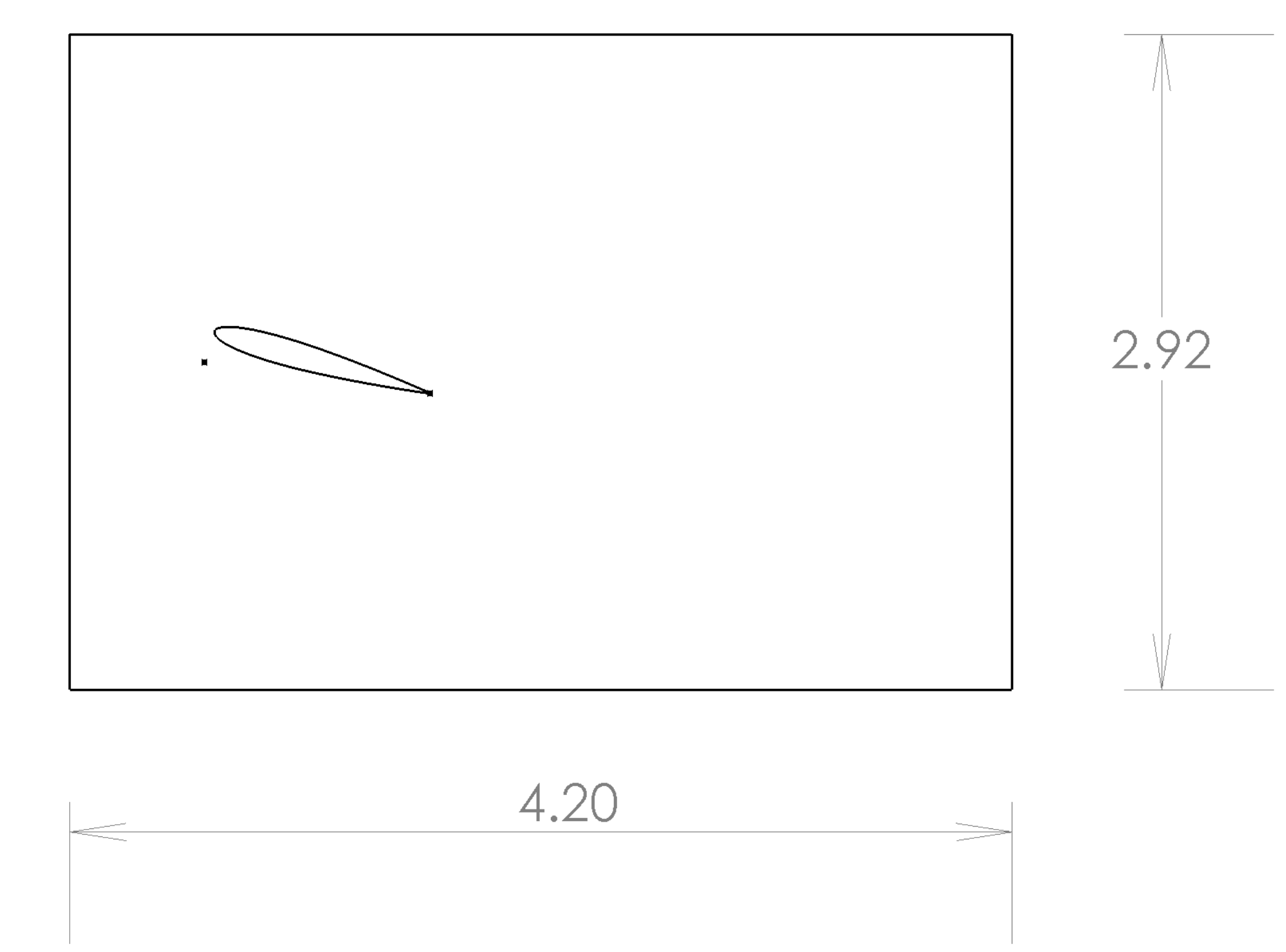}}
\caption{(Left) Computational domain and (right) control window for data extraction.}
\label{fig:computational_domain}
\end{figure}

Fig.~\ref{fig:computational_domain} (left) illustrates the whole computational domain. The Cartesian coordinate system ($x$, $y$) is centred at the point where the leading edge of the aerofoil with zero angle of attack would have been placed. The domain size is $40$ units in the $x$-direction, with the wake region extending $25$ units from the trailing edge. In the $y$-direction, the domain size is $15$ units above and below the trailing edge. 

The numerical simulations were performed with Star-CCM+ Version 13.02. The governing equations are discretised using the finite-volume method. The domain is spatially discretised with an unstructured polyhedral mesh, while an additional prism layer is inserted around the aerofoil to better resolve the near wall region. The mesh is composed of 71,209 cells in total.  The discrete system of equations is integrated in time using a second-order implicit scheme. The normalised tolerance for all equations is set to $10^{-4}$. 

The velocity is set to $\mathbf{u}=\left(1, 0\right)$ at the inlet, no-slip condition is applied on the aerofoil wall, and a pressure boundary condition is imposed at the outlet.   

The simulation starts from a zero field, and after a transient period the flow develops into a vortex shedding limit-cycle, with fundamental frequency $f=0.61$ (the value matches perfectly with the result of \cite{zhang2016biglobal}). The time-step is $\Delta{t}_{DNS}=0.01$, which corresponds to more than 160 steps per period.  We have also compared the spectra at different points in the wake with those of \cite{zhang2016biglobal} and again good matching was found (results not presented for brevity). The simulation continues for a total time of $t=80$, which corresponds to 50 vortex shedding periods. 

\begin{figure}
\subfloat{\includegraphics[width=0.45\textwidth]{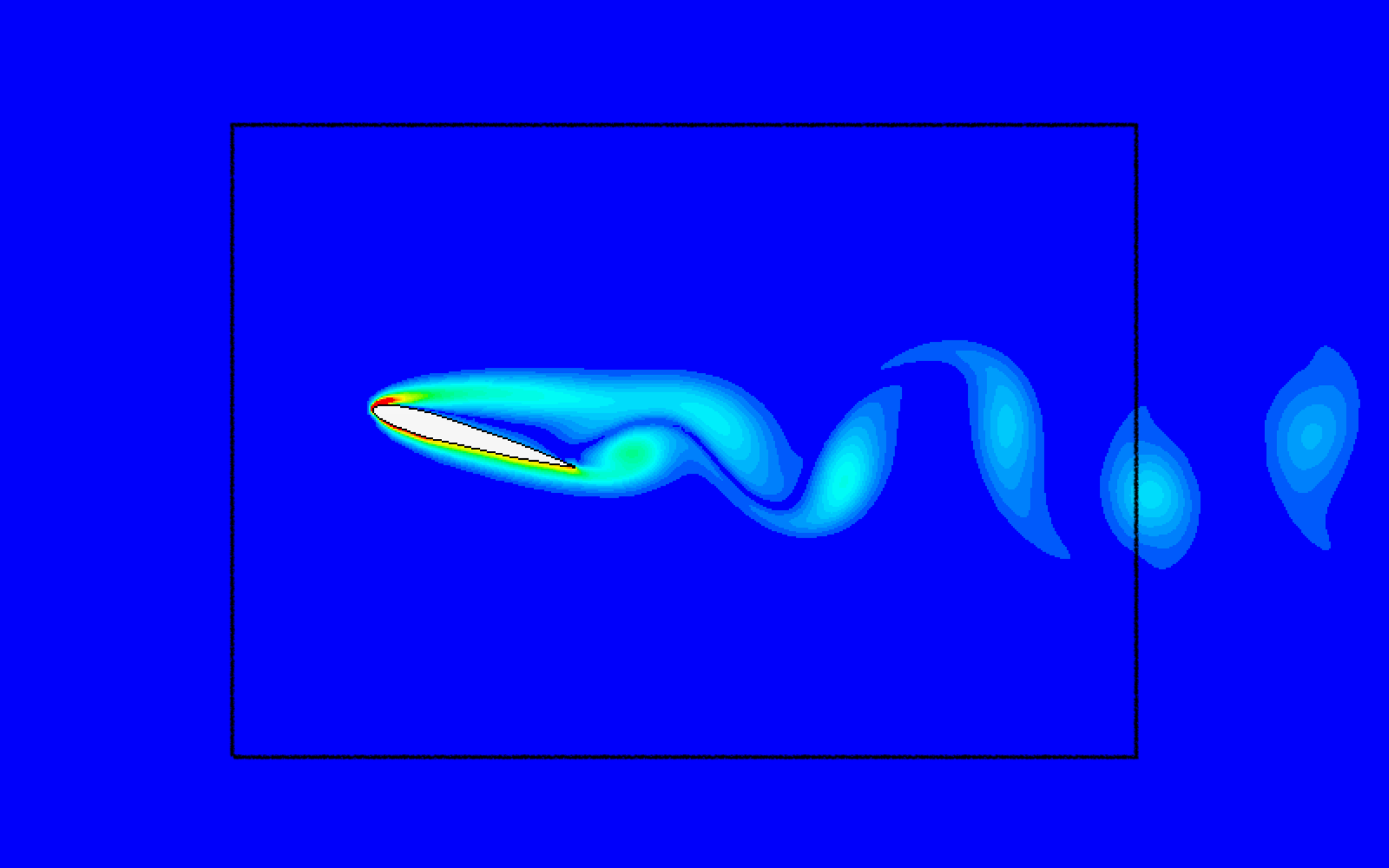}} \quad
\subfloat{\includegraphics[width=0.45\textwidth]{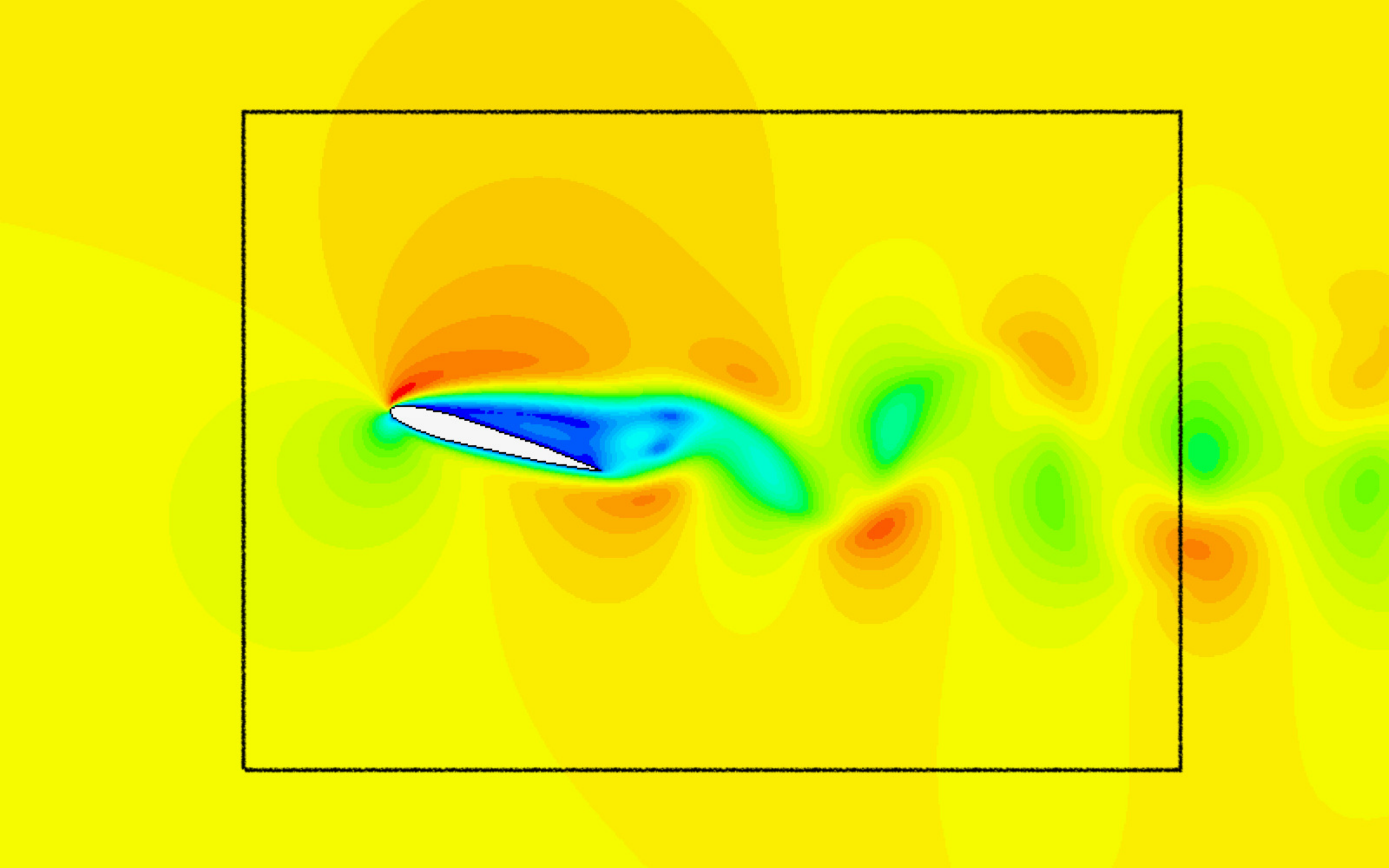}}
\caption{Snapshots of the (left) vorticity field and (right) velocity magnitude. 
The solid black line denotes the outline of the control window.}
\label{fig:CFD_snapshots}
\end{figure}

An instantaneous snapshot of the vorticity and velocity fields is depicted in Fig.~\ref{fig:CFD_snapshots}. Due to the high angle of attack, the flow separates  at the leading edge in the suction side, and vorticity is periodically shed in the wake. In the pressure side, the flow remains attached; this side also periodically releases vortices in the wake. The vortex pattern can be clearly observed in both figures. Due to the flow asymmetry, the vortices emitted from the two sides have different strengths. 

\begin{figure}
    \centering
    \includegraphics[width=0.70\textwidth]{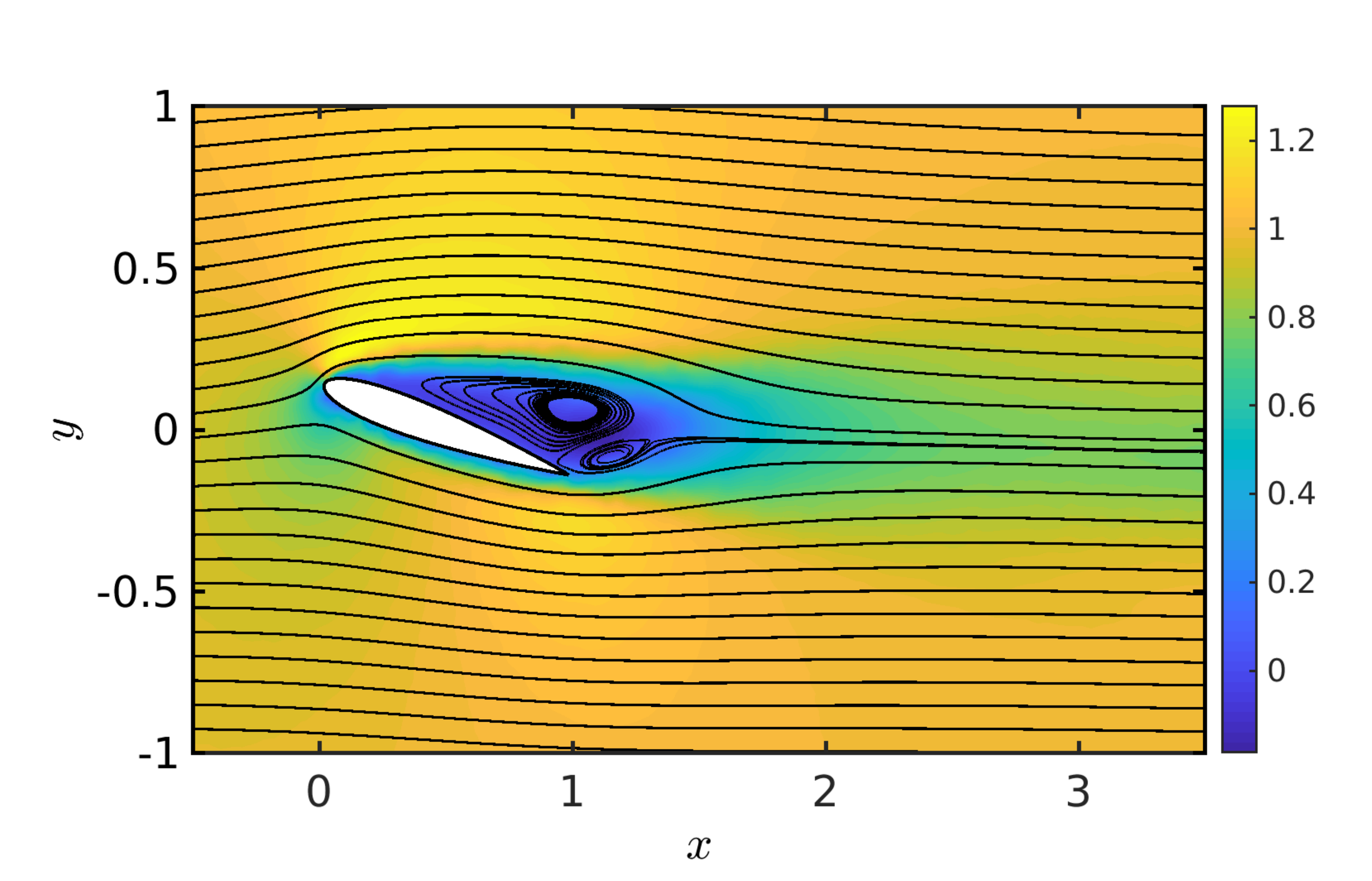}
    \caption{Streamwise component of the mean flow velocity. Black lines denote streamlines.}
    \label{fig:mean_flow}
\end{figure}

In order to compute the time-averaged flow, an initial transient period (of length $\rm{T}=24$) is discarded, and a dataset of length $\rm{T}=56$ (corresponding to 43 vortex shedding periods) is used. The resulting mean flow is plotted in Fig.~\ref{fig:mean_flow}. A low-speed recirculation region, with approximate length $1.5,$ forms in the wake. This region is composed of two bubbles with opposite circulations: (i) a large one encompassing most of the suction side and (ii) a much smaller one close to the trailing edge.  

In order to compute the estimator outlined in Sec.~\ref{sec:Theoretical framework}, velocity data are extracted from points inside the control window  shown in Fig.~\ref{fig:computational_domain}(right). The window is sampled with 100 equidistant points with spacing $\Delta x=0.042$ and $\Delta y=0.0292$ in the $x$ and $y$-directions, respectively. The points located inside the aerofoil are discarded by the software (this applies to 63 points and consequently 9937 nodes output velocity data). The mean flow is subtracted and only the fluctuations around the mean, i.e.\ the oscillatory part, are processed.

\section{\label{sec:results}Results}

\subsection{Reduced-order data}\label{subsec:POD}

The large spatial dimensionality of the data is beyond the current capabilities of most system identification algorithms and needs to be reduced. Let us consider a sequence of $m$ velocity snapshots $\left\{ \mathbf{V}_{\rm{snap}(n)} \right\}_{n=1...m}$ extracted from the simulation. The proper orthogonal decomposition (POD) enables us to compute a ranked orthonormal basis, $\left\{{\mathbf{\Phi}}_{i}\right\}_{i=1...m}$, that satisfy the conditions given by Eqs.~\eqref{eq:orthonormal}. Any velocity field $\mathbf{V}$ can then be projected onto this first $k$ modes to produce the approximate flow field $\mathbf{V}'$ (refer to Eq.~\eqref{eq:projection}). 

The POD decomposition guarantees that the norm-2 of the error $\Vert \mathbf{V} - \mathbf{V}' \Vert^2 = \left\langle \mathbf{V} - \mathbf{V}', \mathbf{V} - \mathbf{V}' \right\rangle $ is minimal for the set of $m$ snapshots and a given order $k.$ The scalar product between two vectors $\left\langle \mathbf{u}^1, \mathbf{u}^2  \right\rangle$, where  $ \mathbf{u}^1=(u^1,v^1)$ and $\mathbf{u}^2=(u^2,v^2)$, is defined as
\begin{equation}
\left\langle \mathbf{u}^1, \mathbf{u}^2  \right\rangle = \int \int_{\Omega} \left( u^1 u^2 + v^1v^2\right)\rm{d} \Omega.
\label{eq:scalar product}
\end{equation}
When $\mathbf{u}^1=\mathbf{u}^2$, the norm is equal to (twice) the kinetic energy in the domain. Based on this scalar product, the POD basis captures the maximum proportion of kinetic energy in the domain $\Omega$ for any given order $k$.  

\begin{figure}
\centering
\subfloat{\includegraphics[width=0.48\textwidth]{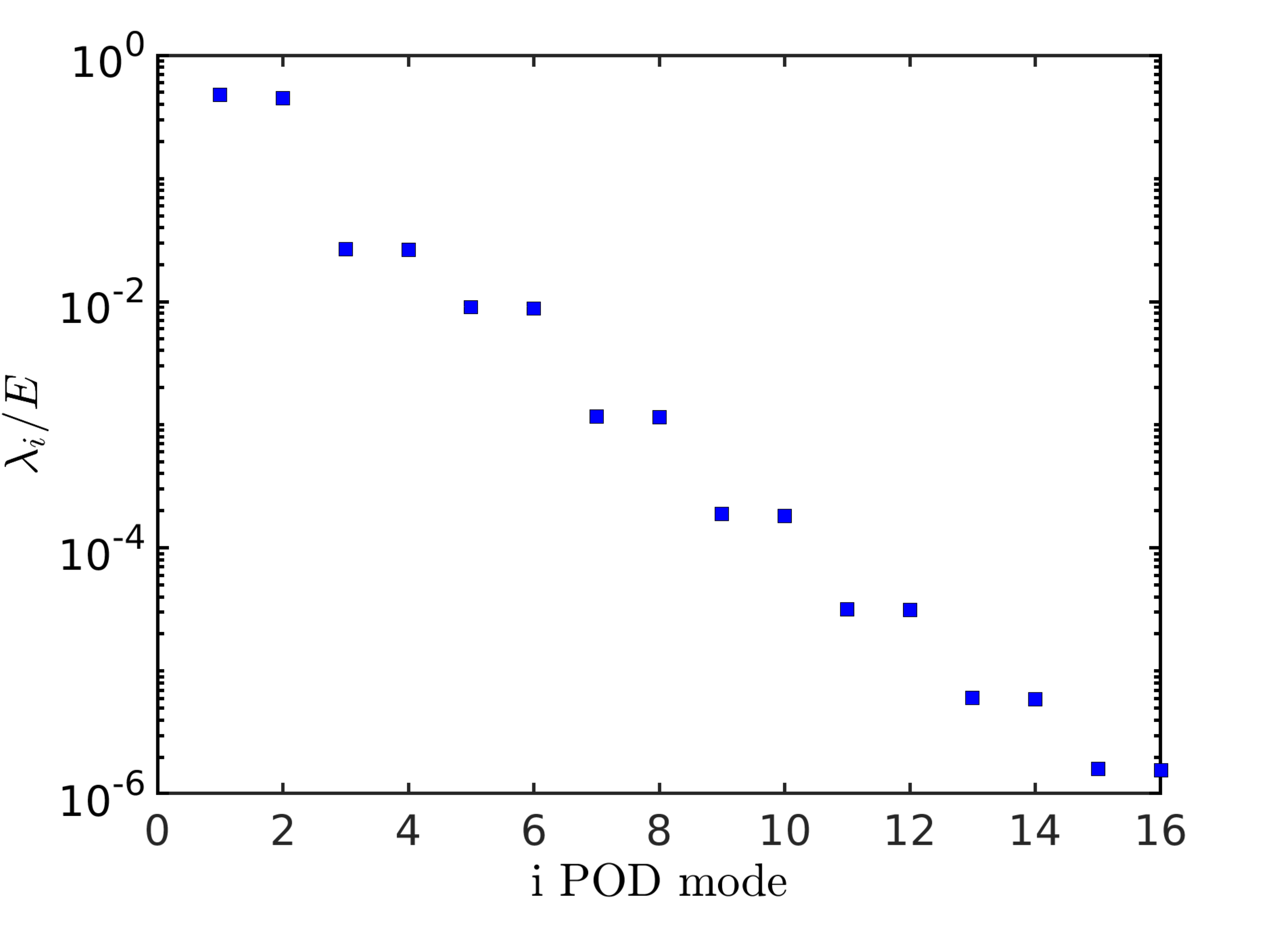}}
\subfloat{\includegraphics[width=0.48\textwidth]{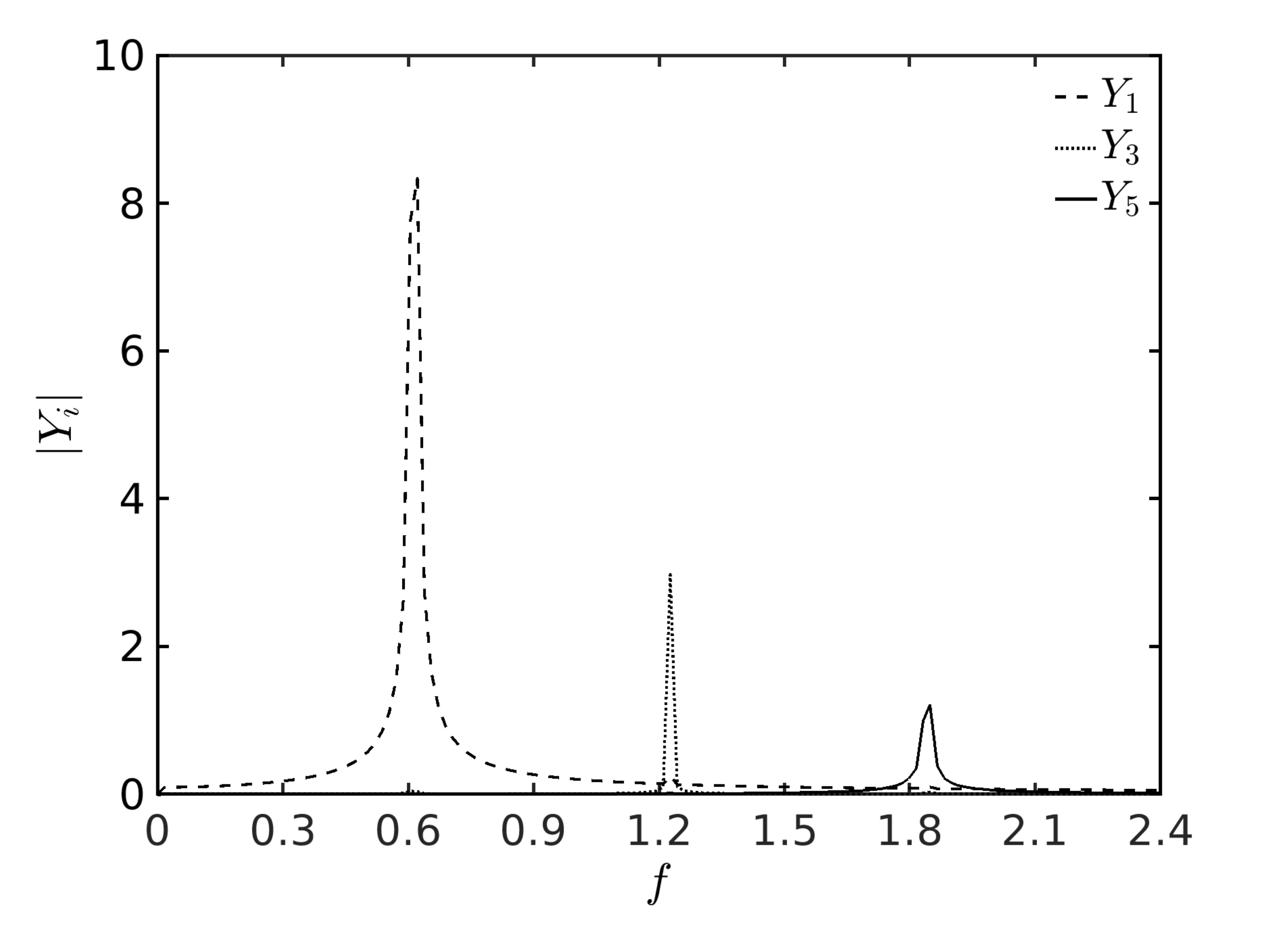}}
\caption{(Left) First 16 POD eigenvalues $\lambda_i$ of the correlation matrix and 
(right) spectrum of the first, third and fifth  POD coefficients.}
\label{fig:KE_fractions}
\end{figure}

\begin{figure}
\centering
\includegraphics[width=\textwidth]{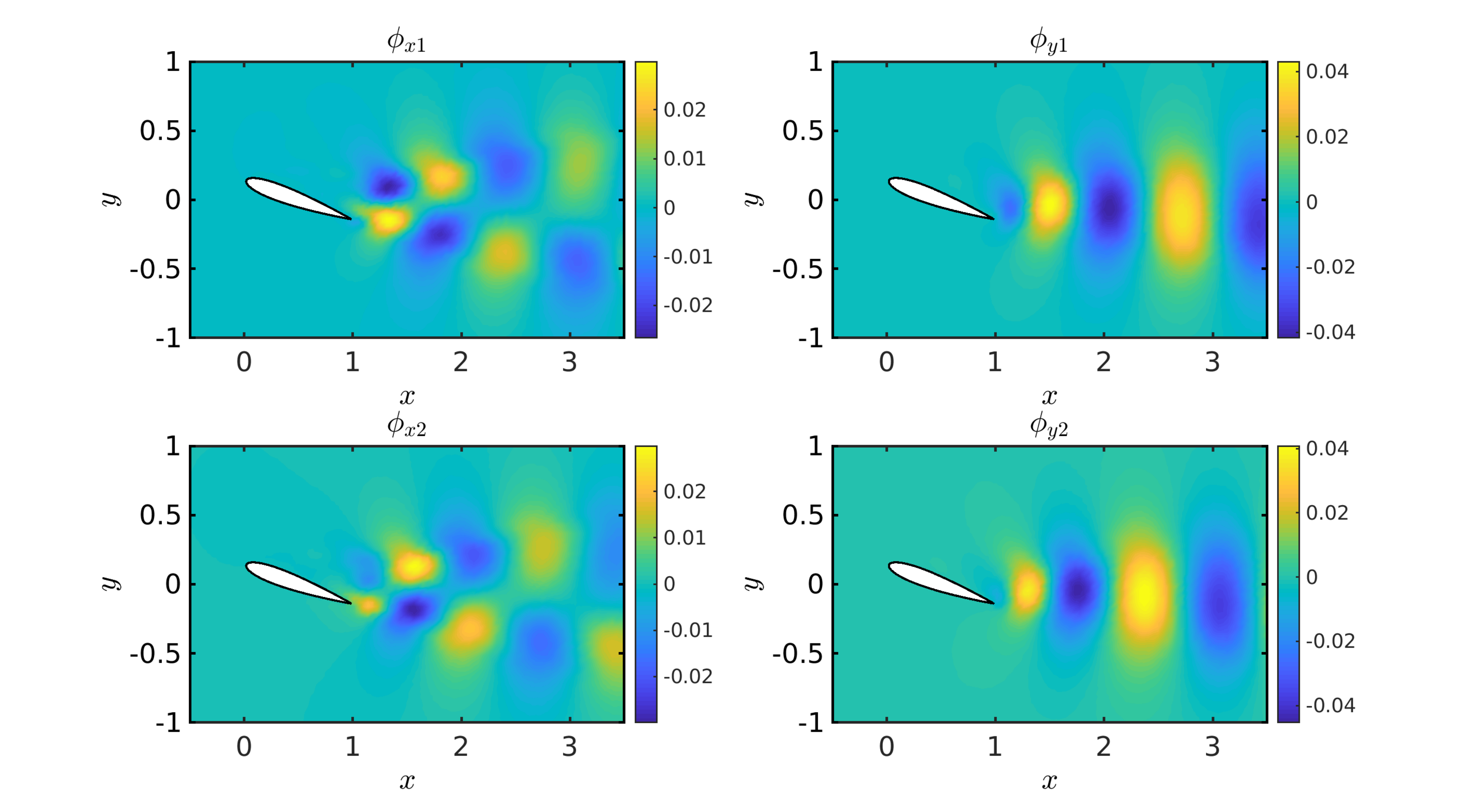}
\caption{Contours of the (left column) streamwise and
(right column) cross-stream velocity components of the (top) first and (bottom) 
second POD-modes.}
\label{fig:POD_modes_12}
\end{figure}

The POD decomposition was applied to the oscillatory part of the same snapshots used to compute the mean flow. Fig.~\ref{fig:KE_fractions}(left) shows the eigenvalues of the correlation matrix (normalised with the total energy of the fluctuating field) for the first 16 POD modes. These eigenvalues represent the energy content carried by each mode. Over 92\% of the oscillatory kinetic energy is captured by the first 2 modes and over 97\% by the first 4. Due to the convective nature of vortex shedding motion, POD modes arise in pairs. 

The spectra of the POD coefficients, shown in Fig.~\ref{fig:KE_fractions}(right),  indicate that the first (and second) POD modes are associated with the fundamental vortex-shedding frequency ($f=0.61$), while the third (and fourth) and fifth (and sixth) modes with the first and second harmonics, respectively. These harmonics are produced by non-linear interactions in the wake. For example, the first harmonic is produced by the interaction of the first two modes (more details can be found in \cite{Sipp_lebedev_2007}). The placement of the sensor point in relation to the spatial location of these interactions affects the performance of the estimator, as will be shown later. 

\begin{figure}
\centering
\includegraphics[width=\textwidth]{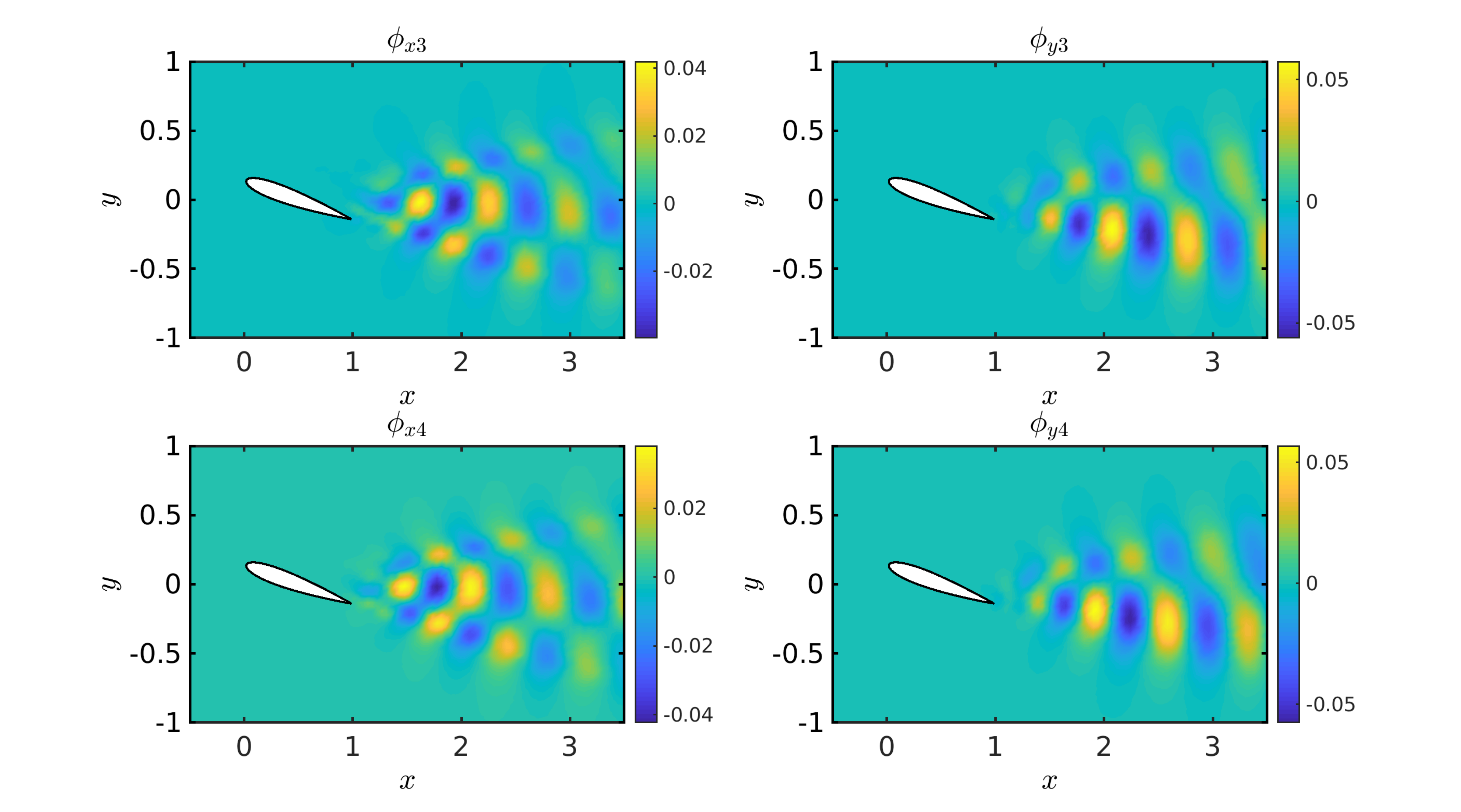}
\caption{Contours of the (left column) streamwise and
(right column) cross-stream velocity components of the (top) third and (bottom) 
fourth POD-modes.}
\label{fig:POD_modes_34}
\end{figure}

Fig.~\ref{fig:POD_modes_12} depicts the streamwise and cross-stream velocity components of the first and second mode. This pair
clearly captures the large coherent structures corresponding to vortex shedding. Note that the two modes have very similar 
structure, but are shifted in space. The coefficients are also shifted in time by $90\deg$ and this combination results in the 
propagation of structures in the wake. Fig.~\ref{fig:POD_modes_34} depicts the velocity components of the third and fourth POD
modes. These structures are smaller compared to those of the first and second mode and are also shifted in space. 

For the application of the system identification algorithm, the time-evolving
POD coefficients corresponding to the $k$ most energetic modes are arranged 
in the reduced order state vector $\textbf{Y}=[y_1, y_2, .., y_k]^\top$
and constitute the output of our system.

\subsection{System identification and validation \label{subsec:Identification}} 

In this section, we seek to build a dynamical system (with the structure given by Eq.~\eqref{eq:td estimator}) that estimates the 
evolution of $\textbf{Y}(n)$ from local measurements. We consider a sensor measuring the streamwise and 
cross-stream components of the oscillatory velocity $s(n)=\left[ u(n), v(n)\right]^\top$ at the
point $(x_{\rm{s}},y_{\rm{s}})=(1.82,-0.04).$ Note that 
for this first application of the identification algorithm, the location of the 
measuring point is arbitrary. A detailed study of the effect of the sensor position on the quality of the identification is
provided later in Sec.~\ref{sec:parameters}.

A time-segment of the simulation, known as learning dataset, is used together with the identification 
algorithm \texttt{N4SID}~\citep{van94} to estimate the unknown  matrices of 
system~\eqref{eq:td estimator}. The reduced-order
model is determined with $k=4$ POD modes and represents the evolution of the vortex-shedding mode and its first harmonic. The number of states 
of the estimator is set to $N_x=k=4$. The length of the learning dataset is $m_{\rm{LD}}=800$ snapshots, with 
time step $\Delta{t}=0.01$, yielding a physical time $\rm{T}_{LD}=8$, which corresponds to 5 vortex-shedding periods. 

\begin{figure}
\centering
\subfloat{\includegraphics[width=\textwidth]{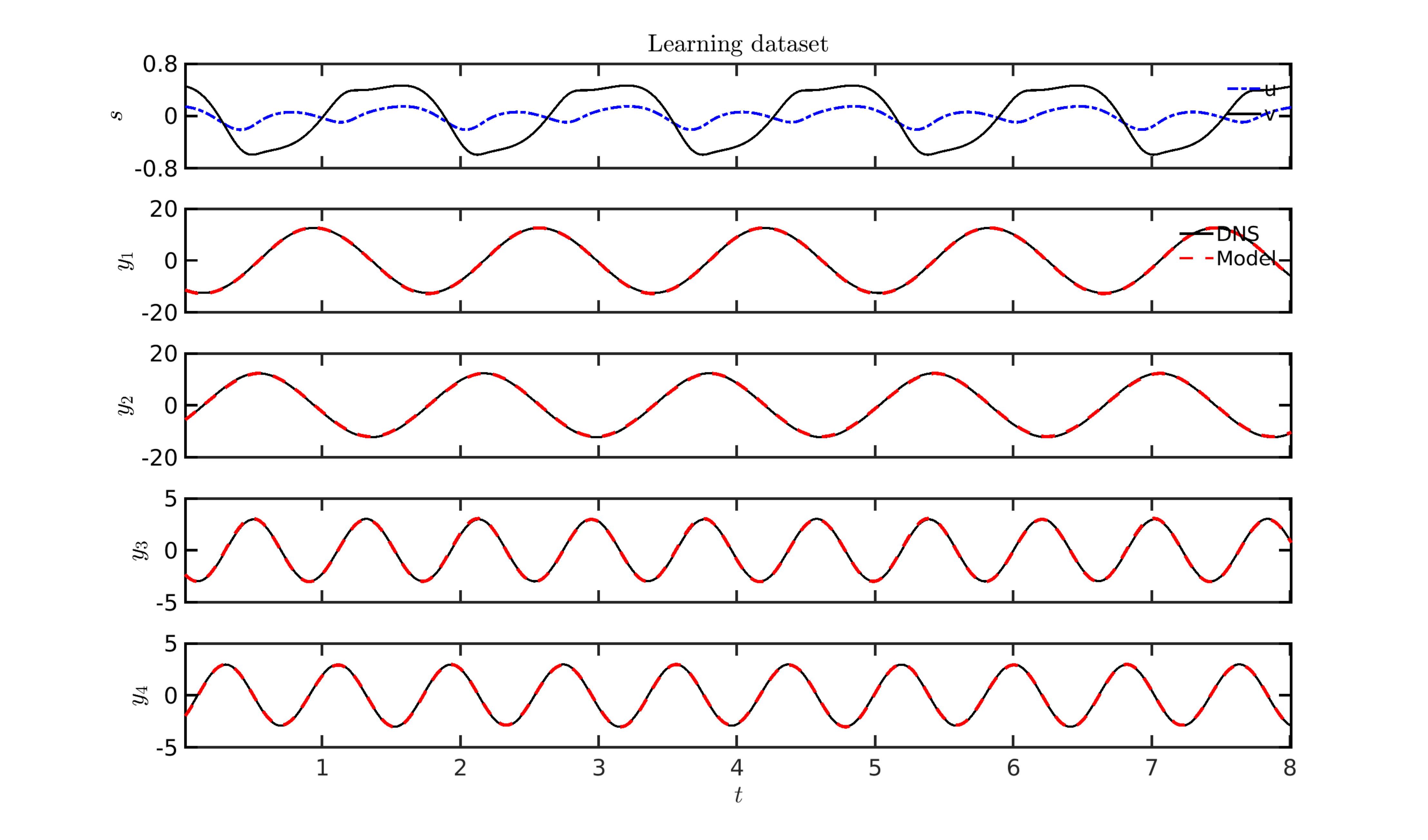}}
\caption{Learning dataset: (top) the measurement signal $s(t)$ and (bottom) first four POD coefficients $y_i(t)$ obtained by projecting the flow field onto the POD modes (solid black) and predicted by the model (dashed red).}
\label{fig:coeffs_learning}
\end{figure}

Fig.~\ref{fig:coeffs_learning} shows a comparison of the true and estimated POD coefficients for the learning dataset. The top plot shows the measurements from the input sensor and the time-evolution of the four POD coefficients is illustrated underneath. For the selected parameters, the algorithm accurately estimates the evolution of the coefficients over the entire learning dataset. 

To quantify the performance of the estimator, we use the fit between the true $i$-th POD coefficient extracted from the DNS and the one predicted defined as
\begin{equation}
  {\mathsf{FIT}}_i[\%] = 100 \times \left( 1 - \frac{\left\| y_i(t) - y_{i,e}(t)
      \right\|}{\left\| y_i(t) - \overline{y_i(t)}\right\|} \right),
  \label{eq:fit}
\end{equation}
where $\left\| \cdot \right\|$ denotes the $2$-norm of a vector. The vector of fits obtained for the learning dataset for the first four POD coefficients is
${\mathsf{FIT}}[\%]=[98.77,\: 97.67,\: 97.27,\: 98.07].$

\begin{figure}
\centering
\includegraphics[width=\textwidth]{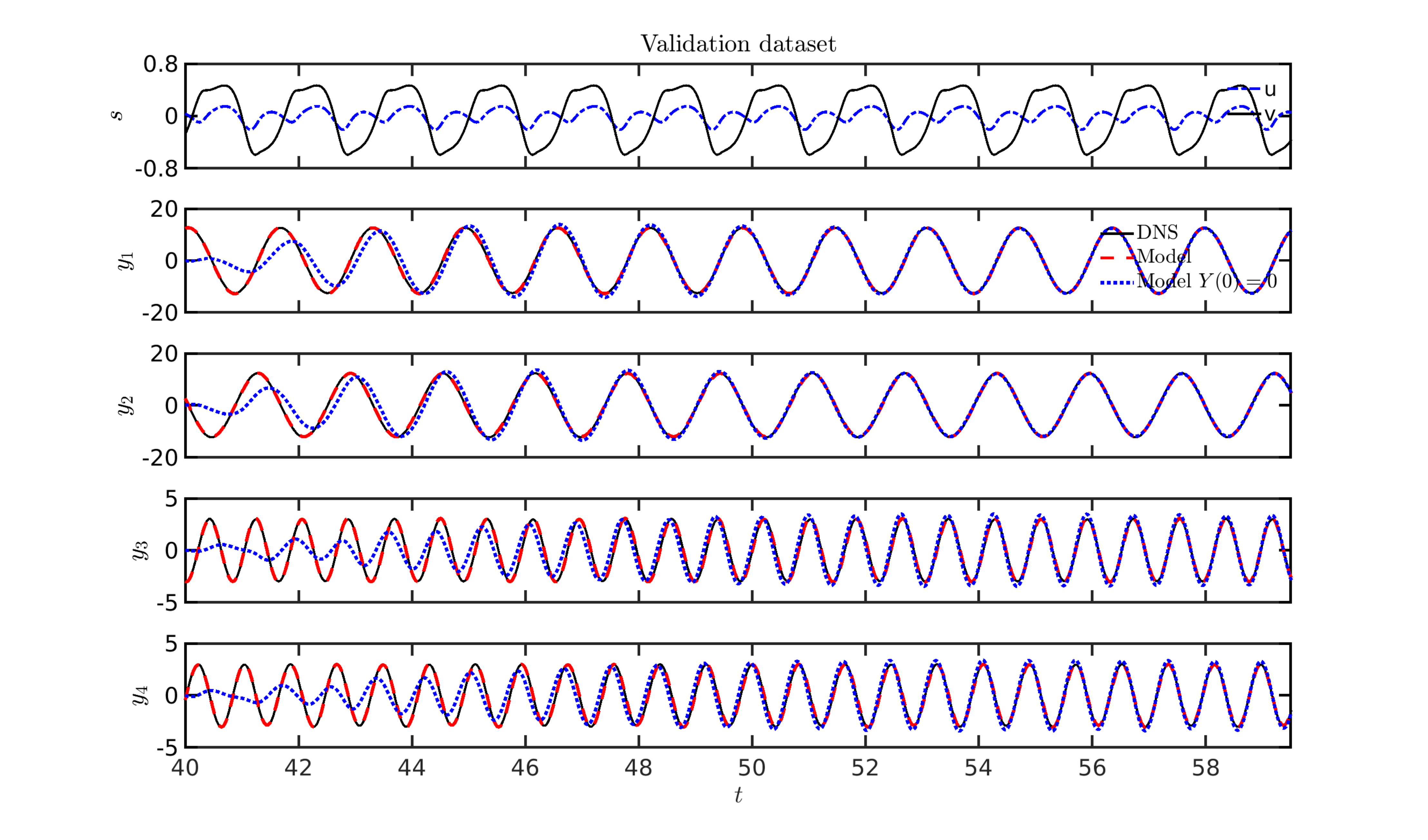}
\caption{Validation dataset: Performance of the system-identified model. (Top) Input data $s$ and (bottom) comparison between the DNS (black solid) and model prediction for four POD coefficients $y_i$. The model is initialised by $\mathbf{X}_e=\mathsf{C}^{\dagger}\mathbf{Y}$ (red dashed) and $\mathbf{X_e}=\mathbf{0}$ (blue dotted) at $t=40.$}\label{fig:coeffs_validation}
\end{figure}

For the learning dataset, the good agreement between the true POD coefficients  and those predicted by the model is expected since the model is obtained by minimising the error $(\mathbf{Y} - \mathbf{Y}_e)$ over that period of time. The validity of the identified model must be confirmed using a different time-segment of the simulation. Fig.~\ref{fig:coeffs_validation} illustrates the performance of the model on a validation dataset. The model states $\mathbf{X}_e$ are initialised using two different values: (i) the actual reduced-order state of the system $\mathbf{X}_e=\mathsf{C}^{\dagger}\mathbf{Y}$ and (ii) $\mathbf{X}_e=\mathbf{0}$. When initialised by the last known state of the system, the prediction of the algorithm accurately matches the evolution of the true POD coefficients over the whole dataset. On the other hand, a transient period of length about $\rm{T}=4$ ($\approx$ 2.4 periods) is required before the model predictions match true
results when the model is initialised with $\mathbf{X}_e=\mathbf{0}$. After the short transient, the coefficients are correctly predicted. The vector of fits obtained for the validation dataset is ${\mathsf{FIT}}[\%]=[98.81,\: 97.61,\: 97.18,\: 98.04]^\top$ when initialised with the actual state.

\begin{figure}
\centering
\includegraphics[width=\textwidth]{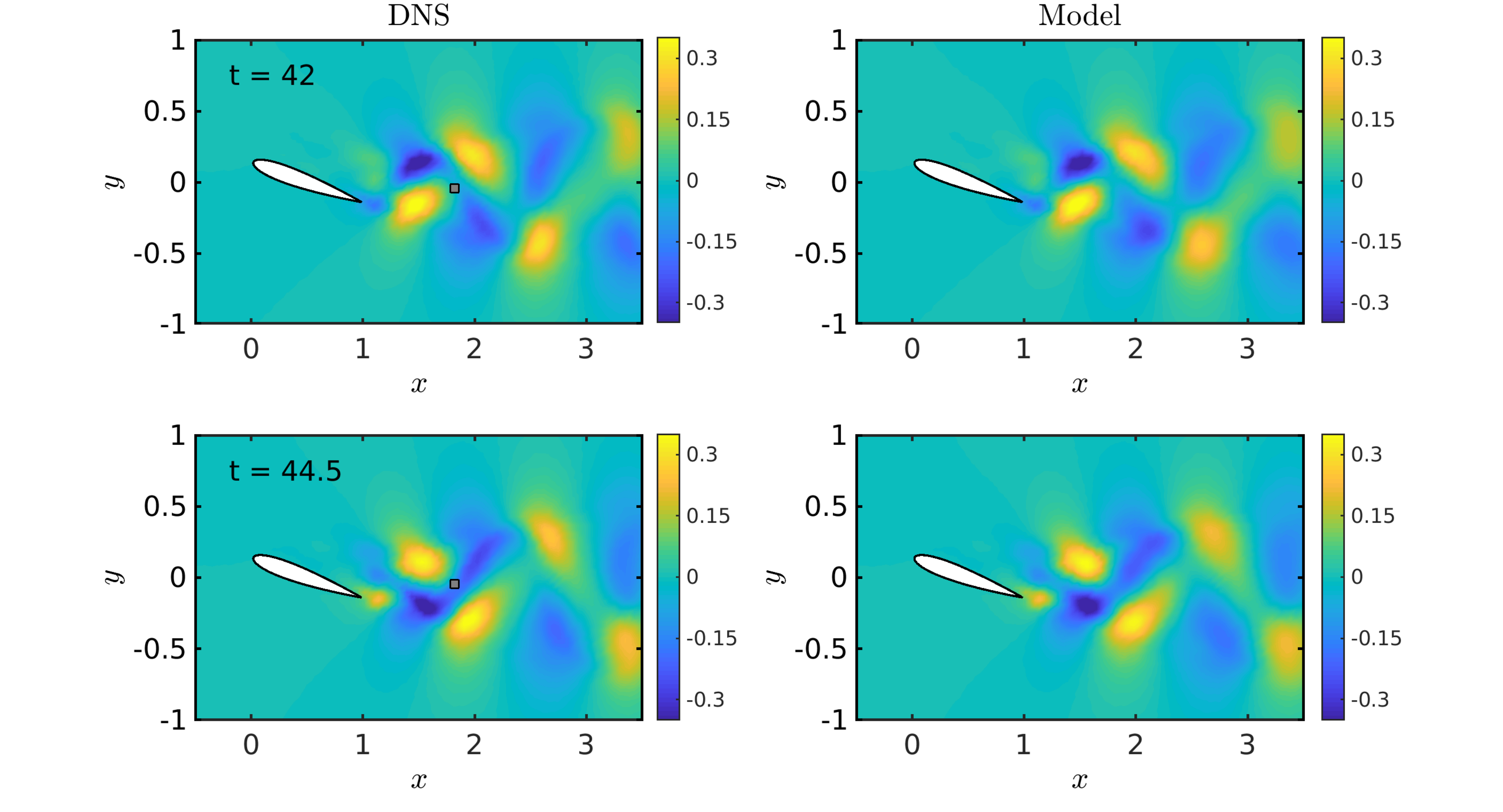}
\caption{Streamwise component of the true oscillatory velocity (left) obtained
from the DNS and (right) model prediction. The grey squares represent the position of the estimation sensor. See supplementary movie 1.}
\label{fig:mag_N4SID_TD}
\end{figure}

Eq.~\eqref{eq:projection} allows us to reconstruct the full velocity 
field from the identified reduced-order model. Fig.~\ref{fig:mag_N4SID_TD} depicts the original and reconstructed streamwise component of the velocity field 
at two arbitrary time instants for the testing dataset. The vortex shedding 
is clearly captured by the model and the reconstructions at both instants appear almost identical. 

\begin{figure}
\centering
\subfloat{\includegraphics[width=0.48\textwidth]{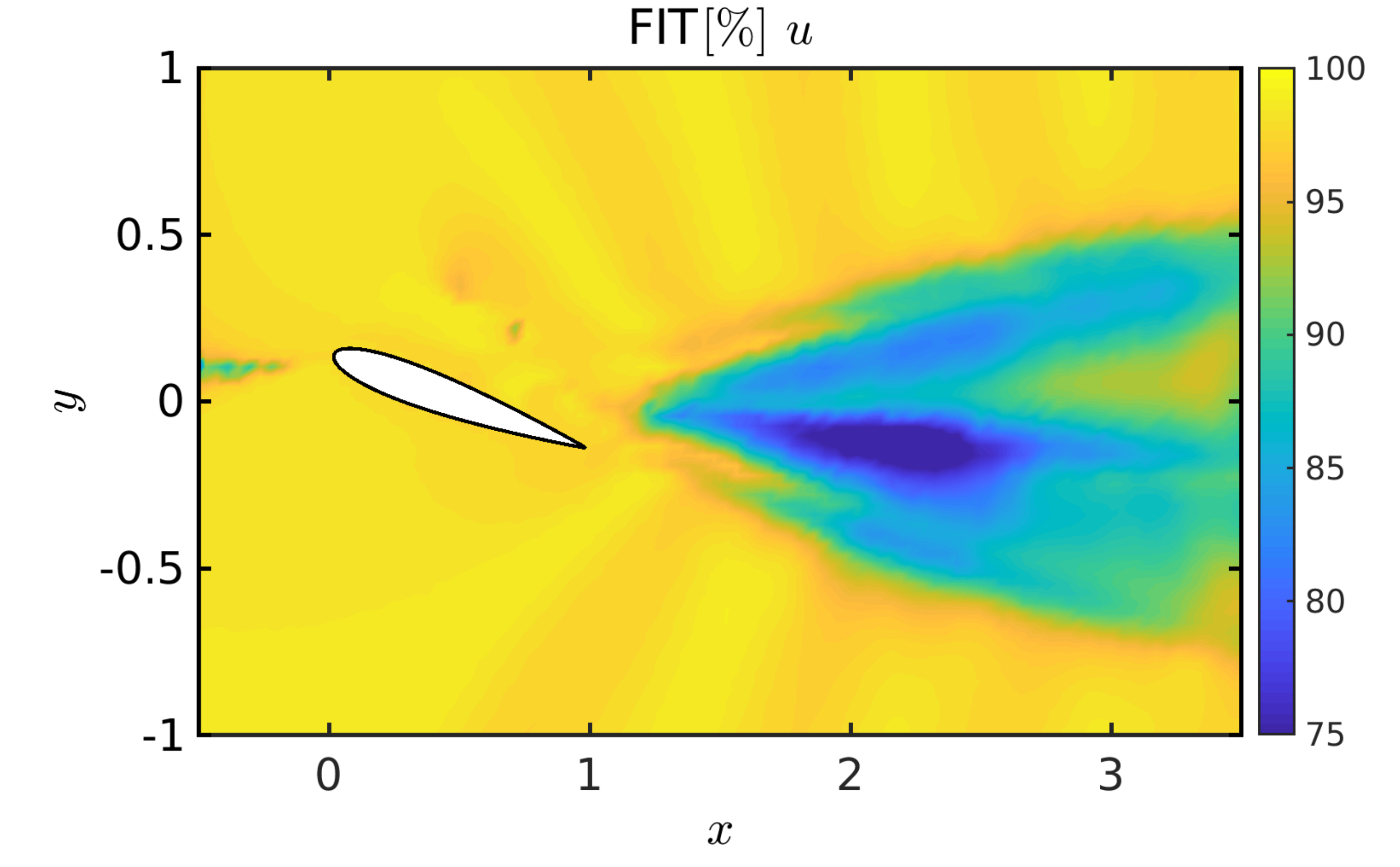}}
\subfloat{\includegraphics[width=0.48\textwidth]{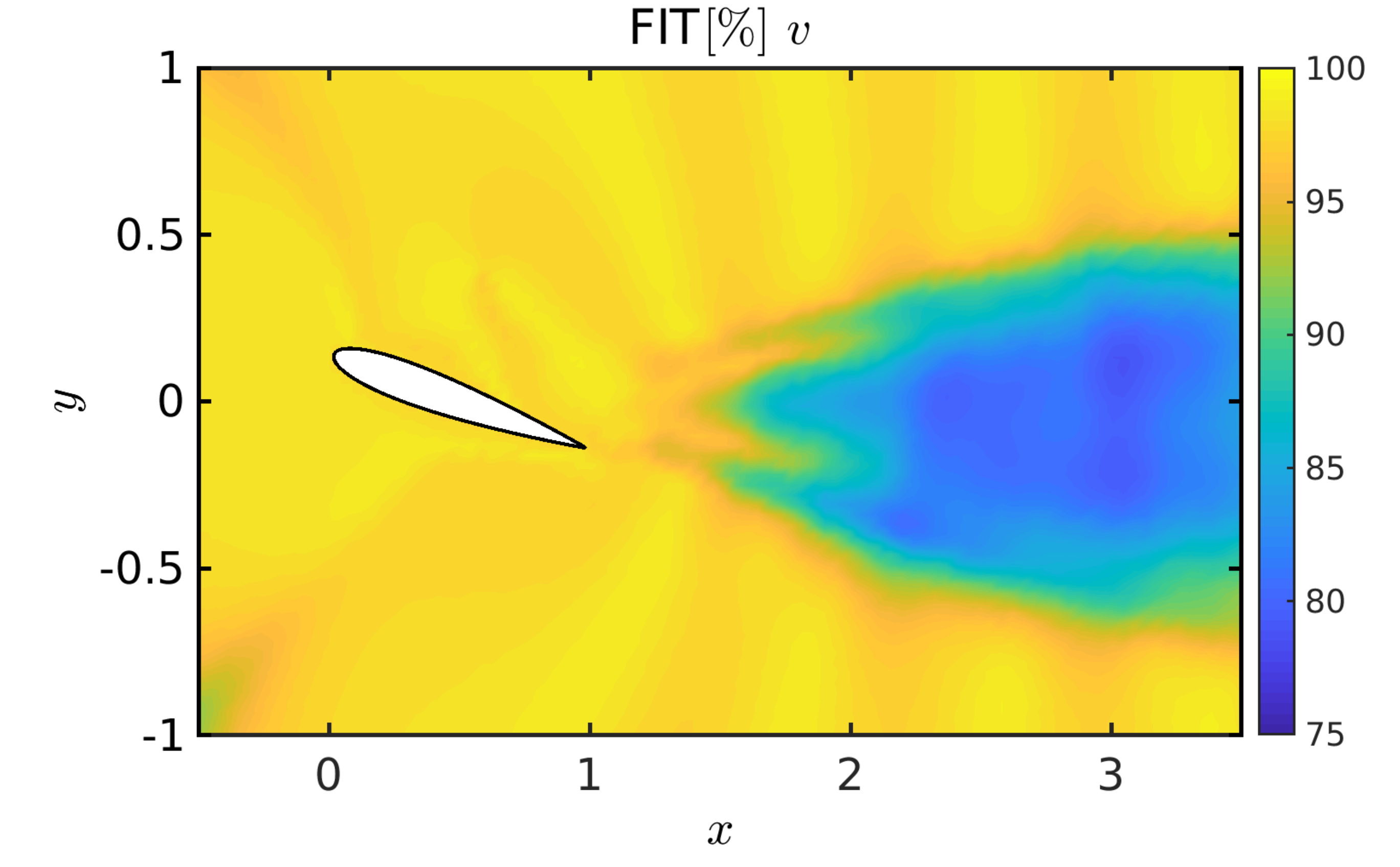}}
\caption{Goodness of fit $\mathsf{FIT}[\%]$ between the actual oscillatory velocity field and the field estimated by the model at each point: (left) streamwise and (right) cross-stream components of the velocity.}
\label{fig:fit_xy}
\end{figure}

We can evaluate the fit between the true perturbation velocity field at each point (computed from the DNS) and the value predicted by the model 
in the same way as we did for the POD coefficients in Eq.~\eqref{eq:fit}. We 
simply need to  replace $y_i(t)$ and $y_{i,e}(t)$ by the true and estimated velocity values respectively at a given point.  Fig.~\ref{fig:fit_xy} shows contour plots of the $\mathsf{FIT}[\%]$ for the validation dataset. A good agreement is found in the whole domain, with $\mathsf{FIT}$ values ranging from 75\% to 100\%. The areas with 
the largest mismatch are in the wake of the aerofoil and coincide with the regions of largest amplitude of the second harmonic, which is not considered in the current estimator model.

\begin{figure}
\centering
\includegraphics[width=\textwidth]{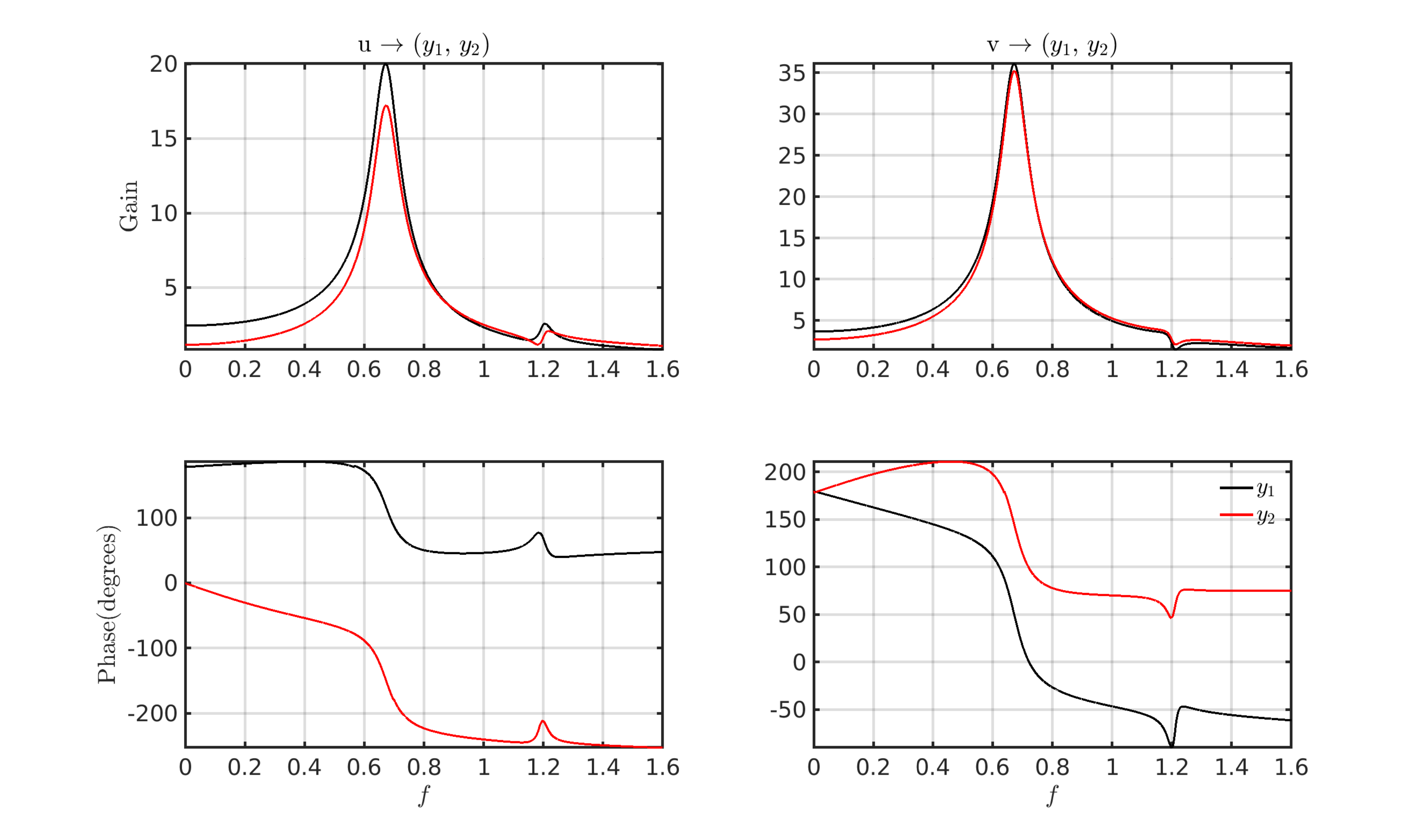}
\caption{Bode plots (gain and phase) of the transfer functions from the two inputs ($u$, $v$) to the two outputs (first and second POD coefficients).}
\label{fig:transfer_function12}
\end{figure}

\begin{figure}
\centering
\includegraphics[width=0.6\textwidth]{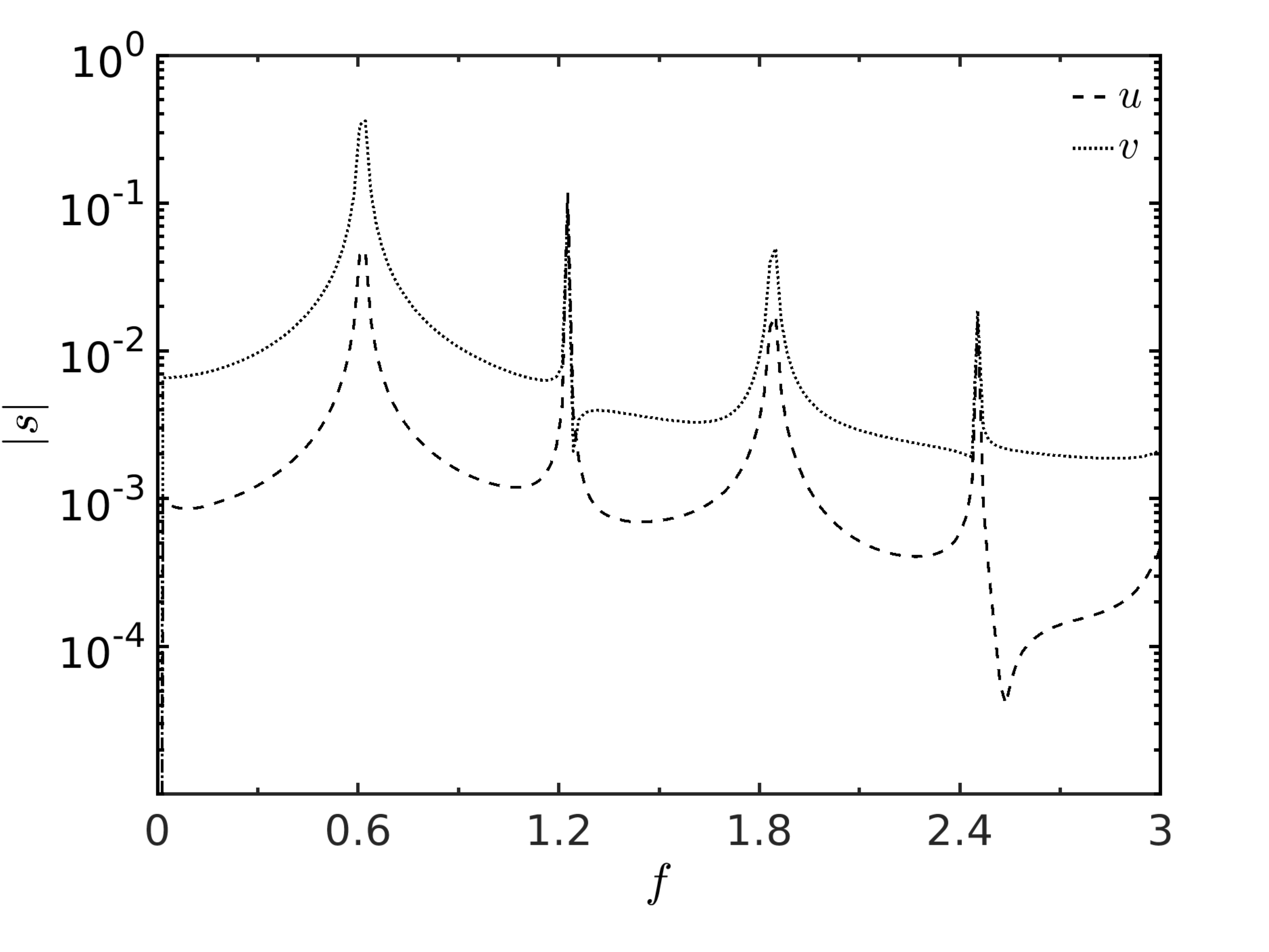}
\caption{Spectra of the streamwise $u$ and cross-stream $v$ components of the oscillatory velocity composing the input $s=[u,v]^\top$.}
\label{fig:input_fourier}
\end{figure}

It is instructive to study also the Bode plots of the transfer function between
the input $s=[u,v]^\top$ and the output $\mathbf{Y}_e(t)=[y_{1,e}(t),y_{2,e}(t),y_{3,e}(t),y_{4,e}(t)]^\top$. The
transfer functions are obtained by taking the
Fourier transform of Eq.~\eqref{eq:tc estimator}. The result 
is $\mathbf{\hat{Y}}_e(\omega) = \mathsf{C}'\left(\omega \mathsf{I}+\mathsf{A}_s'\right)^{-1} \mathsf{L}' \mathbf{\hat{s}}(\omega)$ or
in expanded form

\begin{equation}
   \begin{pmatrix} \hat{y}_{1,e}(\omega) \\  \hat{y}_{2,e}(\omega) \\  \hat{y}_{3,e}(\omega) \\  \hat{y}_{4,e}(\omega)  \end{pmatrix}
    = 
    \begin{pmatrix}
    {M_{u \rightarrow y_1}}(\omega) & {M_{v \rightarrow y_1}}(\omega) \\
    {M_{u \rightarrow y_2}}(\omega) & {M_{v \rightarrow y_2}}(\omega) \\
     {M_{u \rightarrow y_3}}(\omega) & {M_{v \rightarrow y_3}}(\omega) \\
      {M_{u \rightarrow y_4}}(\omega) & {M_{v \rightarrow y_4}}(\omega)
      \end{pmatrix}  \begin{pmatrix} \hat{u}(\omega) \\  \hat{v}(\omega) \end{pmatrix}
\end{equation}

\noindent where the hat ($\mathbf{\hat{}}$) denotes the Fourier-transformed variable. Fig.~\ref{fig:transfer_function12} shows the Bode plots (gain and phase) of $ {M_{u \rightarrow y_1}}(\omega)$ and ${M_{u \rightarrow y_2}}(\omega)$ (left column) as well as  ${M_{v \rightarrow y_1}}(\omega)$ and ${M_{v \rightarrow y_2}}(\omega)$ (right column). The estimator has the form of a second order filter, that has high gain close to the frequency of the 1st and 2nd mode, and damps all the other frequencies. At the mode frequency $0.61$, the gain from $v$ is almost twice as large compared to $u.$ {This result together with the spectra of the input signal (plotted in Fig.~\ref{fig:input_fourier}) indicate that it is mainly the $v$ component that provides the information to extract $y_{1,e}$ and $y_{2,e}$ at the selected point.} Interestingly, the peak value is not exactly at the frequency $0.61$, but slightly displaced. From the phase plot (bottom row) we note also that the estimator captures the correct $90\deg$ phase difference between $y_{1,e}(t)$ and $y_{2,e}(t)$ at the frequency of vortex shedding. 

Fig.~\ref{fig:transfer_function12} shows the Bode plots for the third
and fourth modes ${M_{v \rightarrow y_3/y_4}}(\omega)$(left column) 
and ${M_{v \rightarrow y_3/ y_4}}(\omega)$ (right column). Again the shape is similar to a 
second order filter, centred at the first harmonic, and all the other frequencies are suppressed. Note 
that now both $u$ and $v$ contribute equally to the extraction of the signal $y_3, y_4$, and the estimator 
predicts again the correct phase difference (bottom row).  

\begin{figure}
\centering
\includegraphics[width=\textwidth]{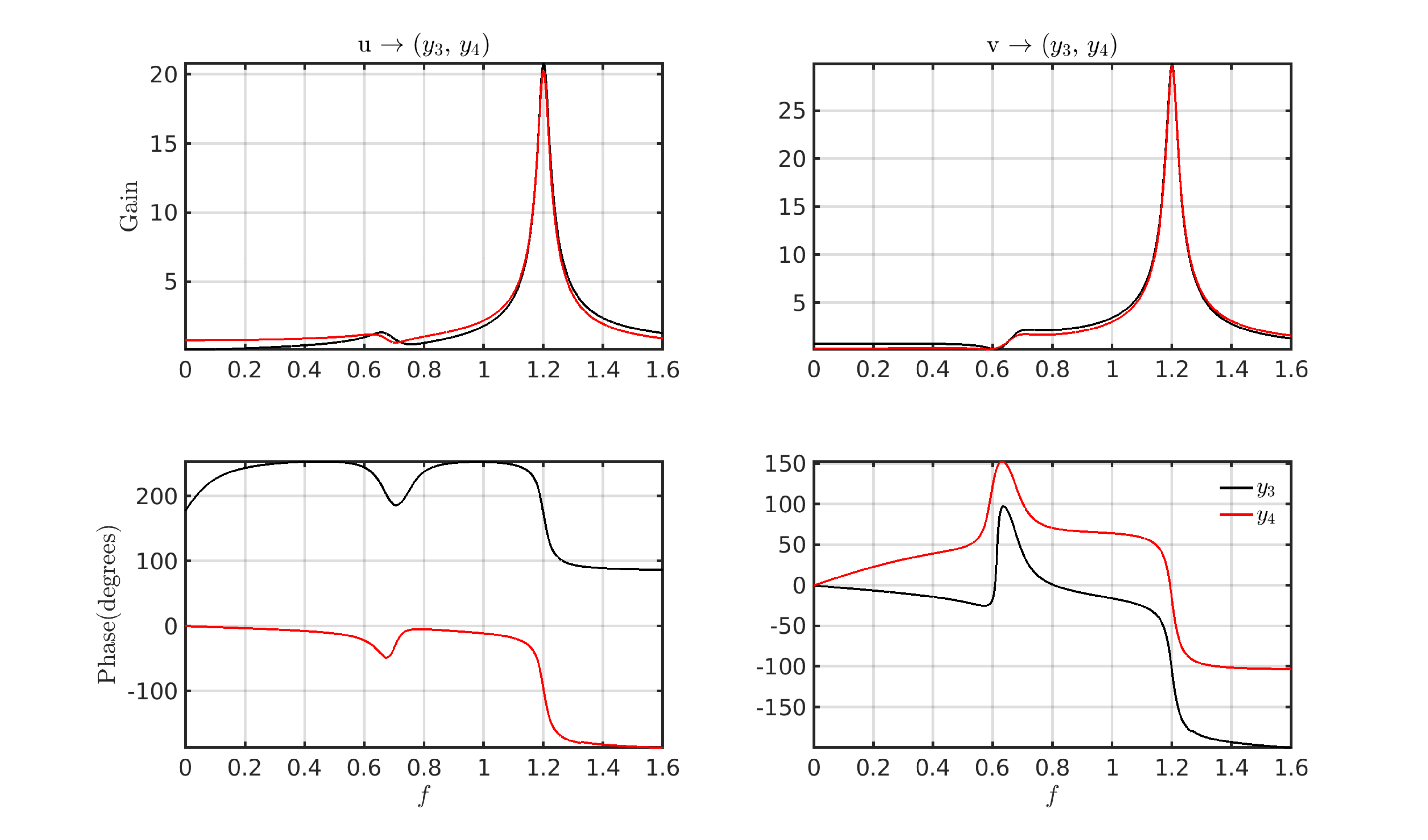}
\caption{Bode plots (gain and phase) of the transfer functions from the two inputs ($u$, $v$) to the two outputs (third and fourth POD coefficients).}
\label{fig:transfer_function34}
\end{figure}


\section{\label{sec:parameters} Influence of different parameters on the identification}

In this section, we explore the effect of the input sensor location on the performance of the estimator. This is significantly influenced by other parameters: (i) the number 
of POD modes ($k$), (ii) the order of the reduced-order system ($N_x$), and (iii) the particular velocity component(s) recorded by the sensor ($u$, $v$, or both). We systematically study the effect of these parameters on the performance of the model. The results are grouped for models composed of $k=2$ (Sec.\ref{subsec:2modes}) and $k=4$ POD modes (Sec.~\ref{subsec:4modes}). 

\subsection{\label{subsec:2modes}2 POD modes}
\begin{figure}
\centering
\subfloat{\includegraphics[width=0.49\textwidth]{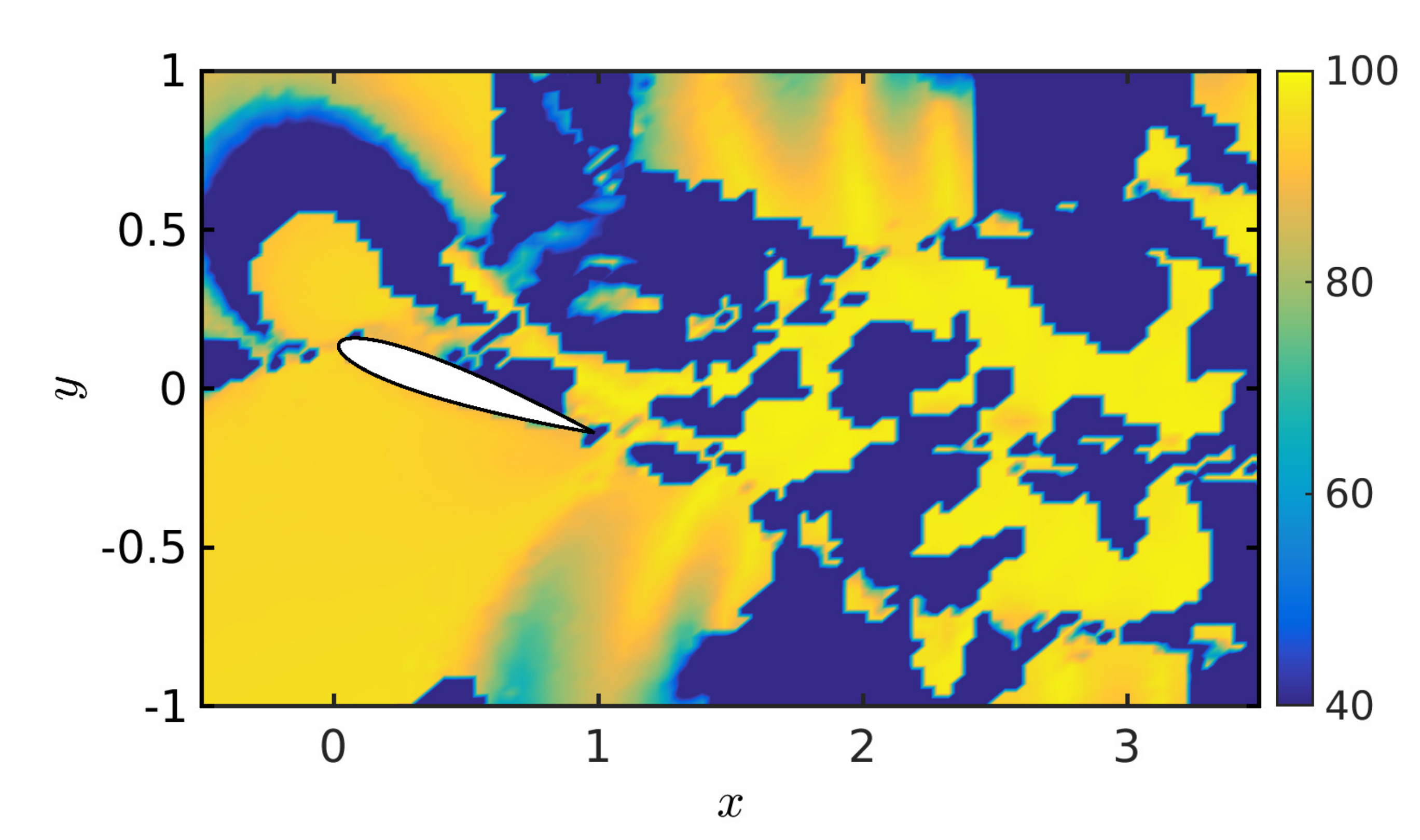}}
\subfloat{\includegraphics[width=0.49\textwidth]{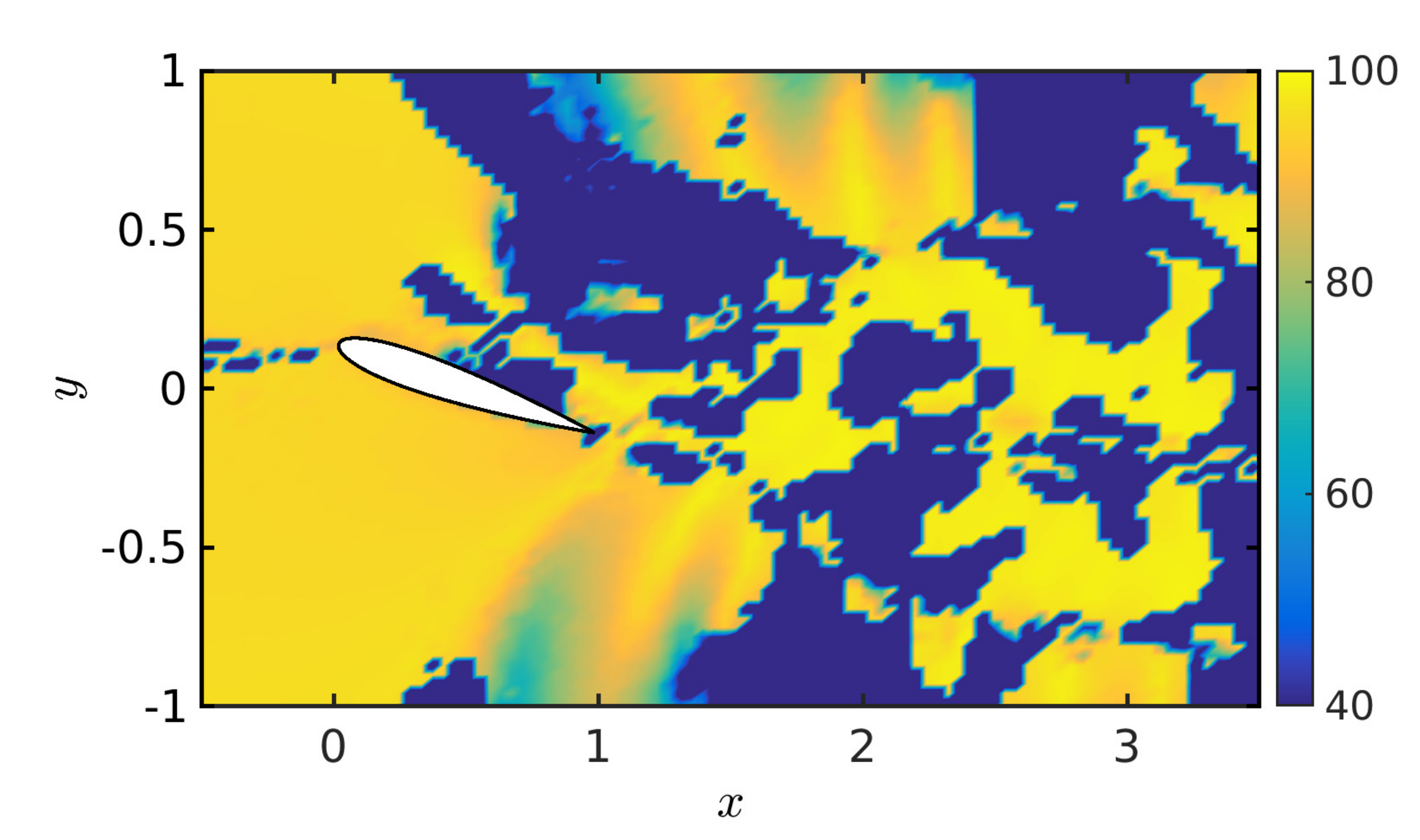}}
\\
\subfloat{\includegraphics[width=0.49\textwidth]{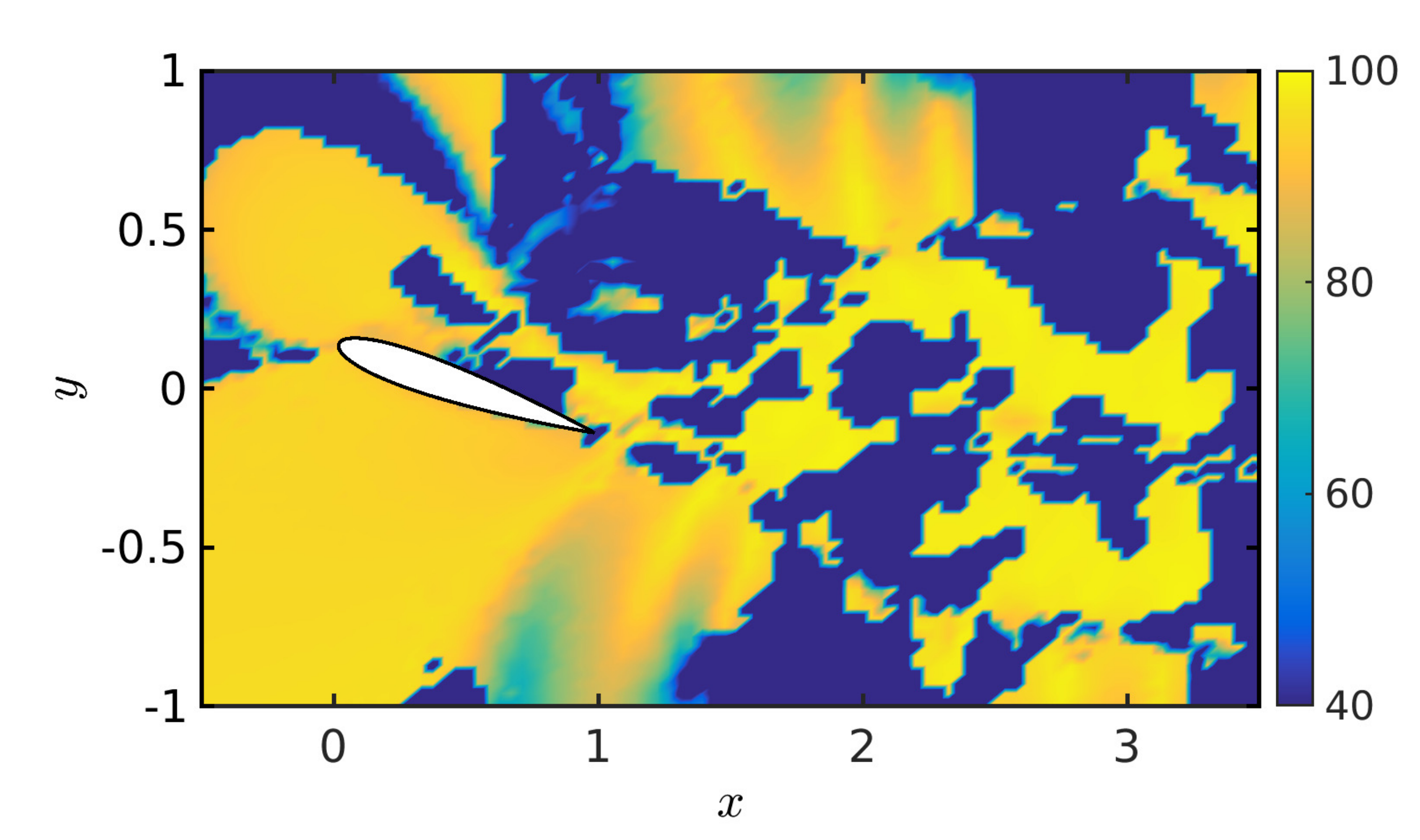}}
\subfloat{\includegraphics[width=0.49\textwidth]{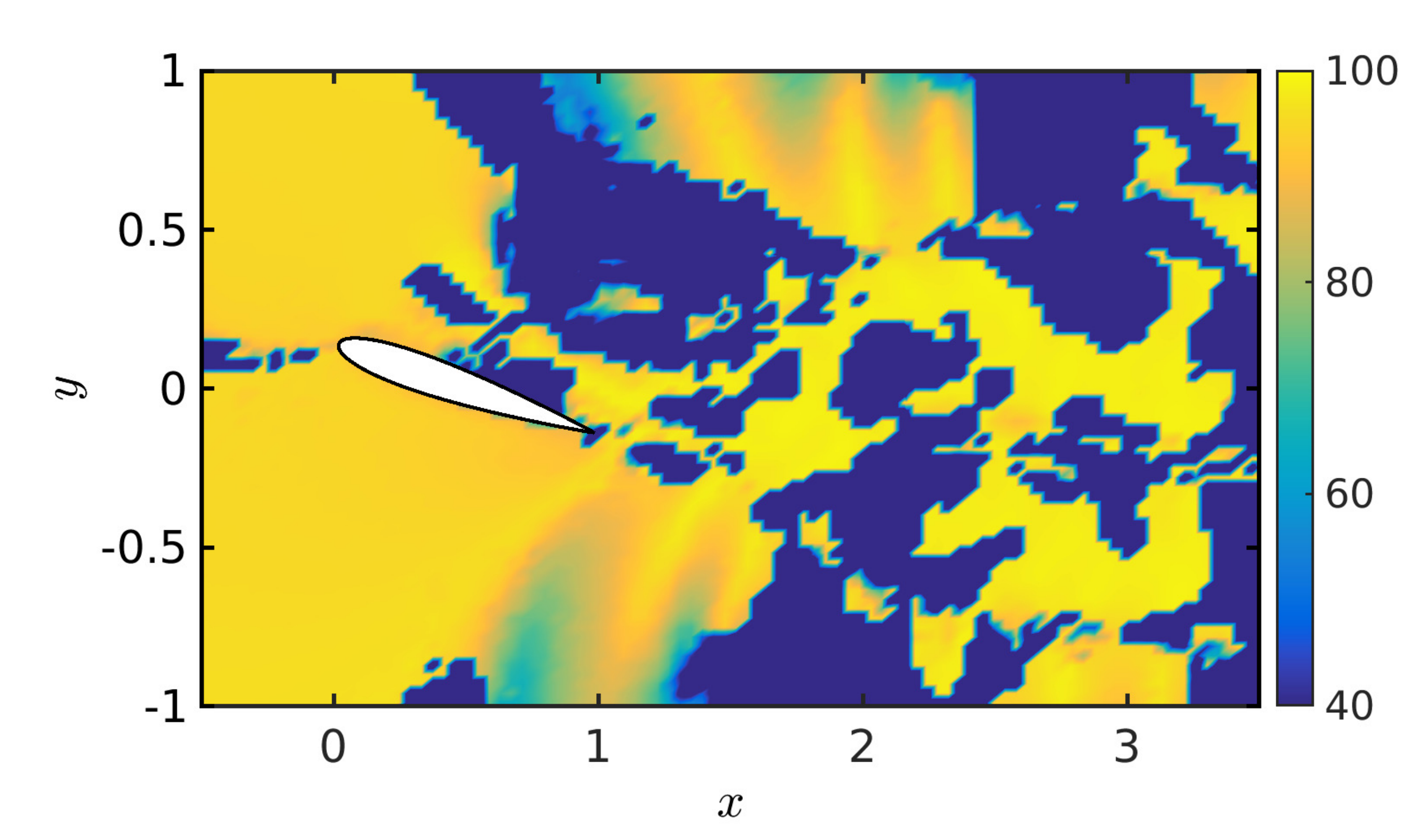}}
\caption{Influence of the position of the estimation sensor on the quality of the identification. $\mathsf{FIT}[\%]$ between the actual first POD coefficient and the coefficients estimated by models composed of 2 POD modes and 2 states obtained from different learning datasets. The datasets lengths are (top) 1000 and 
(bottom) 2000 snapshots, respectively. The input signal is $s=u$.}
\label{fig:FIT_input_U_POD_2_Nx_2}
\end{figure}

We consider first models with input the streamwise velocity recorded at different positions, i.e.\ $s=u(x_s,y_s)$. The number of states is set to $N_x=2$. Fig.~\ref{fig:FIT_input_U_POD_2_Nx_2} displays the fit of the first POD coefficient for a validation dataset {of length $m_{TD}=750$ (initialised by $\mathbf{Y}=\mathsf{C}^{\dagger}\mathbf{X}$).} Yellow (low error) and blue (high error) regions correspond to working and non-working locations for the sensor, respectively. The top row shows the results for a dataset of length 1000 snapshots and the bottom row results for 2000 snapshots. The right column illustrates the effect on the performance when the dataset is shifted in time. Note that the lower limit of the colour bar has been set to 40\%  for visualisation purposes but the actual minimal values can be lower. 

This plot shows that different datasets lead to similar distributions of  working/non-working regions. The pattern downstream of the aerofoil is repeatable in all 4 datasets, except from a region upstream and above  the aerofoil which is more erratic.  Apart from this region, neither the length of the dataset, nor the shift in time has an impact on the performance of the model: the results are therefore robust to such changes. In the following, we fix the dataset length to 2000 snapshots. 

\begin{figure}
\centering
\subfloat{\includegraphics[width=0.49\textwidth]{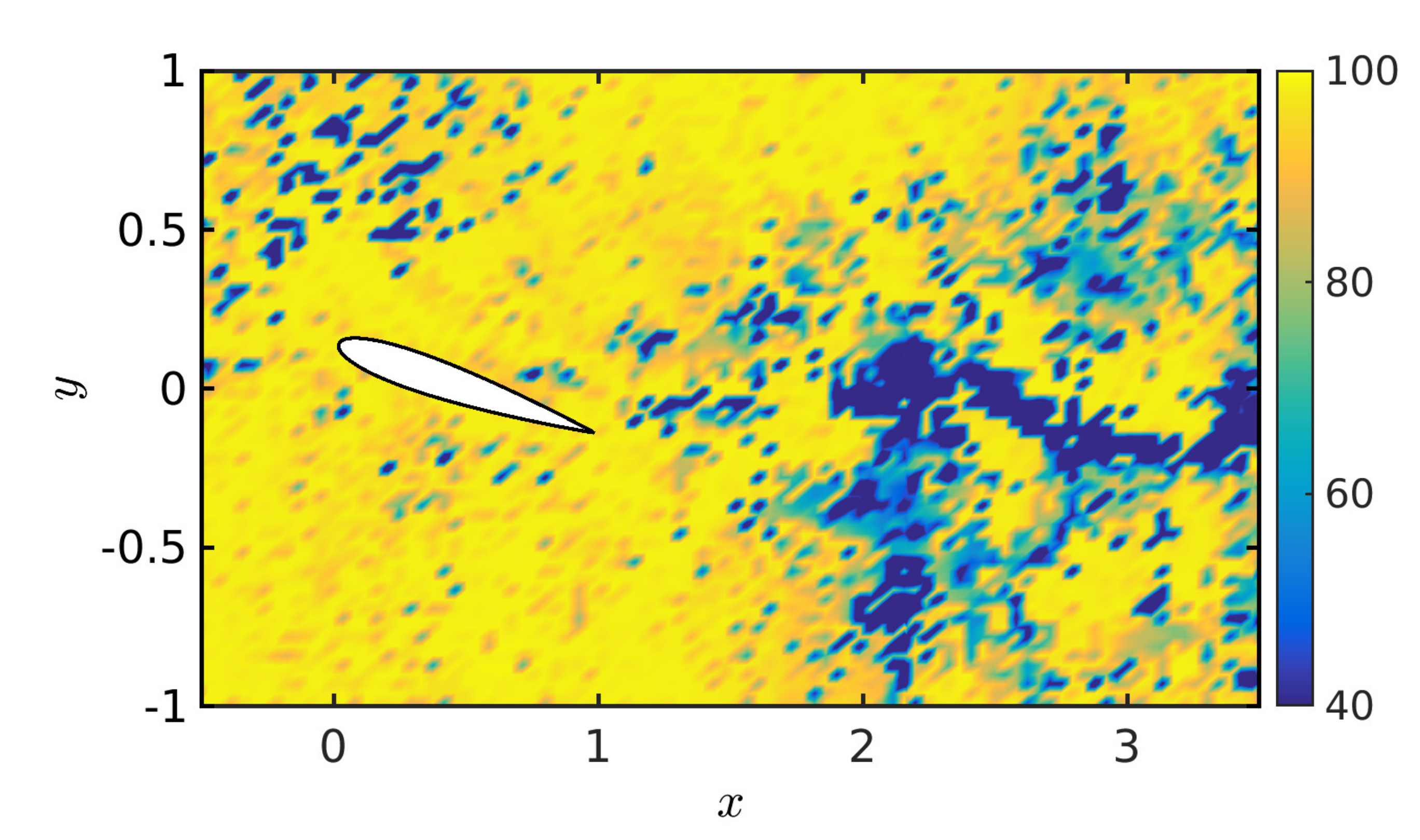}}
\subfloat{\includegraphics[width=0.49\textwidth]{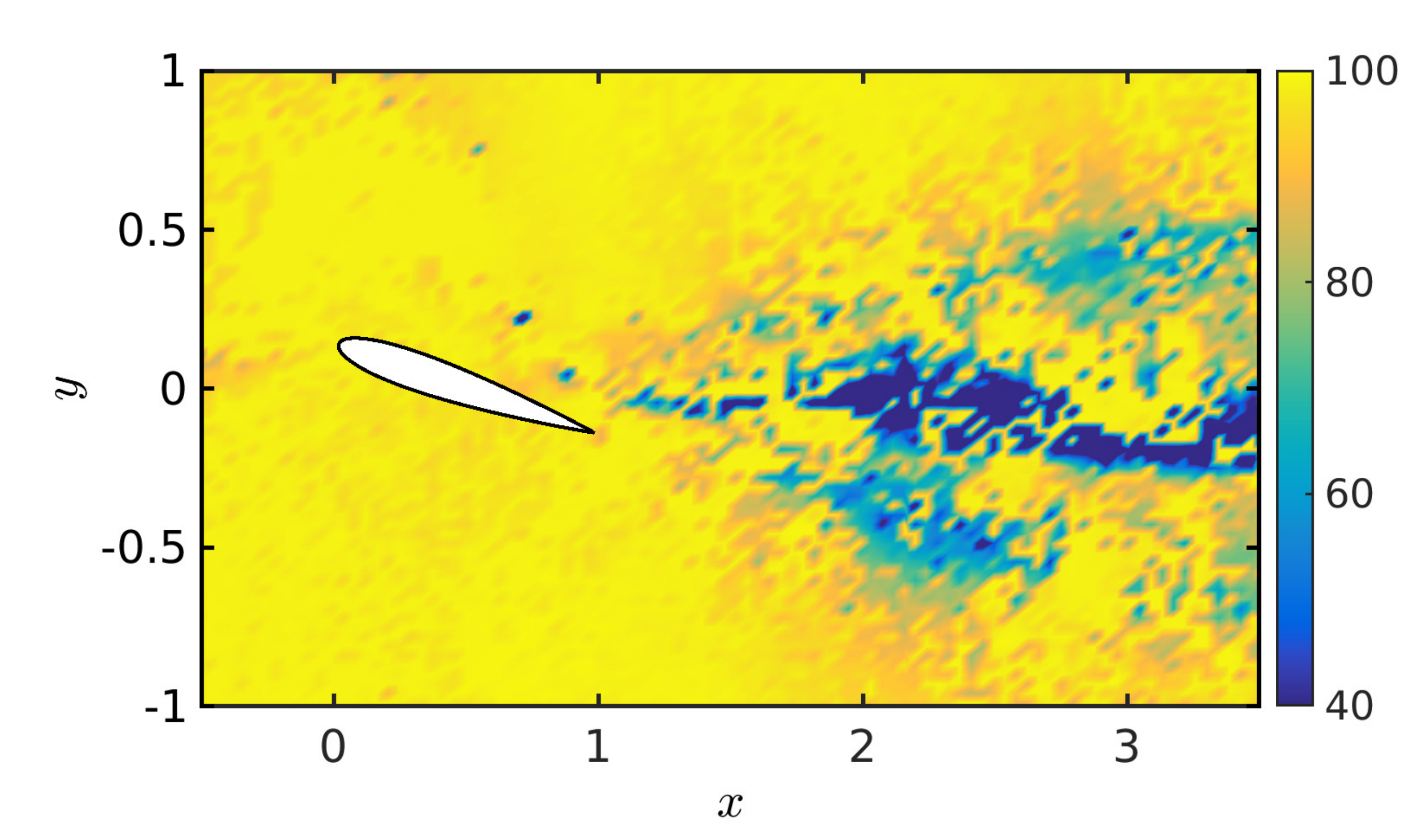}}
\caption{$\mathsf{FIT}[\%]$ between the actual first POD coefficient and the coefficients estimated by models composed of 2 POD modes and (left) 4 and (right) 6 states. The input signal is $s=u$.}
\label{fig:FIT_input_U_POD_2_Nx_4_6}
\end{figure}

We study next the effect of the number of states of the estimator, $N_x$. Fig.~\ref{fig:FIT_input_U_POD_2_Nx_4_6} shows the results for $N_x=4$ (left plot) and  for $N_x=6$ (right plot). In comparison to $N_x=2$, the utilisation of additional states leads to substantial improvement in performance, and a significant enlargement of the region where the estimator is effective. The ability to alter $N_x$ (and thereby improve performance) is an important advantage that dynamic estimators have over static ones, which are very common in literature as explained in the Introduction.  

It is interesting to notice an emerging pattern: as $N_x$ increases, the largest error (lowest fit) is concentrated in areas that become more and more well defined along a narrow region in the wake of the aerofoil and in two patches below and above (centred around $x\approx 2$ and $3,$ respectively). 


\begin{figure}
\centering
\subfloat{\includegraphics[width=0.49\textwidth]{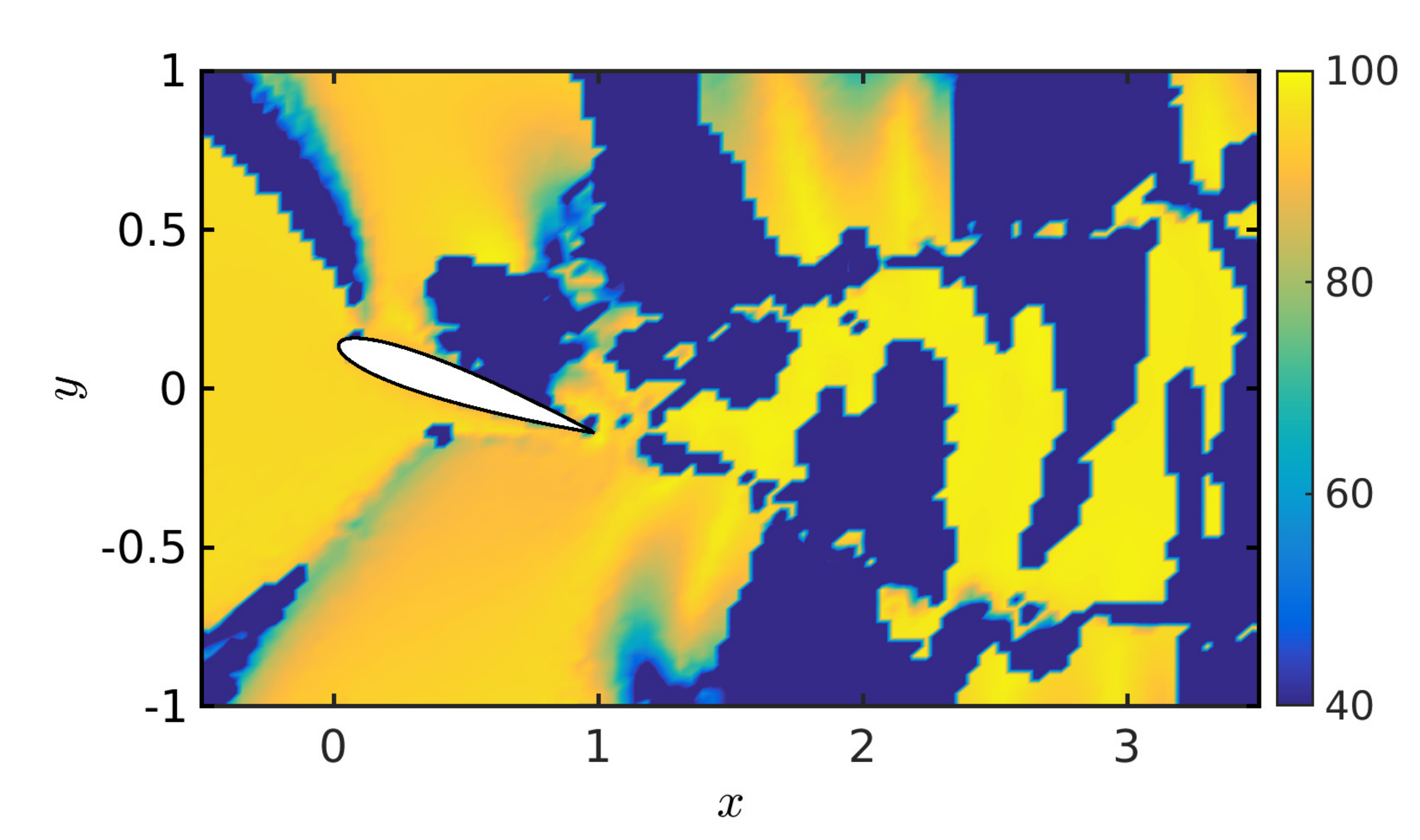}}
\subfloat{\includegraphics[width=0.49\textwidth]{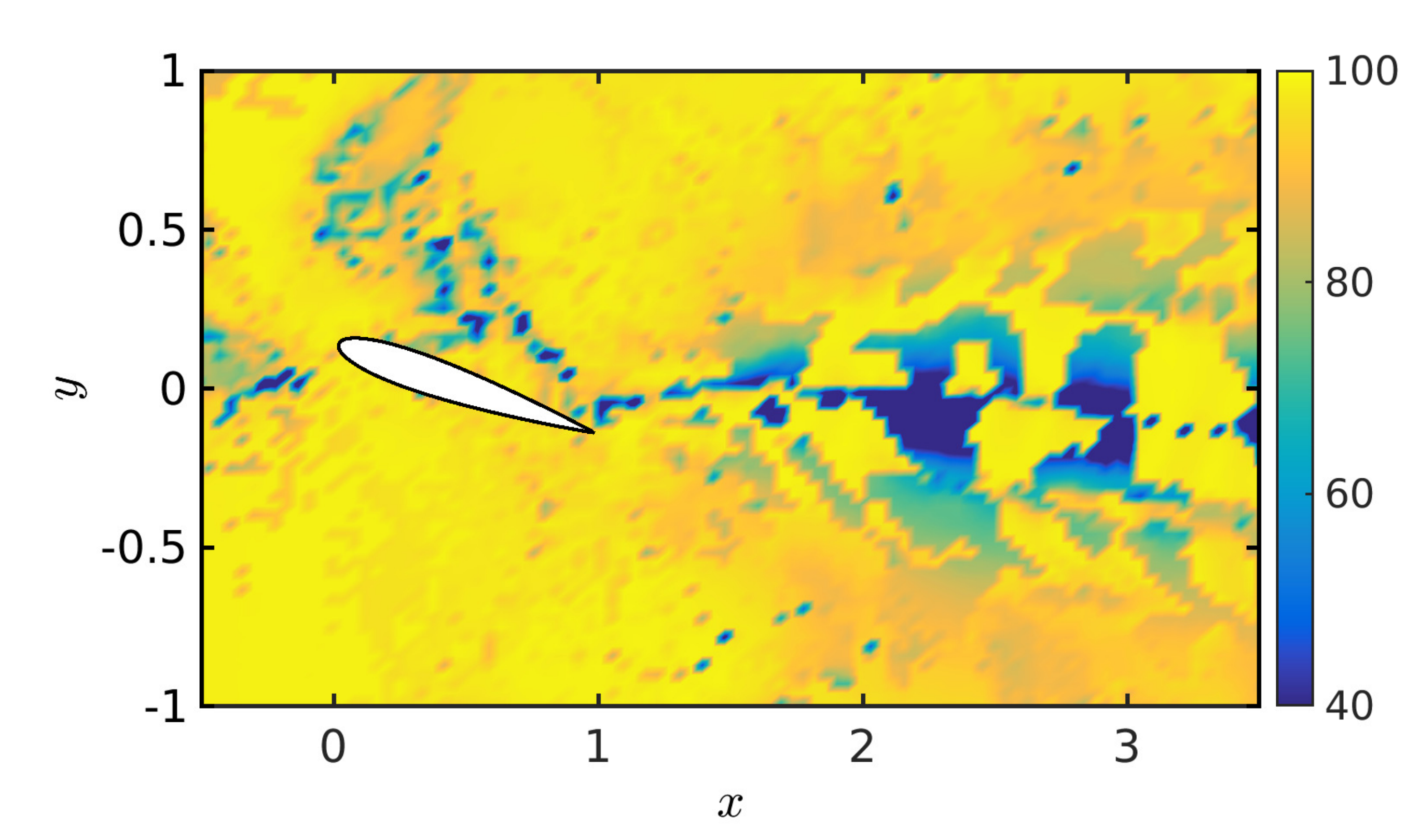}}
\\
\subfloat{\includegraphics[width=0.49\textwidth]{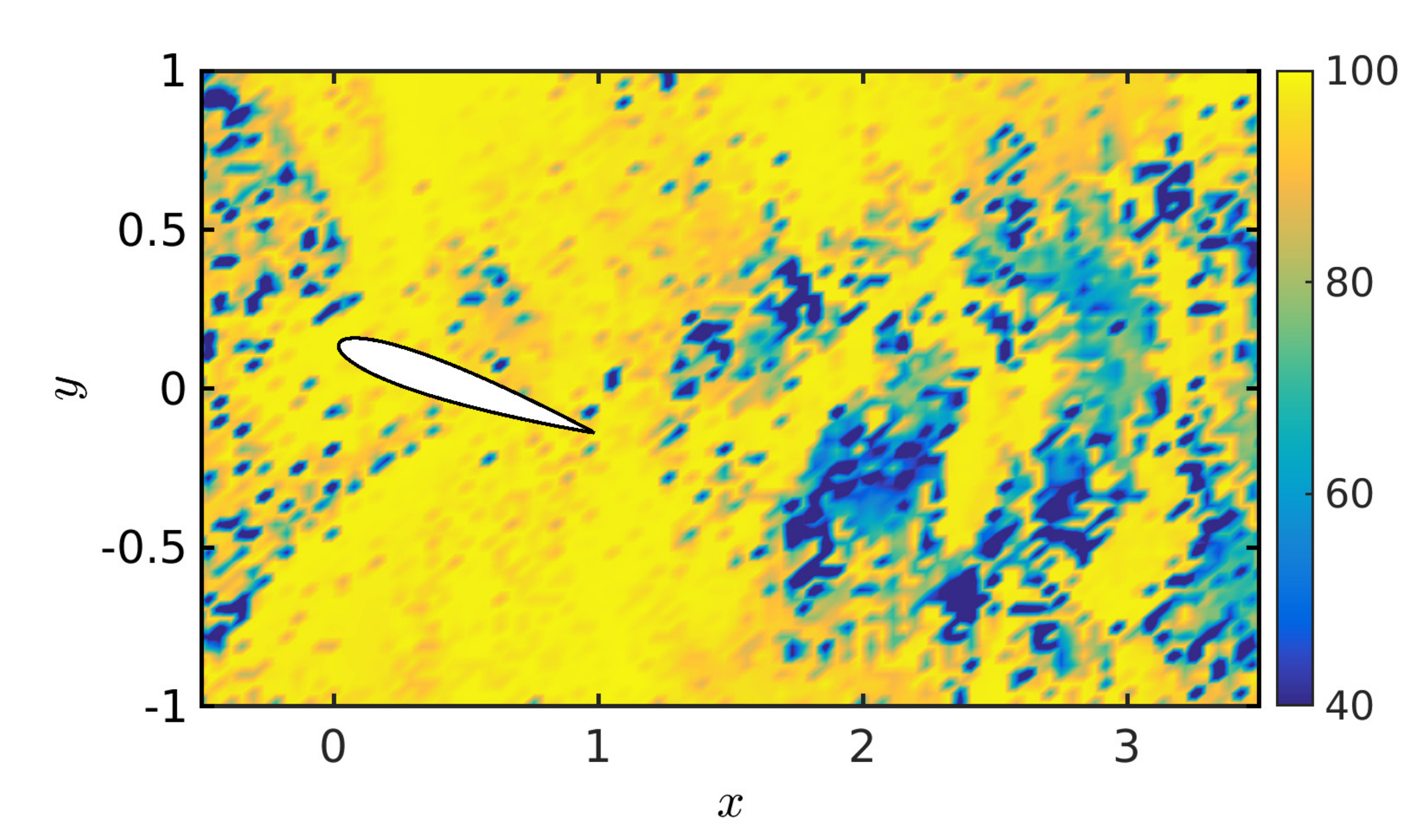}}
\subfloat{\includegraphics[width=0.49\textwidth]{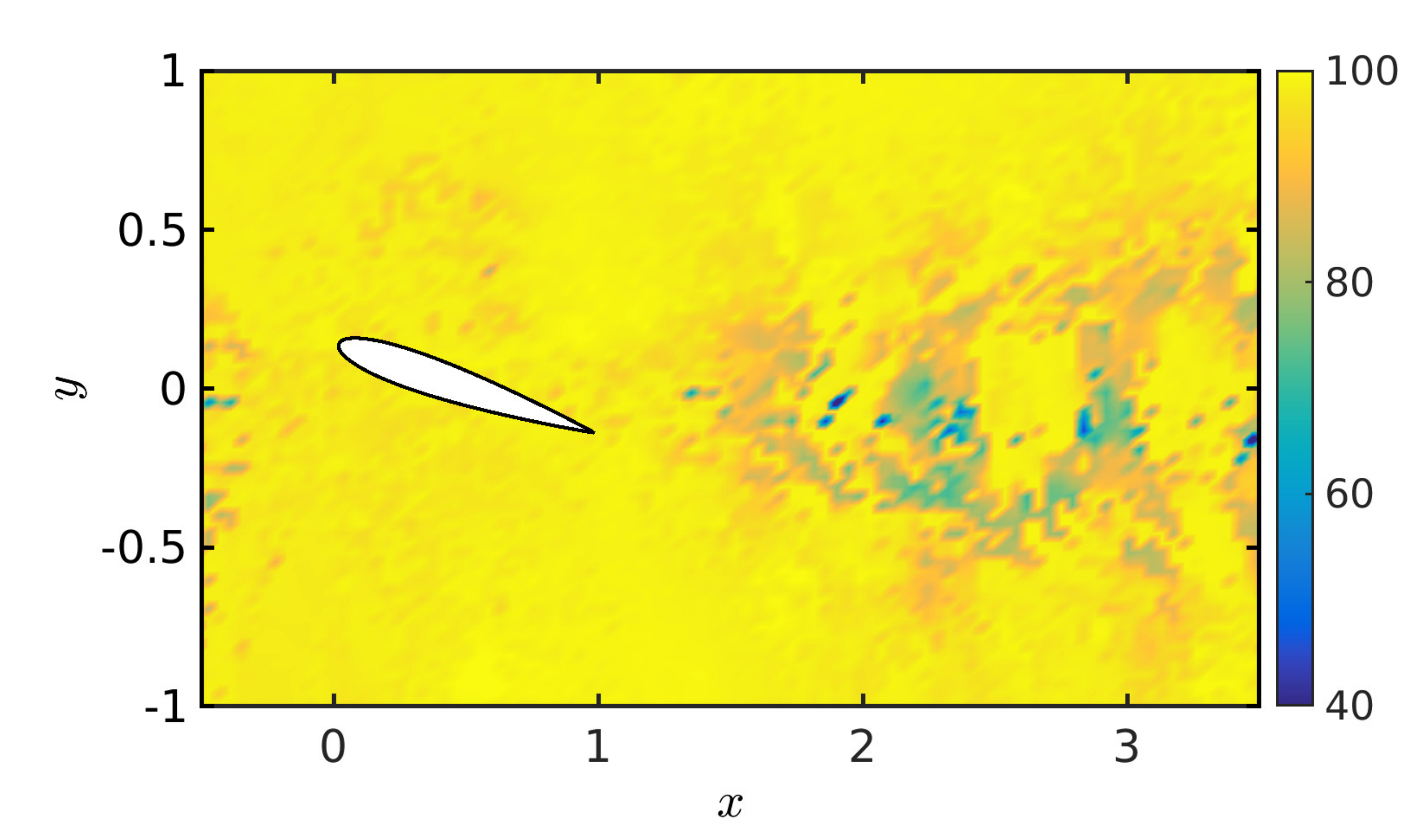}}
\\
\subfloat{\includegraphics[width=0.49\textwidth]{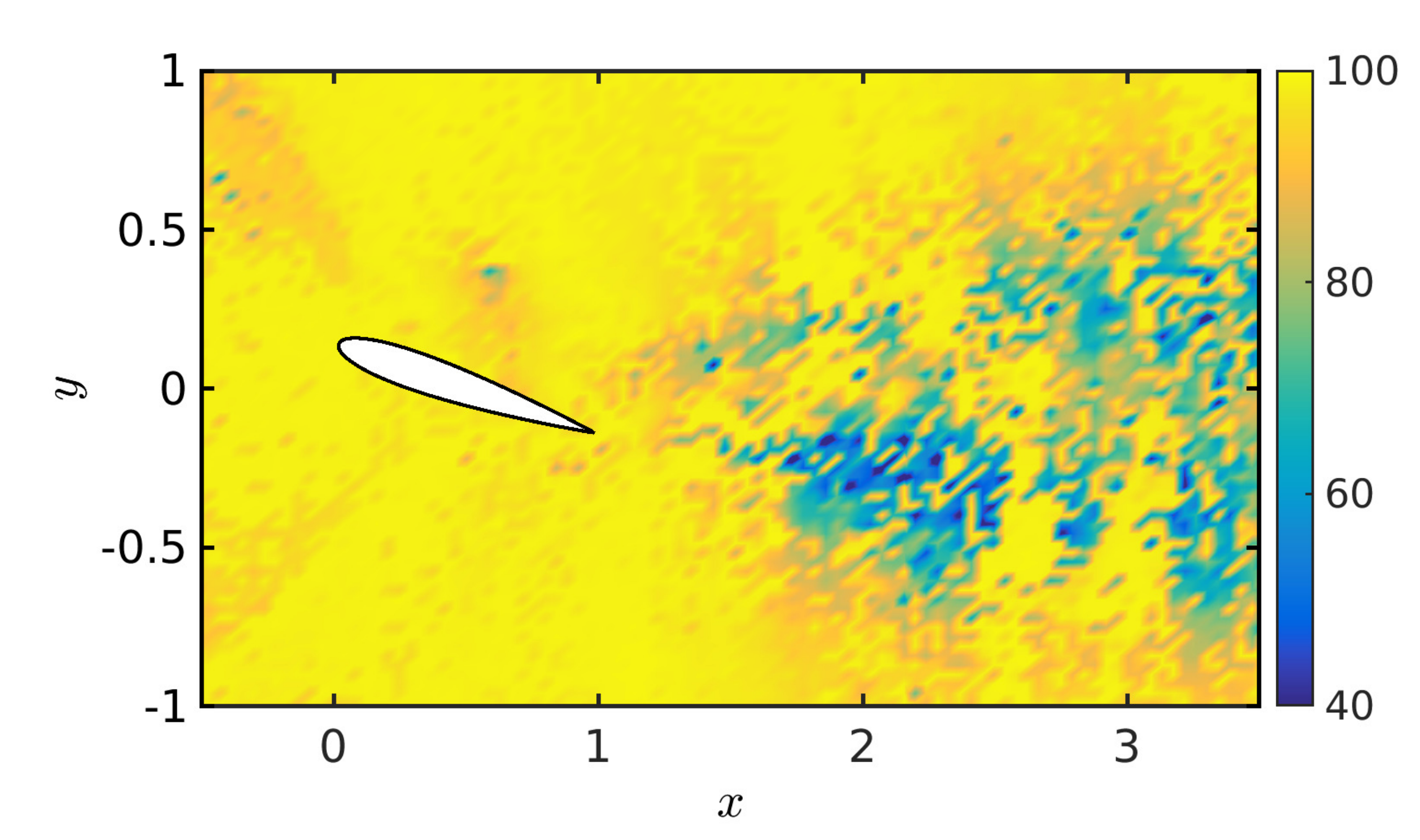}}
\subfloat{\includegraphics[width=0.49\textwidth]{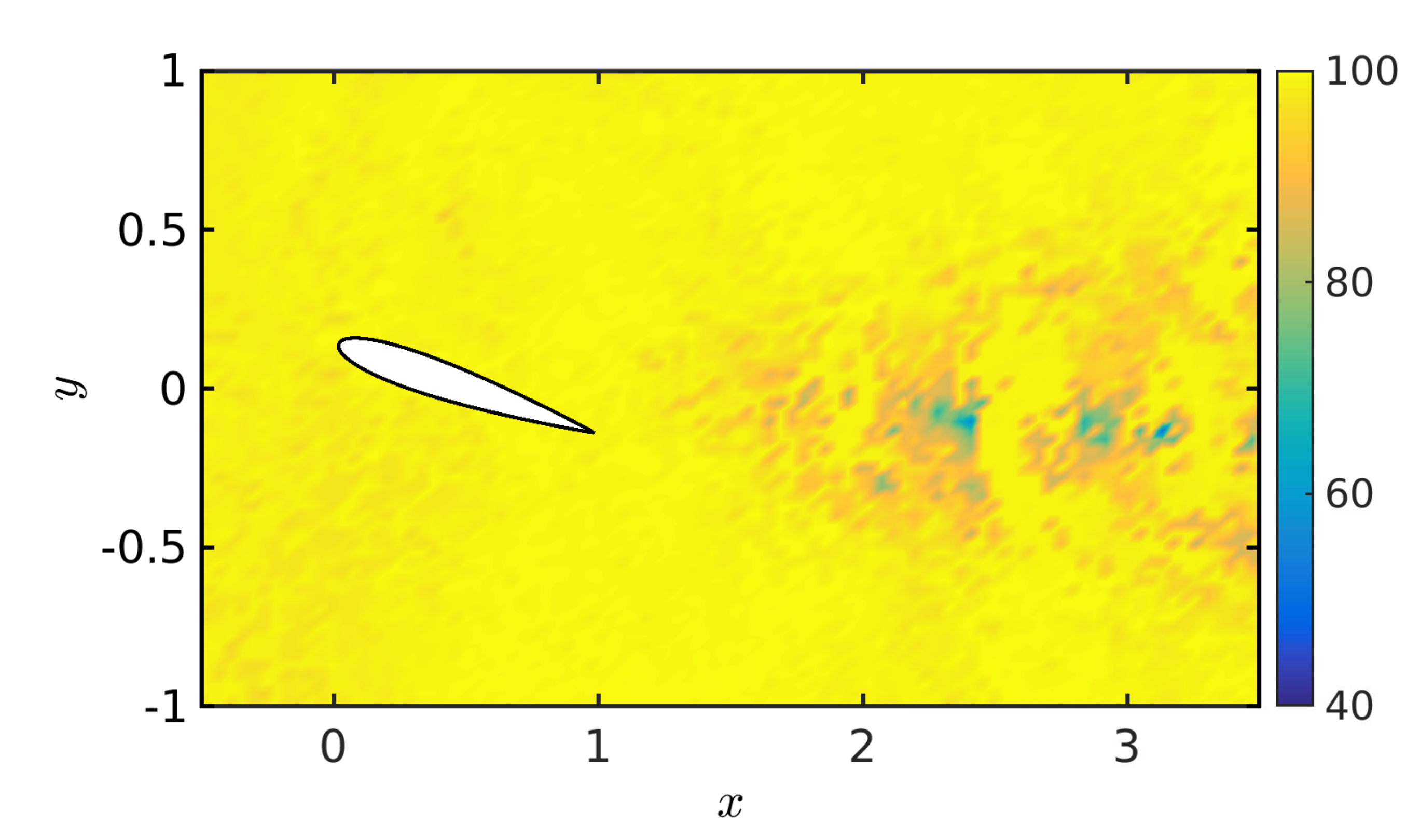}}
\caption{Influence of the input signal on the quality of the model. $\mathsf{FIT}[\%]$ between the actual first POD coefficient and the 
coefficients estimated by models obtained using (left) the cross-stream $s=v,$ and (right) both the streamwise and cross-stream $s=\left[u, v\right]^\top$ components of the oscillatory velocity. The models are composed of 2 POD modes and (top) 2, (middle) 4 and (bottom) 6 states.}
\label{fig:FIT_input_V_UV_POD_2_Nx_2_4_6}
\end{figure}

We now explore the performance when $s=v$ and $s=[u,v]^\top$. The results are depicted in the left and right columns of Fig.~\ref{fig:FIT_input_V_UV_POD_2_Nx_2_4_6} respectively, for three different values of $N_x=2,4,6$. Using the $v$ velocity component appears to yield slightly improved results for all number of states investigated. As before, an increase in the number of states significantly improves the performance of the model.  Again a pattern emerges, and the areas of low performance become sharper. 
When both velocity components are used, i.e.\ $s=[u,v]^\top$, the performance is much improved for all orders $N_x$. Now almost any point can be used for flow reconstruction. 

\begin{figure}
\centering
\subfloat{\includegraphics[width=\textwidth]{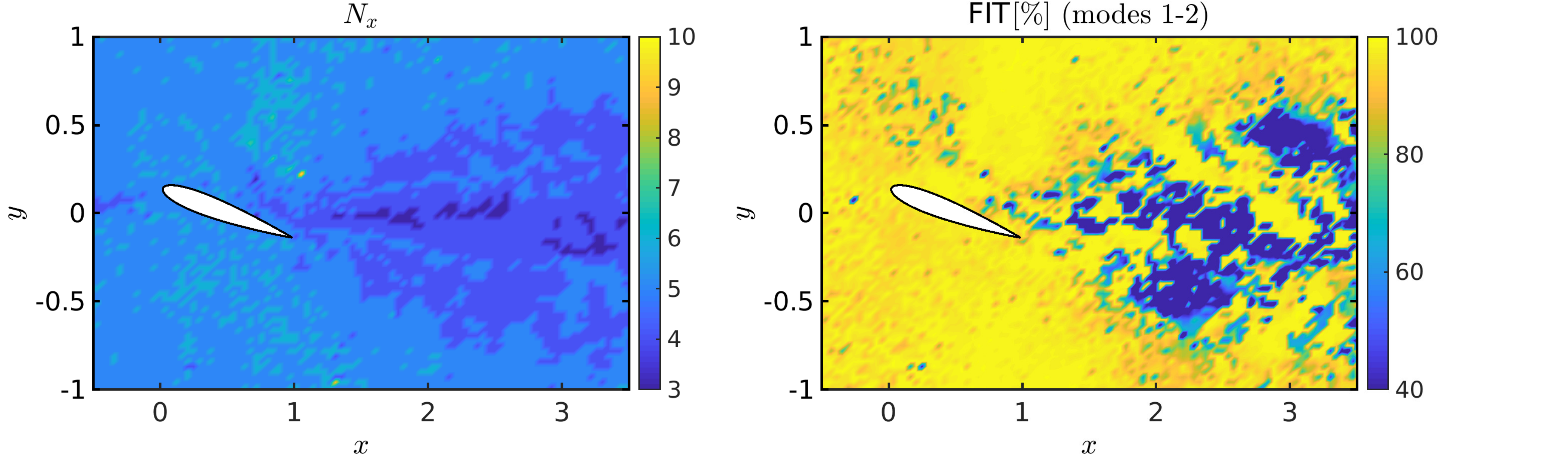}}
\\
\subfloat{\includegraphics[width=\textwidth]{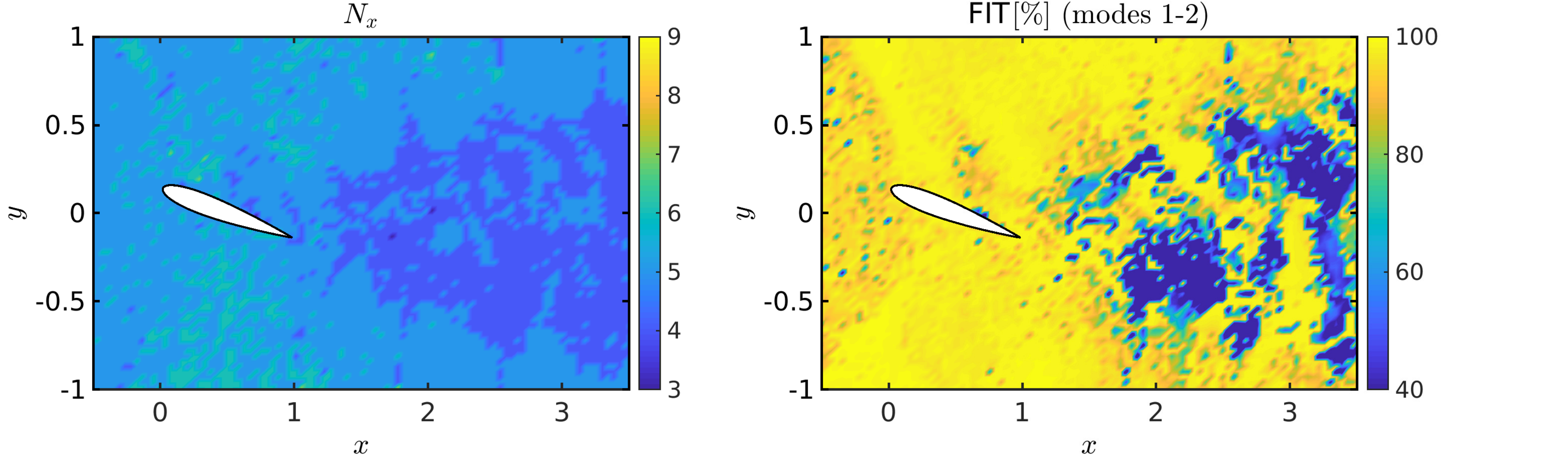}}
\\
\subfloat{\includegraphics[width=\textwidth]{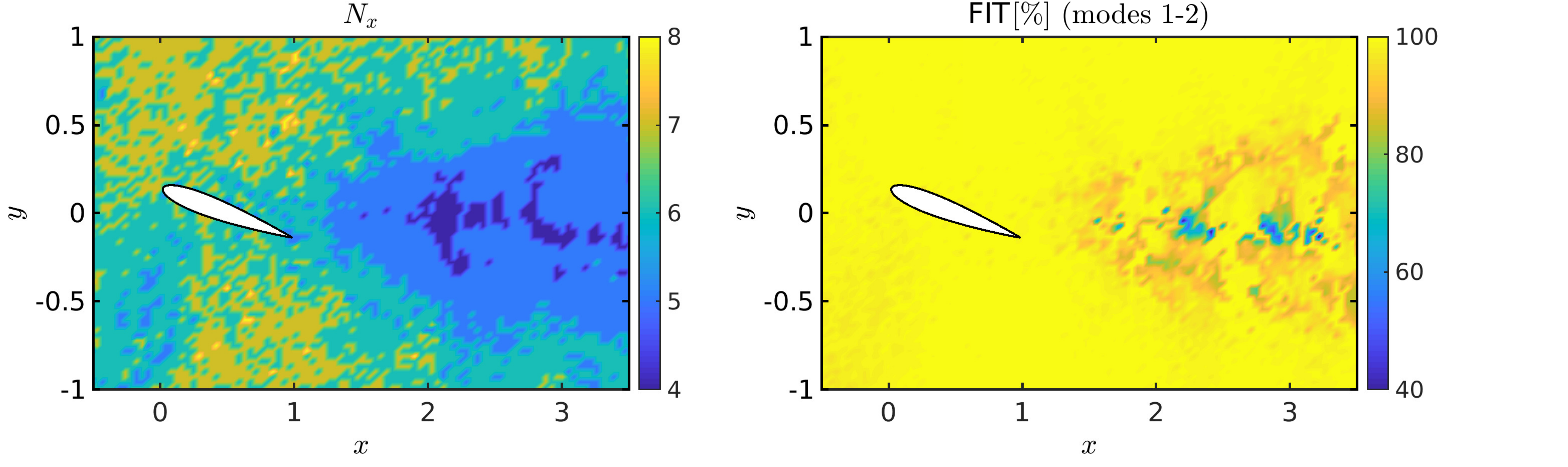}}
\caption{(Left) Optimal number of states estimated by the \texttt{N4SID} algorithm for 2 POD modes and (right) $\mathsf{FIT}[\%]$ between the actual first POD coefficient and the coefficient estimated by such models. The models are obtained using (top) the streamwise $s=u$, (middle) the cross-stream $s=v$ and (bottom) both the streamwise and cross-stream $s=\left[u, v\right]^\top$ components of the oscillatory velocity.}
\label{fig:FIT_POD_2_optimal}
\end{figure}


{It is clear that an increase of the number of
states of the system leads to an increase of the estimator's performance. However, there exists 
a limit beyond which any additional state leads to unnecessary complex models (over-fitting). Different criteria can be used to define if a model is under-fitted or over-fitted (see chapter 16 of the book~\citet{ljung1999system}). Here, we use the optimal number of states of the model.} As mentioned in Sec.~\ref{sec:SysIdent}, the \texttt{N4SID} algorithm can return not only the matrices that best represent the input-output behaviour of the training dataset, but also the optimal order of the estimator (note that the MATLAB implementation limits this value to 10). The order $N_x$ is obtained from the distribution of the singular values of the weighted matrix (constructed by the algorithm using the Hankel matrices of the input-output data). The selection of the dominant singular values is straightforward when the largest ones are well separated from the rest, but it is not so clear--cut when the decay is smooth.  

Fig.~\ref{fig:FIT_POD_2_optimal} illustrates the spatial distribution of the optimal $N_x$ value (in the left column) and the corresponding $\mathsf{FIT}[\%]$ (in the right column). Each row represents different input ($s=u$, $s=v$ and $s=[u,v]^\top$ for the top, middle and bottom rows respectively). When $s=u$, the optimal number of states is between $3-6$, with the smaller numbers localised in the wake. In agreement with the previous figure \ref{fig:FIT_input_U_POD_2_Nx_4_6}, the low fit areas in the wake are again localised along a horizontal region centred around $y\approx 0$ and two patches above and below. This result suggests that an increase in the number of states does not yield a significant change in the low fit areas. It can however affect the $\mathsf{FIT}[\%]$ values. Indeed the fit is better in Fig.~\ref{fig:FIT_input_U_POD_2_Nx_4_6} compared to Fig. \ref{fig:FIT_POD_2_optimal} because $N_x$ was higher. This indicates that the improvement is probably due to over-fitting.  When $s=v$, again the optimal $N_x$ values are smaller in the wake, and the fit pattern is similar to that of the previous figure \ref{fig:FIT_input_V_UV_POD_2_Nx_2_4_6} where $N_x$ was constant. 

When $s=[u,v]^\top$, the results indicate that a model with 4 to 5 states in the wake performs very well. There are some localised spots with low fit, centred around $y\approx 0$, but at these points the computed number of states is minimum (equal to $N_x=4$). It is likely that the algorithm cannot compute accurately the optimal value of $N_x$ at these points, because there is no clear separation in the distribution of singular values (and the boundary between optimal fitting and over-fitting is not clear).  


\begin{figure}
\centering
\subfloat{\includegraphics[width=0.4\textwidth]{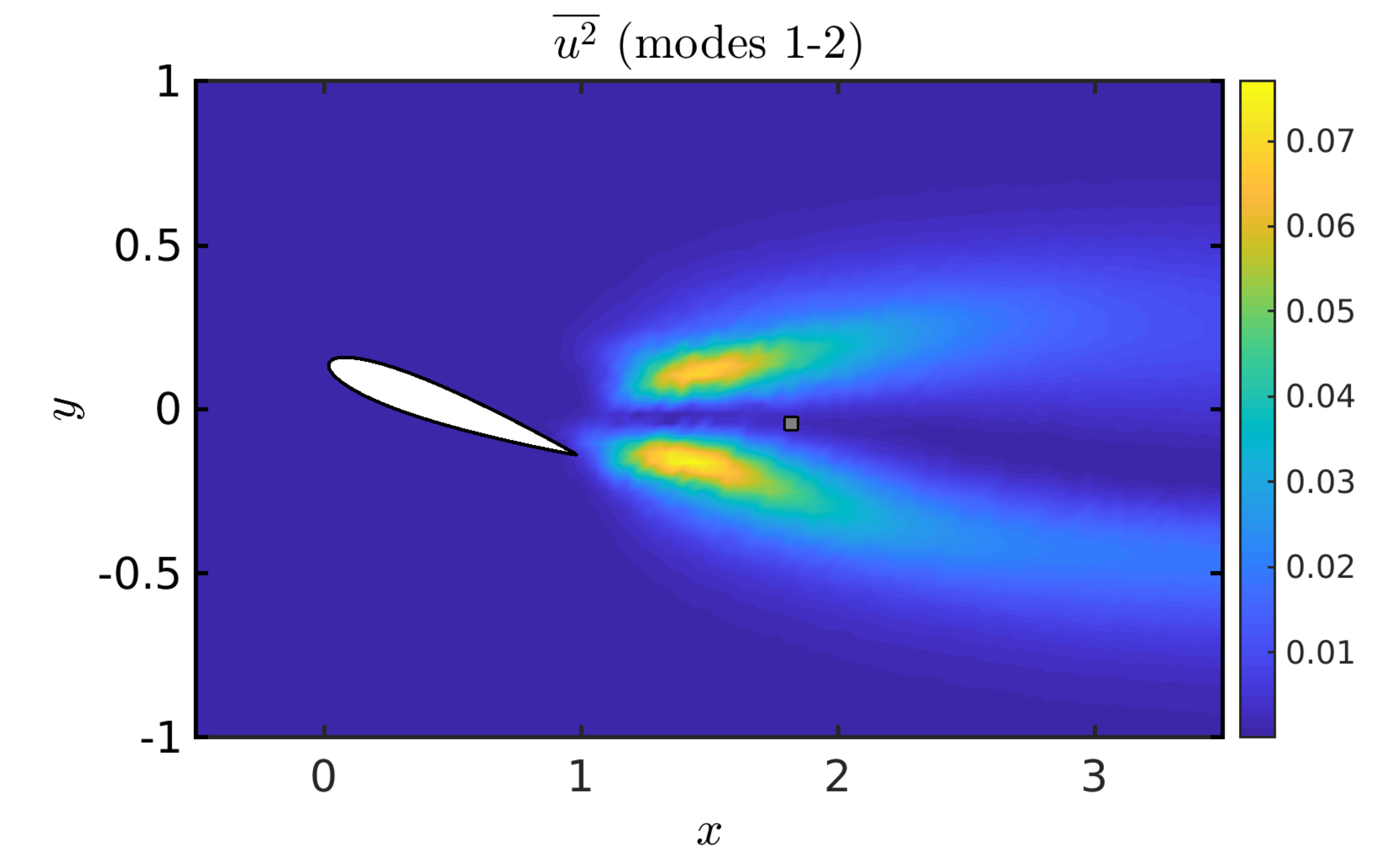}}
\subfloat{\includegraphics[width=0.4\textwidth]{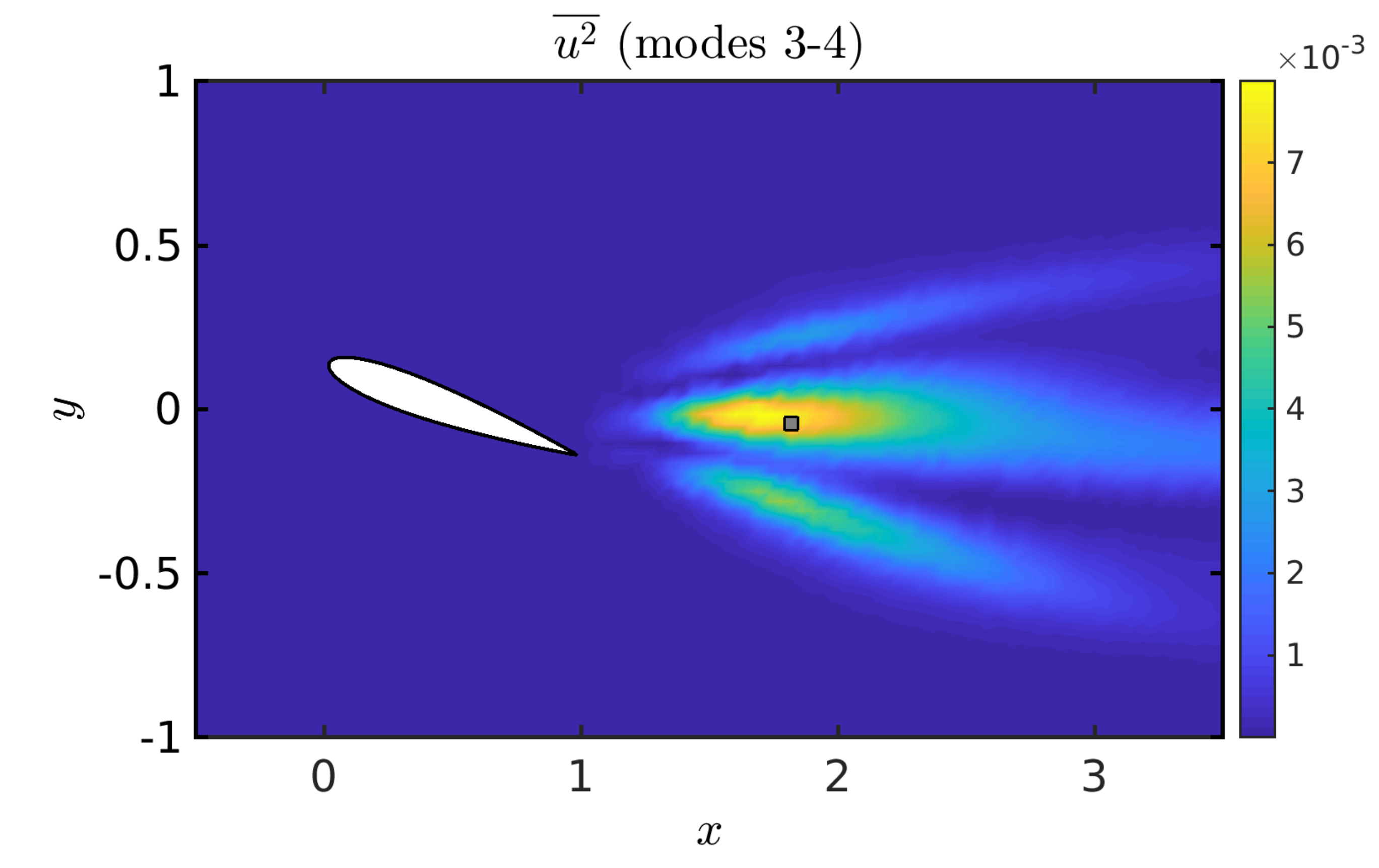}}
\\
\subfloat{\includegraphics[width=0.4\textwidth]{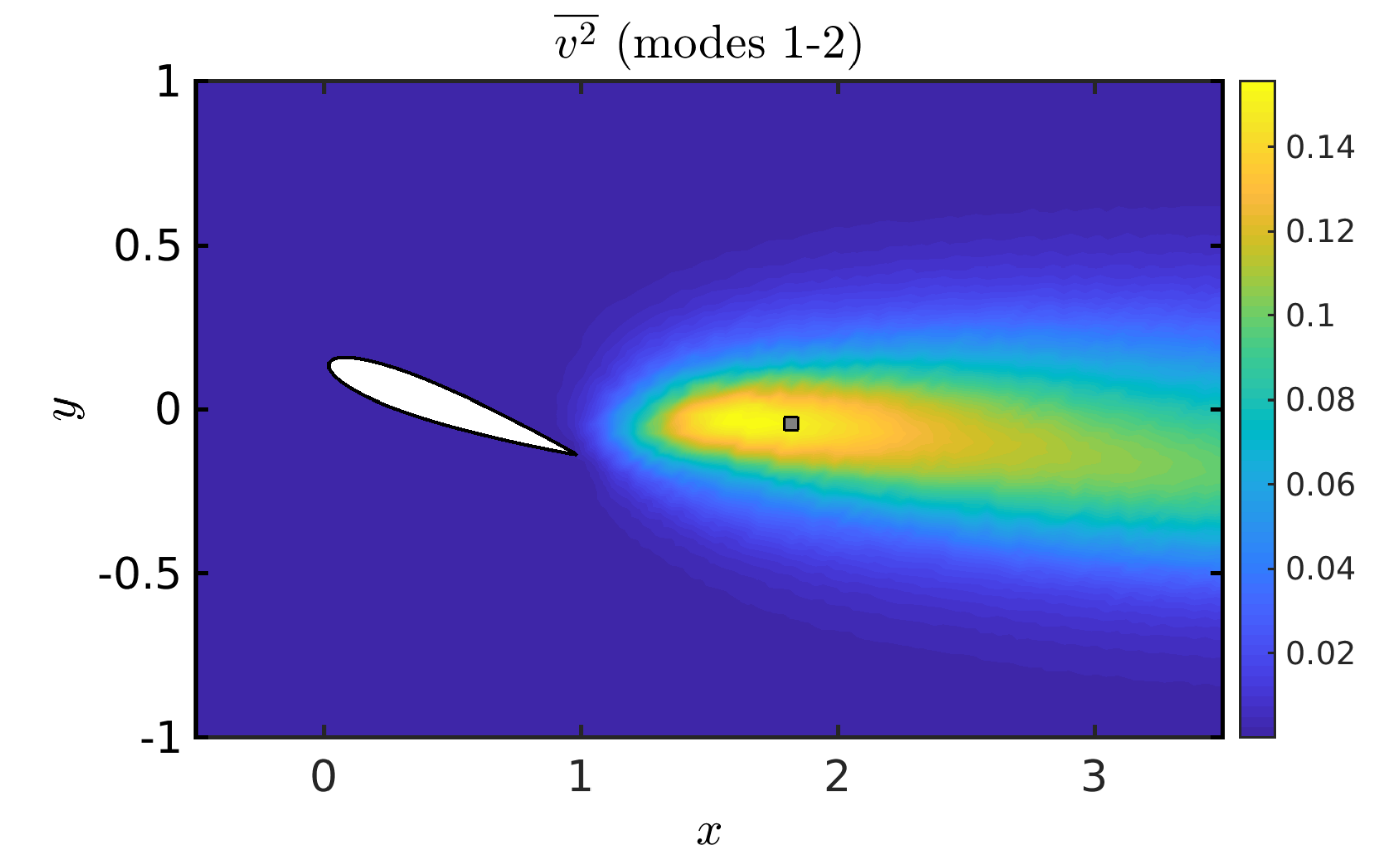}}
\subfloat{\includegraphics[width=0.4\textwidth]{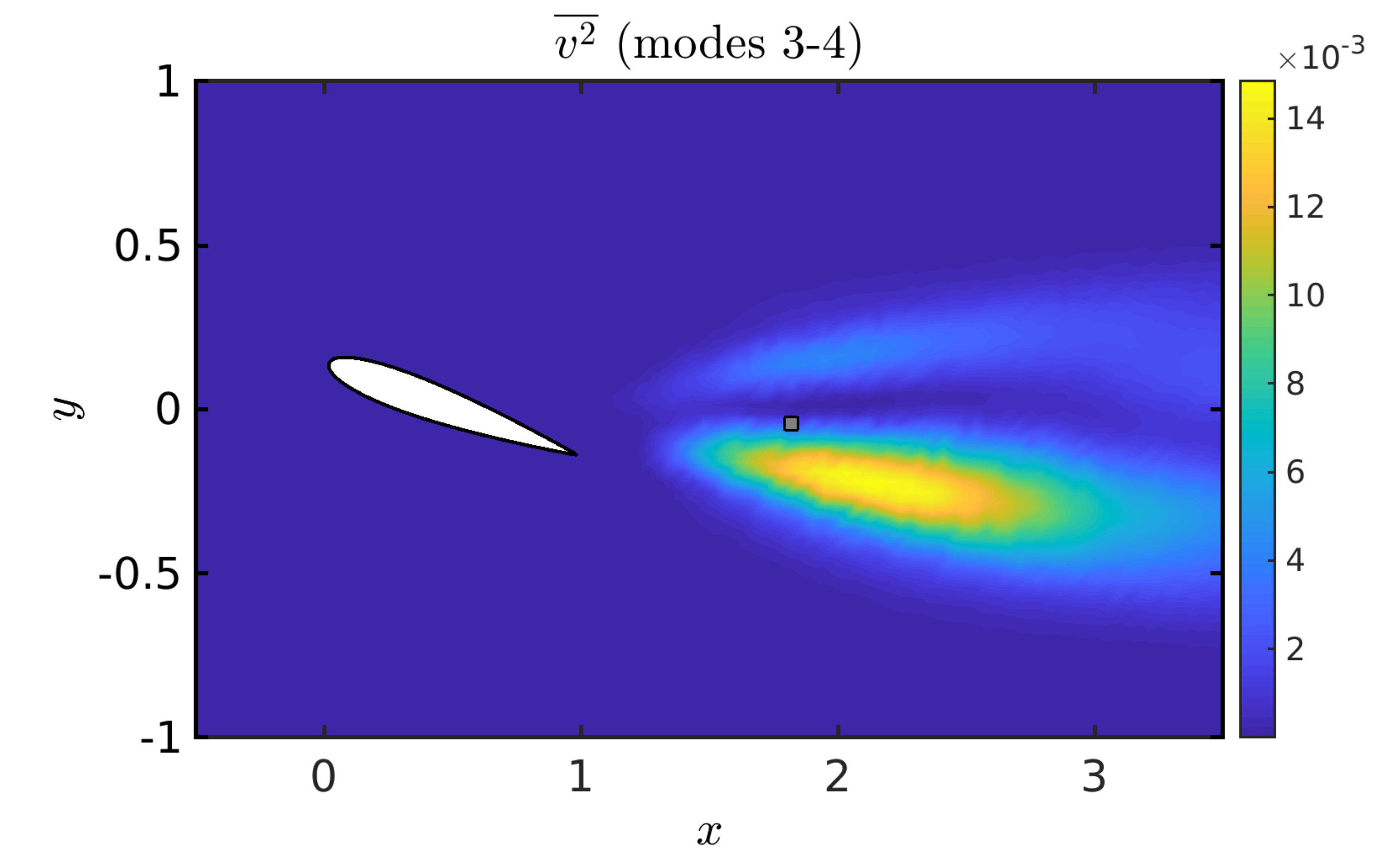}}
\caption{Reynolds stress of the (top) streamwise and (bottom) cross-stream oscillatory velocity projected on (left) the first and second and (right) the third and fourth POD modes, respectively. Point $(x_{\rm{s}},y_{\rm{s}})=(1.82,-0.04)$ is denoted with an open square.}
\label{fig:TKE}
\end{figure}

\begin{figure}
\centering
\subfloat{\includegraphics[width=0.49\textwidth]{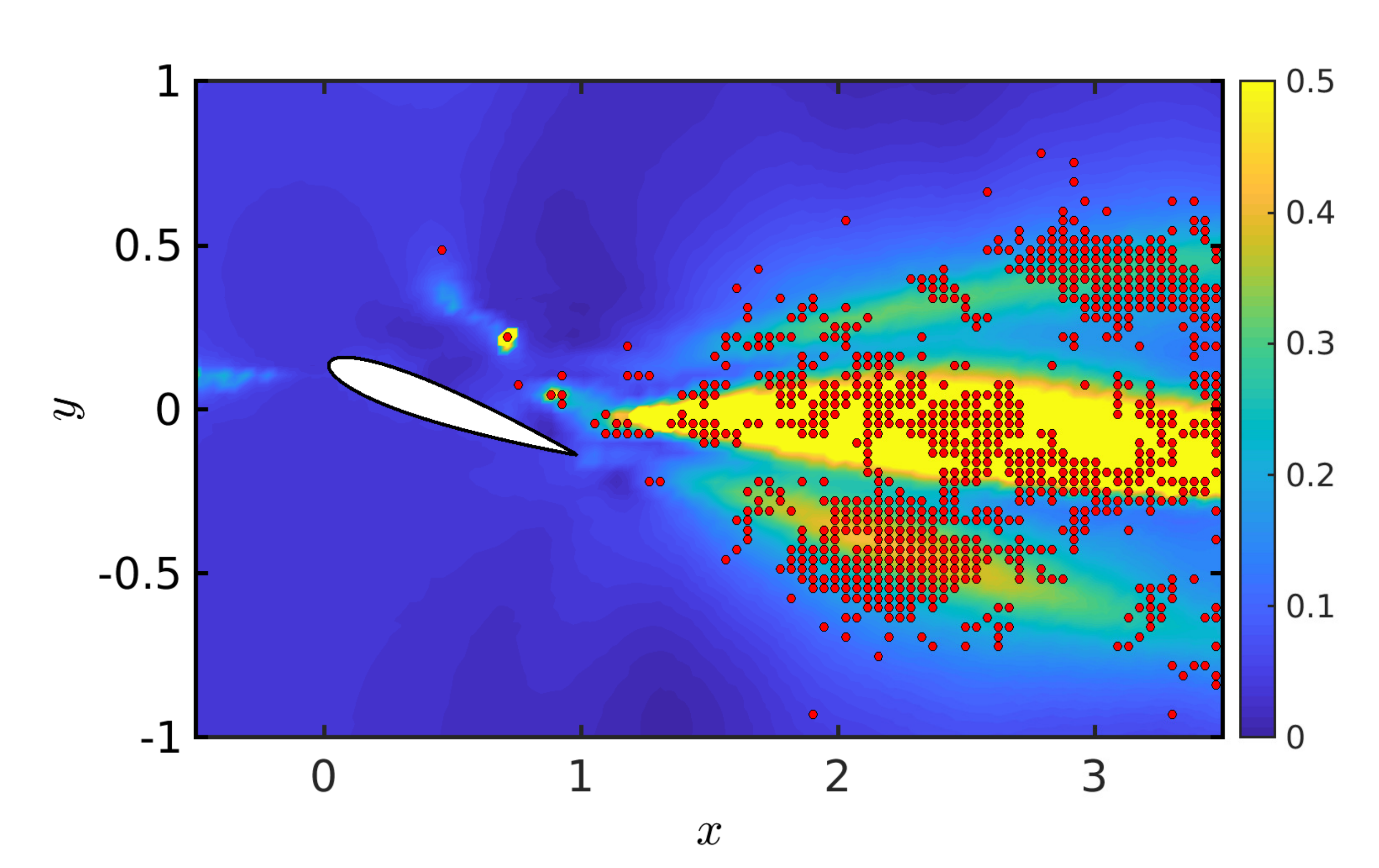}}
\subfloat{\includegraphics[width=0.49\textwidth]{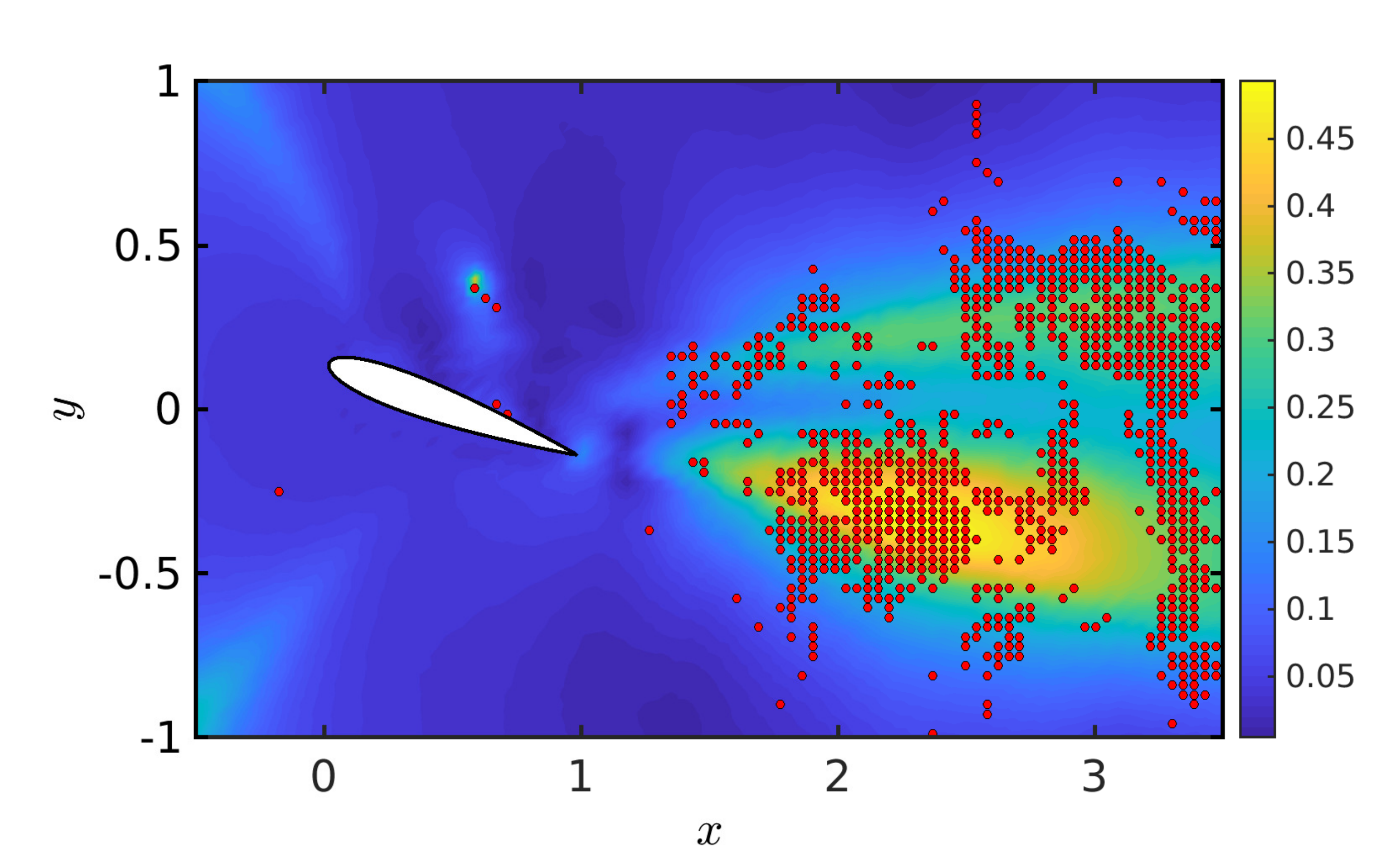}}
\caption{Square root of the ratio of Reynolds stresses contained in modes $3-m$ and $1-2$ superimposed with the locations of the sensor where the $\mathsf{FIT}[\%]$ is less than 60\% (shown as red dots). (left) $\overline{u^2}$ Reynolds stress, (right) $\overline{v^2}$ Reynolds stress.}
\label{fig:TKE ratio}
\end{figure}

To gain further insight into these results, the spatial 
distribution of the time-average 
Reynolds stresses $\overline{u^2}$ and $\overline{v^2}$ of the mode pairs 1-2 and 3-4 are computed.  The left column of Fig.~\ref{fig:TKE} depicts the results for the projection on modes 1-2 (shedding mode) and the right column on modes 3-4 (first harmonic). This figure collects together the areas of spatial dominance of the two most significant mode pairs.

Since each pair of modes is associated with one frequency, Fig.~\ref{fig:TKE} can be used to deduce the dominant frequencies of the signal $s$. For example, if the input is the streamwise $u$ velocity (i.e.\ $s=u$), the vortex shedding frequency ($f=0.6$) is dominant for sensors located along the path of the vortices shed from the pressure and suction sides (top, left figure). If, on the other hand, the $u$ sensor is located in the area sandwiched between these two paths, this main frequency is much weaker, while the first harmonic ($f=1.2$) becomes very strong. The spatial distribution of $\overline{v^2}$ is different. For the main shedding mode, the maximum is found around $y\approx 0$, and is larger compared to $\overline{u^2}$ (by a factor of about 2). If $s=v$, then a sensor located in this region will give a very strong signal at the main frequency. For the first harmonic frequency, the region of maximum values is slanted compared to the free-stream velocity. 

For $k=2$, the estimator needs to filter out any frequency different from the main 
shedding frequency. Therefore, we can assume that the observer's performance will be higher
when the sensor is placed in points where this main frequency is dominant. In order to verify this, we can compare the Reynolds stresses of the velocity field projected on the first pair of modes with the Reynolds stresses for the rest of them. In Fig.~\ref{fig:TKE ratio} (left), we 
show the ratio of Reynolds stresses for modes 3 to $m$ and modes 1-2  $\left[\overline{u^2}\right]_{3-m}/\left[\overline{u^2}\right]_{1-2}$ and we superimpose the locations where $\mathsf{FIT}[\%]$ is less than 60 \% (as red dots). Although there is some scatter, the correlation is clear: in the three regions where the aforementioned ratio is high, the fit is low. This makes intuitive sense: it's difficult for the model to extract the main frequency when the content at this frequency is low and the input signal is dominated by other frequencies. Although in Fig.~\ref{fig:TKE ratio}, the ratio is limited to $0.5$, the actual values are higher. 

\subsection{\label{subsec:4modes}4 POD modes}
\begin{figure}
\centering
\subfloat{\includegraphics[width=\textwidth]{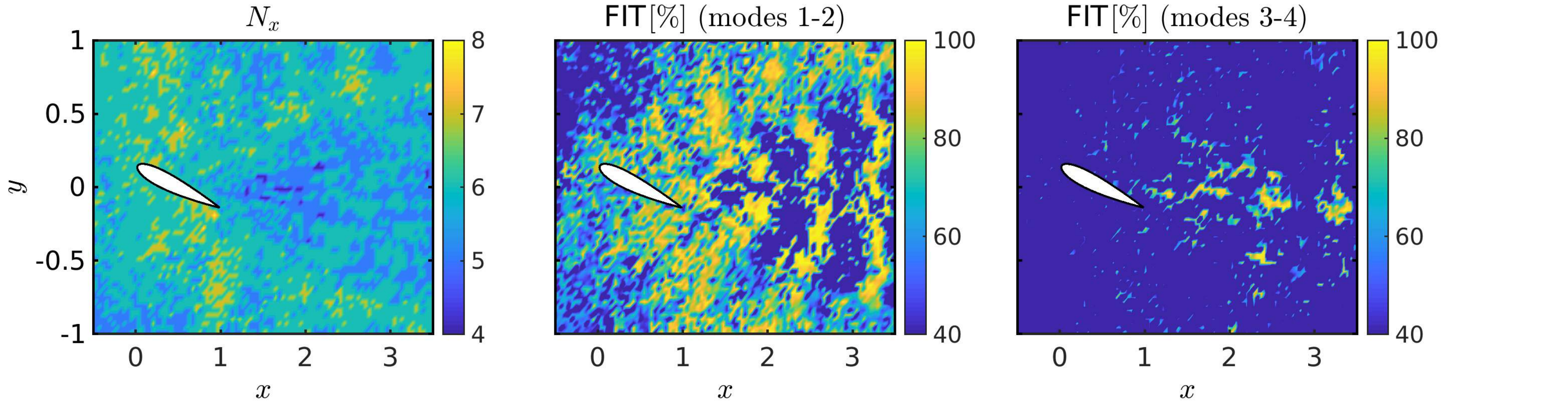}}
\\
\subfloat{\includegraphics[width=\textwidth]{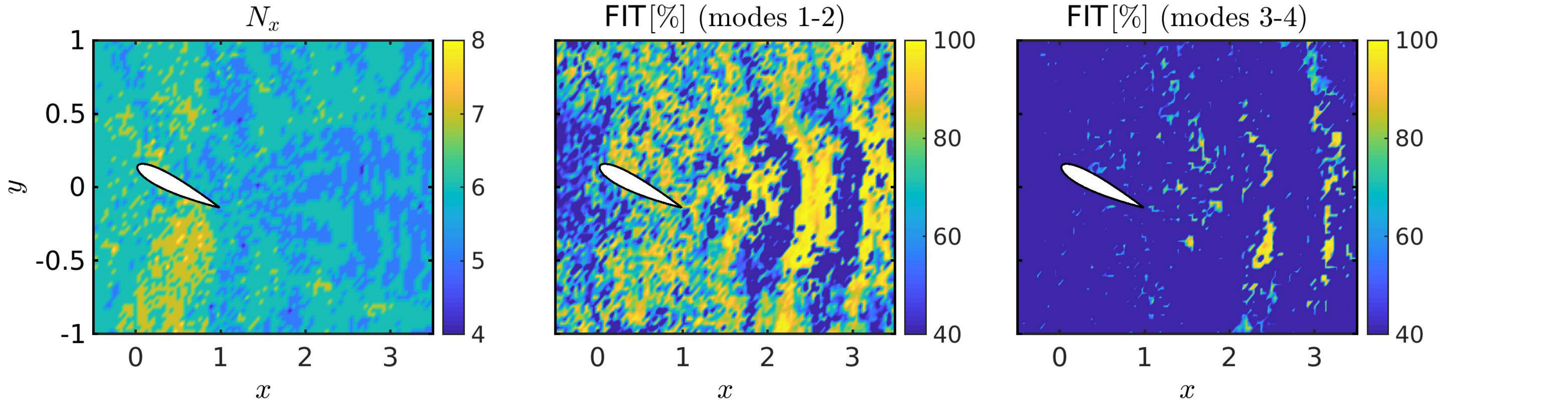}}
\\
\subfloat{\includegraphics[width=\textwidth]{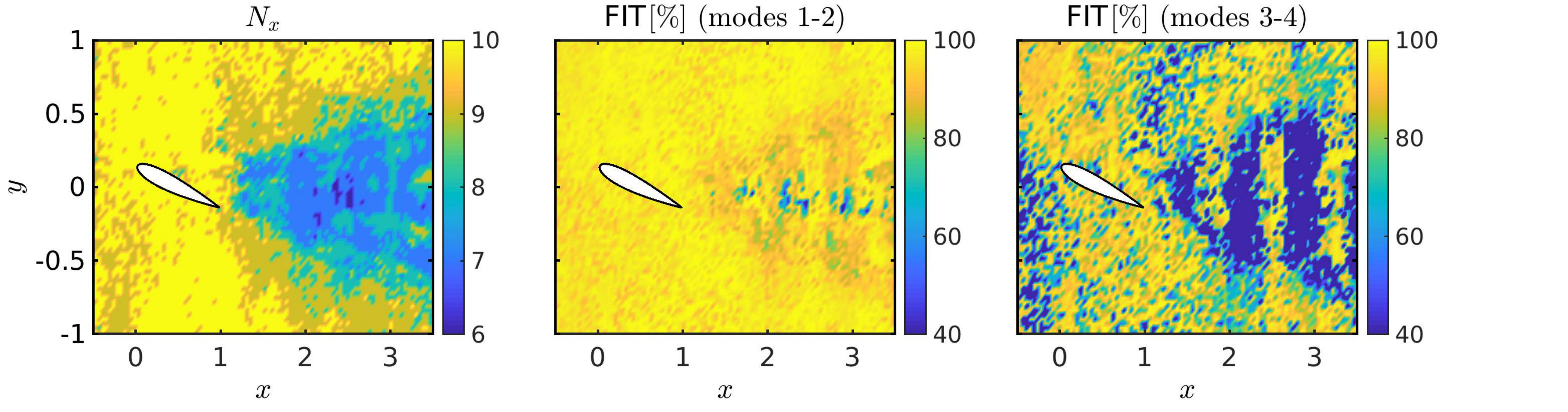}}
\caption{(Left) Optimal number of states estimated by the \texttt{N4SID} algorithm for 4 POD modes and (right) $\mathsf{FIT}[\%]$ for such models. The models are obtained using (top) the streamwise $s=u$, (middle) the cross-streamwise $s=v$ and (bottom) both the streamwise and cross-streamwise $s=\left[u, v\right]$ components of the oscillatory velocity, respectively.}
\label{fig:FIT_POD_4_optimal}
\end{figure}

We now turn our focus to models with $k=4$ POD modes. The task of estimation now becomes 
more difficult. The algorithm is provided with the time signal at just one point in 
the wake, and must extract the vector of 4 
POD coefficients $\mathbf{Y}_e=[y_{1,e},y_{2,e}, y_{3,e},y_{4,e}]^\top$. 
As before, we assessed the effects of dataset length and time shift to make sure that the 
results are independent of the choice of the learning dataset. Indeed, we confirmed 
that this is the case (results not shown for brevity). {Additionally, we confirmed that, as for models with $k=2$ POD modes, adding more states leads to better performance of the estimators. For conciseness, the influence of the sensor location is directly illustrated using models composed of the optimal number of states only.} 

Fig.~\ref{fig:FIT_POD_4_optimal} shows the optimum number of states (left column) as well as the
corresponding fits for the first and third coefficients (middle and right columns, respectively). For a model utilising solely the $u$ or $v$ velocity components, the algorithm selects a lower number of states in the aerofoil wake resulting in relatively low fit; a similar result was also found for a model with 2 POD modes (see Fig.~\ref{fig:FIT_POD_2_optimal}). A higher number of states is selected on top and bottom of the aerofoil. The fit is spotty, with no clear pattern. Overall, few sensor locations achieve a successful estimation when only the streamwise or
cross-stream component of the velocity are considered. The fit is slightly better when the model uses the $v$ velocity component. 

When both velocity components are used, the lower number of states clearly demarcates the wake. The fit for the first mode has high values in almost every point in the domain; there are only a few localised spots in the wake with low values. This pattern is very similar to that of Fig~\ref{fig:FIT_POD_2_optimal} (bottom). The coefficient of the third mode is much
better captured, but a stripy fit pattern is detected. 

{\citet{Overschee_deMoor_1996} realised that the \texttt{N4SID} algorithm is sensitive 
to the scaling of inputs and outputs. To explore this effect, we normalised the inputs-outputs of the system so that their variance was unit. The results
obtained for the normalised case are very similar to the ones reported in Fig.~\ref{fig:FIT_POD_4_optimal}, showing that in this case the algorithm is insensitive to
scaling (results not shown for brevity).}

\begin{figure}
\centering
\subfloat{\includegraphics[width=0.9\textwidth]{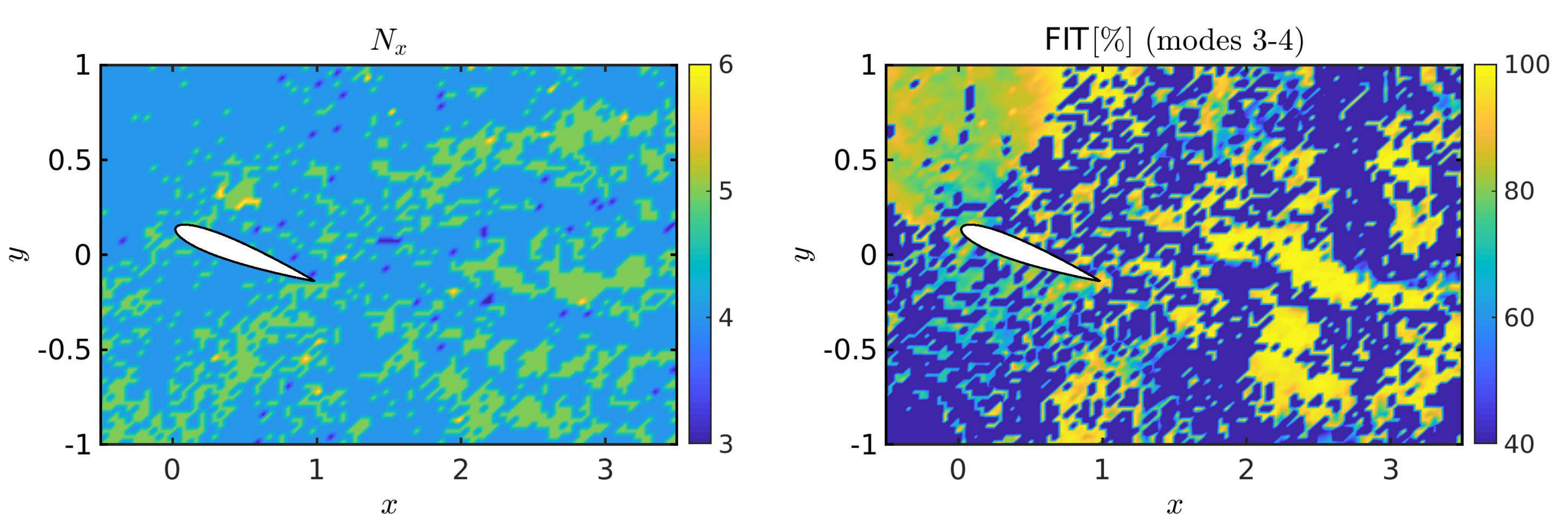}}
\\
\subfloat{\includegraphics[width=0.9\textwidth]{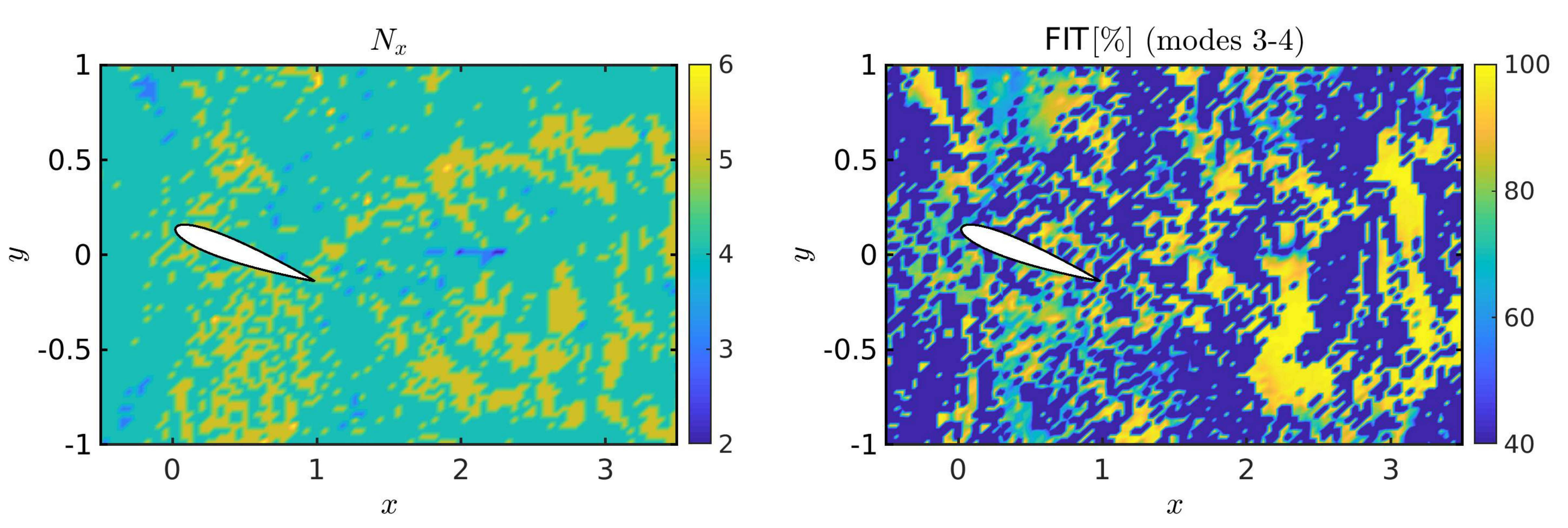}}
\\
\subfloat{\includegraphics[width=0.9\textwidth]{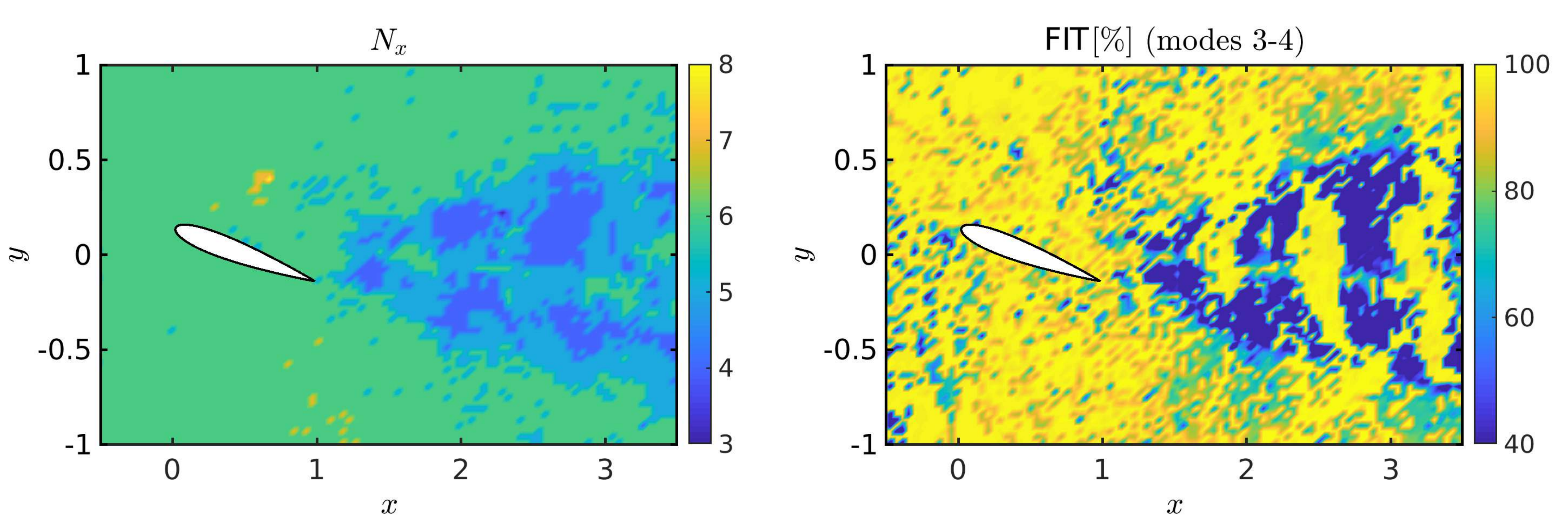}}
\caption{(Left) Optimal number of states estimated by the \texttt{N4SID} algorithm for the 3rd 
and 4th POD modes and (right) $\mathsf{FIT}[\%]$ for such models. The models are obtained using (top) the streamwise $s=u$, (middle) the cross-streamwise $s=v$ and (bottom) both the streamwise and cross-streamwise $s=\left[u, v\right]^\top$ components of the oscillatory velocity, respectively.}
\label{fig:FIT_POD_34_optimal}
\end{figure}

{In a further attempt to improve the performance of the estimator, we have also computed models with output the third and fourth POD coefficients only, 
i.e. $\mathbf{Y}_e=\left[y_{e,3}, y_{e,4}\right]$. We know that any interaction between the main shedding
mode and the harmonics is non-linear. The estimator, being linear, can only capture this interaction through the sensor. The outputs 1-2 and 3-4 of the estimator can therefore be decoupled and can be represented by independent estimators. 

Fig.~\ref{fig:FIT_POD_34_optimal} shows the optimal number of states (left column) and the
corresponding fit for the third coefficient (right column). A clear improvement is obtained 
for models using solely $u$ and $v.$ The working areas are now larger and better defined
compared to Fig.~\ref{fig:FIT_POD_4_optimal}. Additionally, we observe that the working/non-working
locations present a similar pattern to the ones for modes 1-2 (Fig.~\ref{fig:FIT_POD_2_optimal}), but inverted. This can be explained again by reference to  Fig.~\ref{fig:TKE ratio}; when the signature of the main frequency in the sensor is very strong 
compared to the one of the first harmonic, it is more difficult for the estimator to filter it out. For estimators using both components of the velocity field, a stripy pattern is detected again. Although there is a small improvement, the fit distribution shows a very similar pattern to the one obtained for the the coupled case (Fig.~\ref{fig:FIT_POD_4_optimal}). }

\section{\label{sec:conclusions}Summary and conclusions}
The central objective of this paper was to derive a data-driven, dynamic observer that can reconstruct the full 2D velocity field from a single point measurement in the wake of an aerofoil. In the present work, the data were produced by running a CFD simulation, but equally well they could have been obtained using PIV. To reduce the spatial dimensionality of the data, we constructed a reduced order model based on POD. 

The system identification algorithm, \texttt{N4SID}, was then applied in conjunction with the dynamic observer to produce the final model. The effect of a number of parameters on the performance of the algorithm was assessed and the results were related to the flow dynamics and mode interaction in the wake. More specifically, the number of POD modes, length of learning dataset, input velocity component(s), observer initialisation, sensor location and number of states of the estimator, were all investigated. Even when initialising the testing dataset with a state of $\textbf{X}_e(t)=0$, the algorithm was able to estimate the underlying coefficients of the model, though with an added delay. Average fit values of over 90\%  were shown to be achievable, especially for the first and second POD modes, that capture most of the energy. Best results are obtained when both $u$ and $v$ velocity components are measured. 

Overall the algorithm was found to achieve the reconstruction of flow  fields with an impressive degree of accuracy. Considering the unsteady and non-linear nature of the flow, this is especially surprising and opens the approach up to a wide variety of applications~\cite{leclercq2019linear}.





\bibliography{estimation} 

\begin{thebibliography}{37}%
\makeatletter
\providecommand \@ifxundefined [1]{%
 \@ifx{#1\undefined}
}%
\providecommand \@ifnum [1]{%
 \ifnum #1\expandafter \@firstoftwo
 \else \expandafter \@secondoftwo
 \fi
}%
\providecommand \@ifx [1]{%
 \ifx #1\expandafter \@firstoftwo
 \else \expandafter \@secondoftwo
 \fi
}%
\providecommand \natexlab [1]{#1}%
\providecommand \enquote  [1]{``#1''}%
\providecommand \bibnamefont  [1]{#1}%
\providecommand \bibfnamefont [1]{#1}%
\providecommand \citenamefont [1]{#1}%
\providecommand \href@noop [0]{\@secondoftwo}%
\providecommand \href [0]{\begingroup \@sanitize@url \@href}%
\providecommand \@href[1]{\@@startlink{#1}\@@href}%
\providecommand \@@href[1]{\endgroup#1\@@endlink}%
\providecommand \@sanitize@url [0]{\catcode `\\12\catcode `\$12\catcode
  `\&12\catcode `\#12\catcode `\^12\catcode `\_12\catcode `\%12\relax}%
\providecommand \@@startlink[1]{}%
\providecommand \@@endlink[0]{}%
\providecommand \url  [0]{\begingroup\@sanitize@url \@url }%
\providecommand \@url [1]{\endgroup\@href {#1}{\urlprefix }}%
\providecommand \urlprefix  [0]{URL }%
\providecommand \Eprint [0]{\href }%
\providecommand \doibase [0]{http://dx.doi.org/}%
\providecommand \selectlanguage [0]{\@gobble}%
\providecommand \bibinfo  [0]{\@secondoftwo}%
\providecommand \bibfield  [0]{\@secondoftwo}%
\providecommand \translation [1]{[#1]}%
\providecommand \BibitemOpen [0]{}%
\providecommand \bibitemStop [0]{}%
\providecommand \bibitemNoStop [0]{.\EOS\space}%
\providecommand \EOS [0]{\spacefactor3000\relax}%
\providecommand \BibitemShut  [1]{\csname bibitem#1\endcsname}%
\let\auto@bib@innerbib\@empty
\bibitem [{\citenamefont {Holmes}\ \emph {et~al.}(2012)\citenamefont {Holmes},
  \citenamefont {Lumley}, \citenamefont {Berkooz},\ and\ \citenamefont
  {Rowley}}]{holmes_lumley_berkooz_rowley_2012}%
  \BibitemOpen
  \bibfield  {author} {\bibinfo {author} {\bibfnamefont {P.}~\bibnamefont
  {Holmes}}, \bibinfo {author} {\bibfnamefont {J.~L.}\ \bibnamefont {Lumley}},
  \bibinfo {author} {\bibfnamefont {G.}~\bibnamefont {Berkooz}}, \ and\
  \bibinfo {author} {\bibfnamefont {C.~W.}\ \bibnamefont {Rowley}},\
  }\href@noop {} {\emph {\bibinfo {title} {Turbulence, Coherent Structures,
  Dynamical Systems and Symmetry}}},\ \bibinfo {edition} {2nd}\ ed.,\ Cambridge
  Monographs on Mechanics\ (\bibinfo  {publisher} {Cambridge University
  Press},\ \bibinfo {year} {2012})\BibitemShut {NoStop}%
\bibitem [{\citenamefont {Willcox}(2006)}]{Willcox_2006}%
  \BibitemOpen
  \bibfield  {author} {\bibinfo {author} {\bibfnamefont {K.}~\bibnamefont
  {Willcox}},\ }\href {\doibase doi.org/10.1016/j.compfluid.2004.11.006}
  {\bibfield  {journal} {\bibinfo  {journal} {Computers \& Fluids}\ }\textbf
  {\bibinfo {volume} {35}},\ \bibinfo {pages} {208 } (\bibinfo {year}
  {2006})}\BibitemShut {NoStop}%
\bibitem [{\citenamefont {Everson}\ and\ \citenamefont
  {Sirovich}(1995)}]{Everson_Sirovich_1995}%
  \BibitemOpen
  \bibfield  {author} {\bibinfo {author} {\bibfnamefont {R.}~\bibnamefont
  {Everson}}\ and\ \bibinfo {author} {\bibfnamefont {L.}~\bibnamefont
  {Sirovich}},\ }\href@noop {} {\bibfield  {journal} {\bibinfo  {journal} {J.
  Opt. Soc. Am. A}\ }\textbf {\bibinfo {volume} {12}},\ \bibinfo {pages} {1657}
  (\bibinfo {year} {1995})}\BibitemShut {NoStop}%
\bibitem [{\citenamefont {Yildirim}\ \emph {et~al.}(2009)\citenamefont
  {Yildirim}, \citenamefont {Chryssostomidis},\ and\ \citenamefont
  {Karniadakis}}]{YILDIRIM_el_al_2009}%
  \BibitemOpen
  \bibfield  {author} {\bibinfo {author} {\bibfnamefont {B.}~\bibnamefont
  {Yildirim}}, \bibinfo {author} {\bibfnamefont {C.}~\bibnamefont
  {Chryssostomidis}}, \ and\ \bibinfo {author} {\bibfnamefont {G.}~\bibnamefont
  {Karniadakis}},\ }\href {\doibase doi.org/10.1016/j.ocemod.2009.01.001}
  {\bibfield  {journal} {\bibinfo  {journal} {Ocean Modelling}\ }\textbf
  {\bibinfo {volume} {27}},\ \bibinfo {pages} {160 } (\bibinfo {year}
  {2009})}\BibitemShut {NoStop}%
\bibitem [{\citenamefont {Semaan}(2017)}]{SEMAAN_2017}%
  \BibitemOpen
  \bibfield  {author} {\bibinfo {author} {\bibfnamefont {R.}~\bibnamefont
  {Semaan}},\ }\href {\doibase doi.org/10.1016/j.compfluid.2017.10.002}
  {\bibfield  {journal} {\bibinfo  {journal} {Computers \& Fluids}\ }\textbf
  {\bibinfo {volume} {159}},\ \bibinfo {pages} {167 } (\bibinfo {year}
  {2017})}\BibitemShut {NoStop}%
\bibitem [{\citenamefont {McKeon}\ and\ \citenamefont
  {Sharma}(2010)}]{mckeon2010critical}%
  \BibitemOpen
  \bibfield  {author} {\bibinfo {author} {\bibfnamefont {B.~J.}\ \bibnamefont
  {McKeon}}\ and\ \bibinfo {author} {\bibfnamefont {A.~S.}\ \bibnamefont
  {Sharma}},\ }\href@noop {} {\bibfield  {journal} {\bibinfo  {journal} {J.
  Fluid Mech}\ }\textbf {\bibinfo {volume} {658}},\ \bibinfo {pages} {336}
  (\bibinfo {year} {2010})}\BibitemShut {NoStop}%
\bibitem [{\citenamefont {Beneddine}\ \emph {et~al.}(2017)\citenamefont
  {Beneddine}, \citenamefont {Yegavian}, \citenamefont {Sipp},\ and\
  \citenamefont {Leclaire}}]{beneddine2017unsteady}%
  \BibitemOpen
  \bibfield  {author} {\bibinfo {author} {\bibfnamefont {S.}~\bibnamefont
  {Beneddine}}, \bibinfo {author} {\bibfnamefont {R.}~\bibnamefont {Yegavian}},
  \bibinfo {author} {\bibfnamefont {D.}~\bibnamefont {Sipp}}, \ and\ \bibinfo
  {author} {\bibfnamefont {B.}~\bibnamefont {Leclaire}},\ }\href@noop {}
  {\bibfield  {journal} {\bibinfo  {journal} {J. Fluid Mech}\ }\textbf
  {\bibinfo {volume} {824}},\ \bibinfo {pages} {174} (\bibinfo {year}
  {2017})}\BibitemShut {NoStop}%
\bibitem [{\citenamefont {G\'{o}mez}\ \emph {et~al.}(2016)\citenamefont
  {G\'{o}mez}, \citenamefont {Sharma},\ and\ \citenamefont
  {Blackburn}}]{gomez2016estimation}%
  \BibitemOpen
  \bibfield  {author} {\bibinfo {author} {\bibfnamefont {F.}~\bibnamefont
  {G\'{o}mez}}, \bibinfo {author} {\bibfnamefont {A.~S.}\ \bibnamefont
  {Sharma}}, \ and\ \bibinfo {author} {\bibfnamefont {H.~M.}\ \bibnamefont
  {Blackburn}},\ }\href@noop {} {\bibfield  {journal} {\bibinfo  {journal} {J.
  Fluid Mech}\ }\textbf {\bibinfo {volume} {804}} (\bibinfo {year}
  {2016})}\BibitemShut {NoStop}%
\bibitem [{\citenamefont {Thomareis}\ and\ \citenamefont
  {Papadakis}(2018)}]{Thomareis_Papadakis_2018}%
  \BibitemOpen
  \bibfield  {author} {\bibinfo {author} {\bibfnamefont {N.}~\bibnamefont
  {Thomareis}}\ and\ \bibinfo {author} {\bibfnamefont {G.}~\bibnamefont
  {Papadakis}},\ }\href {\doibase 10.1103/PhysRevFluids.3.073901} {\bibfield
  {journal} {\bibinfo  {journal} {Phys. Rev. Fluids}\ }\textbf {\bibinfo
  {volume} {3}},\ \bibinfo {pages} {073901} (\bibinfo {year}
  {2018})}\BibitemShut {NoStop}%
\bibitem [{\citenamefont {Illingworth}\ \emph {et~al.}(2018)\citenamefont
  {Illingworth}, \citenamefont {Monty},\ and\ \citenamefont
  {Marusic}}]{illingworth_monty_marusic_2018}%
  \BibitemOpen
  \bibfield  {author} {\bibinfo {author} {\bibfnamefont {S.~J.}\ \bibnamefont
  {Illingworth}}, \bibinfo {author} {\bibfnamefont {J.~P.}\ \bibnamefont
  {Monty}}, \ and\ \bibinfo {author} {\bibfnamefont {I.}~\bibnamefont
  {Marusic}},\ }\href {\doibase 10.1017/jfm.2018.129} {\bibfield  {journal}
  {\bibinfo  {journal} {J. Fluid Mech}\ }\textbf {\bibinfo {volume} {842}},\
  \bibinfo {pages} {146–162} (\bibinfo {year} {2018})}\BibitemShut {NoStop}%
\bibitem [{\citenamefont {Brunton}\ \emph {et~al.}(2014)\citenamefont
  {Brunton}, \citenamefont {Tu}, \citenamefont {Bright},\ and\ \citenamefont
  {Kutz}}]{Brunton_et_al_2014}%
  \BibitemOpen
  \bibfield  {author} {\bibinfo {author} {\bibfnamefont {S.}~\bibnamefont
  {Brunton}}, \bibinfo {author} {\bibfnamefont {J.}~\bibnamefont {Tu}},
  \bibinfo {author} {\bibfnamefont {I.}~\bibnamefont {Bright}}, \ and\ \bibinfo
  {author} {\bibfnamefont {J.}~\bibnamefont {Kutz}},\ }\href@noop {} {\bibfield
   {journal} {\bibinfo  {journal} {SIAM Journal on Applied Dynamical Systems}\
  }\textbf {\bibinfo {volume} {13}},\ \bibinfo {pages} {1716} (\bibinfo {year}
  {2014})}\BibitemShut {NoStop}%
\bibitem [{\citenamefont {Manohar}\ \emph {et~al.}(2018)\citenamefont
  {Manohar}, \citenamefont {Brunton}, \citenamefont {Kutz},\ and\ \citenamefont
  {Brunton}}]{Manohar_et_al_2018}%
  \BibitemOpen
  \bibfield  {author} {\bibinfo {author} {\bibfnamefont {K.}~\bibnamefont
  {Manohar}}, \bibinfo {author} {\bibfnamefont {B.~W.}\ \bibnamefont
  {Brunton}}, \bibinfo {author} {\bibfnamefont {J.~N.}\ \bibnamefont {Kutz}}, \
  and\ \bibinfo {author} {\bibfnamefont {S.~L.}\ \bibnamefont {Brunton}},\
  }\href {\doibase 10.1109/MCS.2018.2810460} {\bibfield  {journal} {\bibinfo
  {journal} {IEEE Control Systems Magazine}\ }\textbf {\bibinfo {volume}
  {38}},\ \bibinfo {pages} {63} (\bibinfo {year} {2018})}\BibitemShut {NoStop}%
\bibitem [{\citenamefont {Adrian}(1979)}]{Adrian_1979}%
  \BibitemOpen
  \bibfield  {author} {\bibinfo {author} {\bibfnamefont {R.~J.}\ \bibnamefont
  {Adrian}},\ }\href@noop {} {\bibfield  {journal} {\bibinfo  {journal}
  {Physics of Fluids}\ }\textbf {\bibinfo {volume} {22}},\ \bibinfo {pages}
  {2065} (\bibinfo {year} {1979})}\BibitemShut {NoStop}%
\bibitem [{\citenamefont {Adrian}\ and\ \citenamefont
  {Moin}(1988)}]{adrian_moin_1988}%
  \BibitemOpen
  \bibfield  {author} {\bibinfo {author} {\bibfnamefont {R.~J.}\ \bibnamefont
  {Adrian}}\ and\ \bibinfo {author} {\bibfnamefont {P.}~\bibnamefont {Moin}},\
  }\href@noop {} {\bibfield  {journal} {\bibinfo  {journal} {J. Fluid Mech}\
  }\textbf {\bibinfo {volume} {190}} (\bibinfo {year} {1988})}\BibitemShut
  {NoStop}%
\bibitem [{\citenamefont {Kailath}\ \emph {et~al.}(2000)\citenamefont
  {Kailath}, \citenamefont {Hassibi},\ and\ \citenamefont
  {Sayed}}]{Kailath_Hassibi_Sayed_2000}%
  \BibitemOpen
  \bibfield  {author} {\bibinfo {author} {\bibfnamefont {T.}~\bibnamefont
  {Kailath}}, \bibinfo {author} {\bibfnamefont {B.}~\bibnamefont {Hassibi}}, \
  and\ \bibinfo {author} {\bibfnamefont {A.~H.}\ \bibnamefont {Sayed}},\
  }\href@noop {} {\emph {\bibinfo {title} {Linear estimation}}}\ (\bibinfo
  {publisher} {Prentice-Hall International},\ \bibinfo {year}
  {2000})\BibitemShut {NoStop}%
\bibitem [{\citenamefont {Gong}\ \emph {et~al.}(2019)\citenamefont {Gong},
  \citenamefont {Monty},\ and\ \citenamefont {Illingworth}}]{gong2019model}%
  \BibitemOpen
  \bibfield  {author} {\bibinfo {author} {\bibfnamefont {J.}~\bibnamefont
  {Gong}}, \bibinfo {author} {\bibfnamefont {J.~P.}\ \bibnamefont {Monty}}, \
  and\ \bibinfo {author} {\bibfnamefont {S.~J.}\ \bibnamefont {Illingworth}},\
  }\href@noop {} {\bibfield  {journal} {\bibinfo  {journal} {arXiv preprint
  arXiv:1905.00133}\ } (\bibinfo {year} {2019})}\BibitemShut {NoStop}%
\bibitem [{\citenamefont {Stengel}(1994)}]{Stengel_1994}%
  \BibitemOpen
  \bibfield  {author} {\bibinfo {author} {\bibfnamefont {R.~F.}\ \bibnamefont
  {Stengel}},\ }\href@noop {} {\emph {\bibinfo {title} {Optimal Control and
  Estimation}}}\ (\bibinfo  {publisher} {Dover Publications Inc, 2nd edition,
  New York},\ \bibinfo {year} {1994})\BibitemShut {NoStop}%
\bibitem [{\citenamefont {Ljung}(1999)}]{ljung1999system}%
  \BibitemOpen
  \bibfield  {author} {\bibinfo {author} {\bibfnamefont {L.}~\bibnamefont
  {Ljung}},\ }\href@noop {} {\emph {\bibinfo {title} {System identification:
  theory for the user}}}\ (\bibinfo  {publisher} {Prentice-Hall PTR},\ \bibinfo
  {year} {1999})\BibitemShut {NoStop}%
\bibitem [{\citenamefont {Herv{\'e}}\ \emph {et~al.}(2012)\citenamefont
  {Herv{\'e}}, \citenamefont {Sipp}, \citenamefont {Schmid},\ and\
  \citenamefont {Samuelides}}]{herve2012physics}%
  \BibitemOpen
  \bibfield  {author} {\bibinfo {author} {\bibfnamefont {A.}~\bibnamefont
  {Herv{\'e}}}, \bibinfo {author} {\bibfnamefont {D.}~\bibnamefont {Sipp}},
  \bibinfo {author} {\bibfnamefont {P.~J.}\ \bibnamefont {Schmid}}, \ and\
  \bibinfo {author} {\bibfnamefont {M.}~\bibnamefont {Samuelides}},\
  }\href@noop {} {\bibfield  {journal} {\bibinfo  {journal} {J. Fluid Mech}\
  }\textbf {\bibinfo {volume} {702}},\ \bibinfo {pages} {26} (\bibinfo {year}
  {2012})}\BibitemShut {NoStop}%
\bibitem [{\citenamefont {Juillet}\ \emph {et~al.}(2013)\citenamefont
  {Juillet}, \citenamefont {Schmid},\ and\ \citenamefont
  {Huerre}}]{juillet2013control}%
  \BibitemOpen
  \bibfield  {author} {\bibinfo {author} {\bibfnamefont {F.}~\bibnamefont
  {Juillet}}, \bibinfo {author} {\bibfnamefont {P.~J.}\ \bibnamefont {Schmid}},
  \ and\ \bibinfo {author} {\bibfnamefont {P.}~\bibnamefont {Huerre}},\
  }\href@noop {} {\bibfield  {journal} {\bibinfo  {journal} {J. Fluid Mech}\
  }\textbf {\bibinfo {volume} {725}},\ \bibinfo {pages} {522} (\bibinfo {year}
  {2013})}\BibitemShut {NoStop}%
\bibitem [{\citenamefont {Juillet}\ \emph {et~al.}(2014)\citenamefont
  {Juillet}, \citenamefont {McKeon},\ and\ \citenamefont
  {Schmid}}]{juillet2014experimental}%
  \BibitemOpen
  \bibfield  {author} {\bibinfo {author} {\bibfnamefont {F.}~\bibnamefont
  {Juillet}}, \bibinfo {author} {\bibfnamefont {B.~J.}\ \bibnamefont {McKeon}},
  \ and\ \bibinfo {author} {\bibfnamefont {P.~J.}\ \bibnamefont {Schmid}},\
  }\href@noop {} {\bibfield  {journal} {\bibinfo  {journal} {J. Fluid Mech}\
  }\textbf {\bibinfo {volume} {752}},\ \bibinfo {pages} {296} (\bibinfo {year}
  {2014})}\BibitemShut {NoStop}%
\bibitem [{\citenamefont {Gautier}\ and\ \citenamefont
  {Aider}(2014)}]{gautier2014feed}%
  \BibitemOpen
  \bibfield  {author} {\bibinfo {author} {\bibfnamefont {N.}~\bibnamefont
  {Gautier}}\ and\ \bibinfo {author} {\bibfnamefont {J.~L.}\ \bibnamefont
  {Aider}},\ }\href@noop {} {\bibfield  {journal} {\bibinfo  {journal} {J.
  Fluid Mech}\ }\textbf {\bibinfo {volume} {759}},\ \bibinfo {pages} {181}
  (\bibinfo {year} {2014})}\BibitemShut {NoStop}%
\bibitem [{\citenamefont {Guzm{\'a}n-I{\~n}igo}\ \emph
  {et~al.}(2014)\citenamefont {Guzm{\'a}n-I{\~n}igo}, \citenamefont {Sipp},\
  and\ \citenamefont {Schmid}}]{guzman2014dynamic}%
  \BibitemOpen
  \bibfield  {author} {\bibinfo {author} {\bibfnamefont {J.}~\bibnamefont
  {Guzm{\'a}n-I{\~n}igo}}, \bibinfo {author} {\bibfnamefont {D.}~\bibnamefont
  {Sipp}}, \ and\ \bibinfo {author} {\bibfnamefont {P.~J.}\ \bibnamefont
  {Schmid}},\ }\href@noop {} {\bibfield  {journal} {\bibinfo  {journal} {J.
  Fluid Mech}\ }\textbf {\bibinfo {volume} {758}},\ \bibinfo {pages} {728}
  (\bibinfo {year} {2014})}\BibitemShut {NoStop}%
\bibitem [{\citenamefont {Guzm{\'a}n-I{\~n}igo}\ \emph
  {et~al.}(2016)\citenamefont {Guzm{\'a}n-I{\~n}igo}, \citenamefont {Sipp},\
  and\ \citenamefont {Schmid}}]{guzman2016recovery}%
  \BibitemOpen
  \bibfield  {author} {\bibinfo {author} {\bibfnamefont {J.}~\bibnamefont
  {Guzm{\'a}n-I{\~n}igo}}, \bibinfo {author} {\bibfnamefont {D.}~\bibnamefont
  {Sipp}}, \ and\ \bibinfo {author} {\bibfnamefont {P.~J.}\ \bibnamefont
  {Schmid}},\ }\href@noop {} {\bibfield  {journal} {\bibinfo  {journal} {J.
  Fluid Mech}\ }\textbf {\bibinfo {volume} {797}},\ \bibinfo {pages} {130}
  (\bibinfo {year} {2016})}\BibitemShut {NoStop}%
\bibitem [{\citenamefont {Van~Overschee}\ and\ \citenamefont
  {De~Moor}(1994)}]{van94}%
  \BibitemOpen
  \bibfield  {author} {\bibinfo {author} {\bibfnamefont {P.}~\bibnamefont
  {Van~Overschee}}\ and\ \bibinfo {author} {\bibfnamefont {B.}~\bibnamefont
  {De~Moor}},\ }\href@noop {} {\bibfield  {journal} {\bibinfo  {journal}
  {Automatica}\ }\textbf {\bibinfo {volume} {30}},\ \bibinfo {pages} {75}
  (\bibinfo {year} {1994})}\BibitemShut {NoStop}%
\bibitem [{\citenamefont {Qin}(2006)}]{qin06}%
  \BibitemOpen
  \bibfield  {author} {\bibinfo {author} {\bibfnamefont {S.}~\bibnamefont
  {Qin}},\ }\href@noop {} {\bibfield  {journal} {\bibinfo  {journal} {Comp. \&
  Chem. Eng.}\ }\textbf {\bibinfo {volume} {30}},\ \bibinfo {pages} {1502}
  (\bibinfo {year} {2006})}\BibitemShut {NoStop}%
\bibitem [{\citenamefont {Van~Overschee}\ and\ \citenamefont
  {De~Moor}(1996)}]{Overschee_deMoor_1996}%
  \BibitemOpen
  \bibfield  {author} {\bibinfo {author} {\bibfnamefont {P.}~\bibnamefont
  {Van~Overschee}}\ and\ \bibinfo {author} {\bibfnamefont {B.}~\bibnamefont
  {De~Moor}},\ }\href@noop {} {\emph {\bibinfo {title} {Subspace Identification
  for Linear Systems; Theory — Implementation — Applications}}}\ (\bibinfo
  {publisher} {Springer US},\ \bibinfo {year} {1996})\BibitemShut {NoStop}%
\bibitem [{\citenamefont {Loiseau}\ \emph {et~al.}(2018)\citenamefont
  {Loiseau}, \citenamefont {Noack},\ and\ \citenamefont
  {Brunton}}]{loiseau2018sparse}%
  \BibitemOpen
  \bibfield  {author} {\bibinfo {author} {\bibfnamefont {J.~C.}\ \bibnamefont
  {Loiseau}}, \bibinfo {author} {\bibfnamefont {B.~R.}\ \bibnamefont {Noack}},
  \ and\ \bibinfo {author} {\bibfnamefont {S.~L.}\ \bibnamefont {Brunton}},\
  }\href@noop {} {\bibfield  {journal} {\bibinfo  {journal} {J. Fluid Mech}\
  }\textbf {\bibinfo {volume} {844}},\ \bibinfo {pages} {459} (\bibinfo {year}
  {2018})}\BibitemShut {NoStop}%
\bibitem [{\citenamefont {Korda}\ and\ \citenamefont
  {Mezi{\'c}}(2018)}]{korda2018linear}%
  \BibitemOpen
  \bibfield  {author} {\bibinfo {author} {\bibfnamefont {M.}~\bibnamefont
  {Korda}}\ and\ \bibinfo {author} {\bibfnamefont {I.}~\bibnamefont
  {Mezi{\'c}}},\ }\href@noop {} {\bibfield  {journal} {\bibinfo  {journal}
  {Automatica}\ }\textbf {\bibinfo {volume} {93}},\ \bibinfo {pages} {149}
  (\bibinfo {year} {2018})}\BibitemShut {NoStop}%
\bibitem [{\citenamefont {Chaturantabut}\ and\ \citenamefont
  {Sorensen}(2010)}]{chaturantabut2010nonlinear}%
  \BibitemOpen
  \bibfield  {author} {\bibinfo {author} {\bibfnamefont {S.}~\bibnamefont
  {Chaturantabut}}\ and\ \bibinfo {author} {\bibfnamefont {D.~C.}\ \bibnamefont
  {Sorensen}},\ }\href@noop {} {\bibfield  {journal} {\bibinfo  {journal} {SIAM
  Journal on Scientific Computing}\ }\textbf {\bibinfo {volume} {32}},\
  \bibinfo {pages} {2737} (\bibinfo {year} {2010})}\BibitemShut {NoStop}%
\bibitem [{\citenamefont {Fosas~de Pando}\ \emph {et~al.}(2016)\citenamefont
  {Fosas~de Pando}, \citenamefont {Schmid},\ and\ \citenamefont
  {Sipp}}]{fosas2016nonlinear}%
  \BibitemOpen
  \bibfield  {author} {\bibinfo {author} {\bibfnamefont {M.}~\bibnamefont
  {Fosas~de Pando}}, \bibinfo {author} {\bibfnamefont {P.~J.}\ \bibnamefont
  {Schmid}}, \ and\ \bibinfo {author} {\bibfnamefont {D.}~\bibnamefont
  {Sipp}},\ }\href@noop {} {\bibfield  {journal} {\bibinfo  {journal} {J. Comp.
  Phys.}\ }\textbf {\bibinfo {volume} {324}},\ \bibinfo {pages} {194} (\bibinfo
  {year} {2016})}\BibitemShut {NoStop}%
\bibitem [{\citenamefont {Penrose}(1955)}]{penrose1955generalized}%
  \BibitemOpen
  \bibfield  {author} {\bibinfo {author} {\bibfnamefont {R.}~\bibnamefont
  {Penrose}},\ }\href@noop {} {\bibfield  {journal} {\bibinfo  {journal}
  {Mathematical proceedings of the Cambridge philosophical society}\ }\textbf
  {\bibinfo {volume} {51}},\ \bibinfo {pages} {406} (\bibinfo {year}
  {1955})}\BibitemShut {NoStop}%
\bibitem [{\citenamefont {Antoulas}(2005)}]{antoulas05}%
  \BibitemOpen
  \bibfield  {author} {\bibinfo {author} {\bibfnamefont {A.}~\bibnamefont
  {Antoulas}},\ }\href@noop {} {\emph {\bibinfo {title} {Approximation of
  large-scale dynamical systems}}}\ (\bibinfo  {publisher} {SIAM Publishing},\
  \bibinfo {year} {2005})\BibitemShut {NoStop}%
\bibitem [{\citenamefont {Van~Overschee}\ and\ \citenamefont
  {De~Moor}(1995)}]{van95}%
  \BibitemOpen
  \bibfield  {author} {\bibinfo {author} {\bibfnamefont {P.}~\bibnamefont
  {Van~Overschee}}\ and\ \bibinfo {author} {\bibfnamefont {B.}~\bibnamefont
  {De~Moor}},\ }\href@noop {} {\bibfield  {journal} {\bibinfo  {journal}
  {Automatica}\ }\textbf {\bibinfo {volume} {31}},\ \bibinfo {pages} {1853}
  (\bibinfo {year} {1995})}\BibitemShut {NoStop}%
\bibitem [{\citenamefont {Zhang}\ and\ \citenamefont
  {Samtaney}(2016)}]{zhang2016biglobal}%
  \BibitemOpen
  \bibfield  {author} {\bibinfo {author} {\bibfnamefont {W.}~\bibnamefont
  {Zhang}}\ and\ \bibinfo {author} {\bibfnamefont {R.}~\bibnamefont
  {Samtaney}},\ }\href@noop {} {\bibfield  {journal} {\bibinfo  {journal}
  {Physics of Fluids}\ }\textbf {\bibinfo {volume} {28}},\ \bibinfo {pages}
  {044105} (\bibinfo {year} {2016})}\BibitemShut {NoStop}%
\bibitem [{\citenamefont {Sipp}\ and\ \citenamefont
  {Lebedev}(2007)}]{Sipp_lebedev_2007}%
  \BibitemOpen
  \bibfield  {author} {\bibinfo {author} {\bibfnamefont {D.}~\bibnamefont
  {Sipp}}\ and\ \bibinfo {author} {\bibfnamefont {A.}~\bibnamefont {Lebedev}},\
  }\href@noop {} {\bibfield  {journal} {\bibinfo  {journal} {J. Fluid Mech}\
  }\textbf {\bibinfo {volume} {593}},\ \bibinfo {pages} {333–358} (\bibinfo
  {year} {2007})}\BibitemShut {NoStop}%
\bibitem [{\citenamefont {Leclercq}\ \emph {et~al.}(2019)\citenamefont
  {Leclercq}, \citenamefont {Demourant}, \citenamefont {Poussot-Vassal},\ and\
  \citenamefont {Sipp}}]{leclercq2019linear}%
  \BibitemOpen
  \bibfield  {author} {\bibinfo {author} {\bibfnamefont {C.}~\bibnamefont
  {Leclercq}}, \bibinfo {author} {\bibfnamefont {F.}~\bibnamefont {Demourant}},
  \bibinfo {author} {\bibfnamefont {C.}~\bibnamefont {Poussot-Vassal}}, \ and\
  \bibinfo {author} {\bibfnamefont {D.}~\bibnamefont {Sipp}},\ }\href@noop {}
  {\bibfield  {journal} {\bibinfo  {journal} {J. Fluid Mech}\ }\textbf
  {\bibinfo {volume} {868}},\ \bibinfo {pages} {26} (\bibinfo {year}
  {2019})}\BibitemShut {NoStop}%
\end{thebibliography}%


%
\end{document}